\documentclass[a4paper,12pt,twoside,english]{article}
\usepackage[latin1]{inputenc}
\usepackage{parskip}               
\usepackage{graphicx,color}
\usepackage{wrapfig,rotating}
\usepackage{array}
\usepackage{hyperref}
\usepackage{amsfonts}       
\usepackage{amsmath}        
\usepackage{amssymb}        
\usepackage{hhline}
\usepackage{feynmp}
\usepackage{multirow}
\usepackage{subfigure}
\usepackage{cite}                  
\usepackage[ , ,bf,it]{caption}    
\usepackage{hyperref}

\newlength{\dinwidth}
\newlength{\dinmargin}
\graphicspath{
  {figs/}
}

\newcommand{\ltsim}{\protect\raisebox{-0.5ex}{$\:\stackrel{\textstyle <}
	{\sim}\:$}}

\newcommand{\e}[1][]{\mathrm{e}^{#1}}   

\newcommand{\ifb}{\,\mathrm{\,fb}^{-1}}

\newcommand{\MeV}{\ensuremath{{\mathrm{MeV}}}}
\newcommand{\GeV}{\ensuremath{{\mathrm{GeV}}}}

\newcommand{\tQ}{t}

\newcommand{\Rslash}   {\ensuremath{ R\kern-0.5em\slash   }}

\definecolor{dblack}   {rgb}{0.00, 0.00, 0.00}  
\definecolor{mgrey}    {rgb}{0.45, 0.45, 0.50}  
\definecolor{rred}     {rgb}{1.00, 0.00, 0.00}  
\definecolor{dgreen}   {rgb}{0.05, 0.65, 0.00}  
\definecolor{bblue}    {rgb}{0.10, 0.10, 0.96}  
\newcommand{\eeto}    {\mbox{$ {\, \mathrm e}^+ {\mathrm e}^- \to             $}}
\def\MXN#1{\mbox{$ M_{\tilde{\chi}^0_#1}                                $}}
\def\MXC#1{\mbox{$ M_{\tilde{\chi}^{\pm}_#1}                            $}}

\def\XPM#1{\mbox{$ \tilde{\chi}^{\pm}_#1                                $}}
\def\XN#1{\mbox{$ \tilde{\chi}^0_#1                                     $}}

\def\p#1{\mbox{$ \mbox{\bf p}_1                                         $}}

\newcommand{\smur}    {\mbox{$ \tilde{\mu}_{\mathrm R}                     $}}

\newcommand{\snu}     {\mbox{$ \tilde\nu                                   $}}
\newcommand{\msnu}    {\mbox{$ m_{\tilde\nu}                               $}}

\newcommand{\stau}    {\mbox{$ \tilde{\tau}                                $}}
\newcommand{\stone}   {\mbox{$ \tilde{\tau}_1                              $}}

\newcommand{\MW}{\ensuremath{{M_W}}}

\newcommand{\mt}{\ensuremath{{m_t}}}
\newcommand{\sweff}{\ensuremath{\sin^2{\theta^\ell}_{{\rm eff}}}}
\newcommand{\als}{\ensuremath{\alpha_s}}

\def\Author#1{\begin{center}{ \sc #1} \end{center}}

\def\HU{Institut f\"ur Physik, Humboldt-Universit\"at zu Berlin, 12489 Berlin, Germany}
\def\ICREA{ICREA, 08010 Barcelona, Spain and IFAE, BIST, 08193 Bellaterra, Barcelona, Spain (on leave)}
\def\kek{High Energy Accelerator Research Organization (KEK), Tsukuba,
  Ibaraki, JAPAN  }

\def\Tsinghua{Center for High Energy Physics, Tsinghua University, Beijing, CHINA}
\def\Toyama{Department of Physics, University of Toyama, Toyama 930-8555, JAPAN}
\def\Seoul{Department of Physics and Astronomy, Seoul National
  University, Seoul 151-747,  \\ \hskip 0.4in   KOREA}
\def\DESY{DESY, Notkestrasse 85, 22607 Hamburg, GERMANY}
\def\Cornell{Laboratory for Elementary Particle Physics, Cornell
  University, Ithaca, NY 14853, \\ \hskip 0.4in USA  }
\def\Oklahoma{Department of Physics and Astronomy, University of Oklahoma, 
Norman, OK, 73019, USA}
\def\Oregon{Center for High Energy Physics, 1274 University of Oregon, 
Eugene, Oregon 97403-1274, USA}
\def\Orsay{LAL, Centre Scientifique d'Orsay, Universit\'e Paris-Sud, 
  F-91898 Orsay CEDEX, \\ \hskip 0.4in FRANCE }

\def\MPP{Max-Planck-Institut f\"ur Physik, F\"ohringer Ring 6, 80805 Munich, GERMANY}
\def\Tokyo{ICEPP, University of Tokyo, Hongo, Bunkyo-ku, Tokyo,
  113-0033, JAPAN}
\def\UTA{Department of Physics, University of Texas, Arlington, TX 76019, USA}
\def\Madrid{Campus of International Excellence UAM+CSIC, Cantoblanco, 28049, Madrid, Spain\\
Instituto de F\'isica Te\'orica (UAM/CSIC), Universidad Aut\'onoma de Madrid, Cantoblanco, 28049, Madrid, Spain\\
Instituto de F\'isica de Cantabria (CSIC-UC), 39005, Santander, Spain}
\def\Michigan{Michigan Center for Theoretical Physics, University of
  Michigan, Ann Arbor, \\ \hskip 0.4in MI 48109, USA}
\def\Berkeley{Department of Physics, University of California,
  Berkeley, CA 94720, USA}
\def\LBL{Theoretical Physics Group, Lawrence Berkeley National
  Laboratory, Berkeley, \\  \hskip 0.4in   CA 94720, USA}
\def\IPMU{Kavli Institute for the Physics and Mathematics of the
  Universe, \\  \hskip 0.4in   University of Tokyo, Kashiwa 277-8583, JAPAN}

\def\SLAC{SLAC,
    Stanford University, Menlo Park, CA 94025, USA}
\def\Kansas{Department of Physics and Astronomy, University of Kansas, Lawrence, KS 66045, USA}

\begin{document}
\begin{titlepage}
  \begin{flushleft}
    {\tt DESY 17-012} \\
    {\tt KEK Preprint 2016-60} \\
    {\tt SLAC-PUB-16916} \\
    {\tt LAL 17-017} \\
    {\tt MPP-2017-5} \\
    {\tt IFT-UAM/CSIC-17-008} \\
  \end{flushleft}

\vfill
  \begin{center}
    \begin{Large}
      {\bfseries \boldmath The Potential of the ILC\\
 for Discovering New Particles} \\
 \vspace{0.5cm}
           {\normalsize Document Supporting the ICFA Response Letter to the ILC Advisory Panel} 
    \end{Large}
  \end{center}

\vfill
\Author{LCC Physics Working Group}
\bigskip
\Author{Keisuke Fujii$^1$, Christophe
Grojean$^{2,3,4}$, Michael E. Peskin$^5$(conveners); 
Tim Barklow$^5$, Yuanning Gao$^6$,
Shinya Kanemura$^7$, Hyungdo Kim$^8$, Jenny List$^2$,
Mihoko Nojiri$^{1,9}$, Maxim Perelstein$^{10}$, Roman P\"oschl$^{11}$,  
J\"urgen Reuter$^2$, Frank Simon$^{12}$, 
Tomohiko Tanabe$^{13}$,
James D. Wells$^{14}$, Jaehoon Yu$^{15}$; Howard Baer$^{16}$, Mikael Berggren$^2$, Sven Heinemeyer$^{17}$, Suvi-Leena Lehtinen$^2$,  
Junping Tian$^{13}$, Graham Wilson$^{18}$, Jacqueline Yan$^{1}$; 
Hitoshi Murayama$^{9,19,20}$, 
James Brau$^{21}$}

\vfill
  \begin{abstract}
This paper addresses the question of whether the International Linear
Collider has the capability of discovering new particles that have not
already been discovered at the CERN Large Hadron Collider. We summarize 
the various paths to discovery offered by the ILC, and discuss them in
the context of three different scenarios: 1.~LHC does not discover any 
new particles, 2.~LHC discovers some new low mass states and 3.~LHC discovers
new heavy particles. We will show that in each case, ILC plays a critical
role in discovery of new phenomena and in pushing forward the frontiers of 
high-energy physics as well as our understanding of the universe in a manner 
which is highly complementary to that of LHC.

For the busy reader, a two-page executive summary is provided at the beginning
of the document.

  \end{abstract}
  
 \vfill

\newpage

\begin{raggedright}
\noindent $^1$ \kek \\
$^2$  \DESY \\
$^3$  \HU \\
$^4$ \ICREA \\
$^5$ \SLAC\\
$^6$ \Tsinghua \\
$^7$  \Toyama\\
$^8$ \Seoul \\
$^9$ \IPMU \\ 
$^{10}$ \Cornell\\
$^{11}$  \Orsay\\
$^{12}$ \MPP \\
$^{13}$  \Tokyo\\
$^{14}$  \Michigan \\
$^{15}$ \UTA\\
$^{16}$  \Oklahoma \\
$^{17}$  \Madrid \\
$^{18}$  \Kansas \\
$^{19}$ \Berkeley \\
$^{20}$ \LBL  \\
$^{21}$  \Oregon\\
\end{raggedright} 
\end{titlepage}

\section*{Executive Summary}
\label{sec:exsumm}
In this Executive Summary, we give the main conclusions of this report on the potential of the ILC for the discovery of new phenomena.
Throughout this report, numerical estimates of precision or reach are based on the 20-year plan for ILC operation presented in~\cite{Barklow:2015tja}, including $4000$\,fb$^{-1}$ of luminosity at $500$\,GeV.

The ILC discovery program exploits the joint power of direct searches with the potential to produce new particles and  precision measurements  which are able to detect virtual effects of new particles at higher mass scales. The latter will shed light on the structure of physics arising from 
beyond the Standard Model even if the associated new particles are too heavy to be produced directly at LHC or ILC. 
All these measurements will rely on the clean operating environment, 
low backgrounds, and adjustable beam energy and polarization provided by the ILC. 

We will discuss the discovery of new interactions in the following programs:
\begin{enumerate}
\item {\bf New properties of the Higgs boson:}
The ILC will be a {\it Higgs boson factory} which offers 
absolute, model-independent measurements of the Higgs boson couplings 
to Standard Model (SM) fermions and gauge bosons, 
most of them to better than 1\% precision.
These measurements would probe for modifications of the Higgs interactions arising from composite structure of the Higgs or from mixings with new particles, including new, heavier Higgs bosons.
In addition, the self-coupling of the Higgs boson can be measured to an accuracy of 27\% via reactions such as $e^+e^-\rightarrow ZHH$, improving to 10\% via $\nu\bar{\nu}HH$ at $1$\,TeV. 
This measurement is a critical test of the theory of electroweak baryogenesis, a leading contender for explaining the cosmic
matter-antimatter asymmetry.

\item {\bf New properties of the Top quark:} 
The ILC will be a precision top quark factory. 
Scans of the production threshold of $e^+e^-\rightarrow t\bar{t}$  
can determine the top quark mass to a precision of 
$50$\,MeV or better, including all theoretical uncertainties. 
A precision determination of $m_t$ plays a central role in global 
electroweak fits which are indirectly 
sensitive to new particles and new interactions.
Using polarized beams, the ILC can determine separately the top-quark left- and right-handed couplings to gauge bosons to 
the sub-percent level. Such measurements offer a huge discovery potential for
a variety of composite Higgs or extra-dimensional new physics models, even if
the scale of the new physics is in the tens of TeV range.

\item {\bf New force carriers:} By measuring distributions in $e^+e^-\rightarrow f\bar{f}$
production (where $f$ stands for different SM fermions), the ILC is 
sensitive to new force particles $Z^\prime$ with masses as high as $12$\,TeV. Via fermion pair production, either directly or by virtual effects, the ILC can also explore for other new physics resonances that occur in composite Higgs or extra-dimensional models.
These measurements are not unlike the first indirect observation of the
$Z$ boson at {\sc Petra} and {\sc Tristan} via virtual effects in fermion pair production.

\item {\bf Additional Higgs bosons:} In addition to the possibility to discover relatives of the Higgs boson
via studying the properties of the $125$-GeV particle, the ILC offers unique opportunities to discover
additional lighter Higgs bosons -- or, more generally, any weakly interacting light scalar or pseudo-scalar particle -- by their direct production.

\item {\bf Supersymmetric sisters of the Higgs boson:}
If supersymmetry (SUSY) is the way nature has chosen to generate the symmetry-breaking potential of the Higgs boson, then light sister particles of the Higgs boson -- higgsinos are required. It is very possible, even theoretically preferred, that these particles have masses in the range of $\sim 100$-$300$\,GeV (the lighter the better), while all other supersymmetric particles are heavier.
Such light higgsinos are difficult, perhaps impossible, to observe at 
LHC, but their discovery would be straightforward at ILC. In that case, ILC  would  be a {\it higgsino factory}, providing quantitative tests of the hypothesis of supersymmetry and its implications for unification of forces.

\item {\bf SUSY without loop-holes:}
A central prediction of supersymmetry is that sparticles couple with the same strength 
as their SM partners. 
Thus, the rates for production of SUSY particles are well predicted as a function of mass. 
Then the clean and well-defined conditions at the ILC guarantee either discovery or exclusion. 
This is not true at the LHC, where many scenarios with light SUSY particles can evade detection.

\item {\bf Discovering dark matter particles:} It is possible that particles of dark matter are being produced 
copiously at accelerators but are invisible to their detectors. 
To search for pair-production of invisible particles, one must hunt for an associated photon or 
gluon from initial-state radiation.  
Such searches at the ILC are complementary to those at the LHC, since they probe dark matter couplings 
to leptons rather than quarks.  Because of the simplicity and calculability of background 
reactions at the ILC, the ILC mass reach for discovery of dark matter particles is 
similar to that of LHC despite the difference in center of mass energy. 

\item {\bf Identifying the nature of dark matter:} The precision capabilities of the ILC are optimal to uncover which mechanisms are responsible for generating dark matter in the early universe: 
Is it thermal or non-thermal? Does entropy-dilution play a role? 
Are there super-WIMPs? Is there a WIMP-axion admixture? 
Dark matter production in the early universe may be 
much more intricate than the simple thermal WIMP miracle scenario 
and ILC can play a key role in gaining a more complete understanding.

\item {\bf Neutrinos:} 
Though some models for neutrino mass invoke particles with masses well beyond the energy of any 
realistic accelerator, other models generate neutrino masses through new physics effects at 
energies that ILC and LHC can access.  
We give examples of models in which the ILC will discover new particles and measure properties 
that are simply related to the neutrino mixing angles.

\end{enumerate}

We conclude that the physics case for the ILC is very strong, independently of future
findings at the LHC. We illustrate this with a detailed discussion of consequences 
for ILC physics 1. if LHC discovers no new particles, 
2. if LHC discovers some new low-mass states and 
3. if LHC discovers new heavy particles. 
Under each scenario, ILC will play a critical role in the 
discovery of new phenomena. ILC will push forward our knowledge of 
high energy physics and our understanding of the universe.

\clearpage
\newpage
\vspace{5cm}
\begin{center}
{\large\bf Prelude}
\end{center}
\bigskip
\begin{verse}
I saw no Way-- the heavens were stitched--\\
I felt the Columns close--\\
The Earth reversed her Hemispheres--\\
I touched the Universe--

And back it slid-- and I alone--\\
A speck upon a Ball--\\
Went out upon Circumference--\\
Beyond the Dip of Bell--

\footnotesize{Emily Dickinson, 1862}
\end{verse}

\bigskip
\bigskip

\begin{quotation}
Herein the poetess, seeing no way forward, achieves an unexpected breakthrough 
which transports her to realms beyond the limits of space and time$\cdots$
\end{quotation}
\clearpage
\section{Introduction}
\label{sec:intro}
%
%

This paper addresses the question of whether the International Linear
Collider has the capability of discovering new particles that have not
already been discovered at the CERN Large Hadron Collider.   The basic
conclusions of this paper have already  been enunciated in a letter
sent in December 2015 from ICFA to the MEXT ILC Advisory Panel.   This
paper provides an appendix to that letter that summarizes the
physics studies underlying
 its conclusions.

The most important question in particle physics today is the validity
of the Standard Model of particle physics that describes the strong,
weak, and electromagnetic interactions.   This model has been tested
in precision experiments at and around the $Z$ resonance in the 1990's 
at LEP and SLC and more
recently in the measurement of complex processes at the 
Tevatron and LHC involving
the production of single and multiple $W$ and $Z$ bosons and top
quarks. So far, the Standard Model has
 passed every test.  

Still, the Standard Model is visibly inadequate as an ultimate theory
of nature.   It does not contain the dark matter and dark energy that
make up 95\% of the energy content of the universe.   It does not
explain the fact that the visible universe is made up of matter but no
antimatter. Though the Standard Model successfully predicted the
appearance and general properties of the Higgs boson, essentially
every property of this particle is determined by a parameter adjusted
by hand.   The Standard Model theory of the Higgs boson field provides
no understanding of the mass of this particle or the masses that it
 produces for the matter particles of the Standard Model, the quarks and leptons.  

Most troubling, the Standard Model does not explain the most
mysterious aspect of the Higgs field, the fact that it fills space with
a nonzero value, producing an ordered thermodynamic state like the
superconductors and superfluids of condensed matter physics.   This
state, crucial to all other properties of the Higgs field, appears
only because a particular parameter of the model is given a negative
sign. In the history of the theory of superconductivity, we passed
through a similar stage of partial understanding, in which Landau and
Ginzburg gave a quantitative description of the properties of the
superconducting state using a model with adjustable
parameters~\cite{LandauGinzburg}.  The fundamental explanation of
superconductivity came almost a decade later, from a beautiful and
subtle theory of the interactions of electrons in metals by 
 Bardeen, Cooper, and Schrieffer~\cite{BCS}.

What causes the Higgs field to behave as it does?  No interaction of
the Standard Model can account for this.  We require new interactions,
hidden at short distances or high energies.  In quantum field theory,
such new interactions also require new particles, outside those of the
Standard Model. So, particle physicists ask, where are these new
particles or forces?  
How can they be discovered?

The search for these new particles has now become the primary goal of
the experiments at the LHC.  Taking advantage of the high energy and
large reaction rates possible for proton-proton collisions, the LHC
experiments have searched for a wide variety of new particles
predicted by theories that extend the Standard Model.  So far, this
search has come up
 empty.

We will continue the search for new particle production at the LHC.
Another factor of 100 in data will be available by the end of the LHC
program in the 2030's, giving the opportunity to extend current
searches to new particle masses higher by a factor 1.3--3. Still, it is
not too early to ask whether other approaches are needed to discover
new interactions at high 
energy.

One of these approaches is the precision study of the heaviest
particles of the Standard Model --- the Higgs boson and the top quark.
New interactions that explain the behavior of the Higgs boson
necessarily alter its properties. Such new interactions can also
alter the interactions of the top quark, through its strong coupling
to the Higgs boson that is responsible for its large mass. It is
well documented that measurement of the couplings of the Higgs boson
and the top quark to the percent level can be sensitive to new interactions
at high energies, beyond the ability of the LHC to search directly.
But also,  the pattern of deviations seen in these couplings from the
predictions of the Standard Model gives information on the properties
of the new interactions that induce them~\cite{Fujii:2015jha,Kanemura:2014bqa}.
This idea was recently confirmed by an examination of the models
proposed for the 750~GeV resonance suggested by the 2015 LHC data. It
was shown that precision Higgs boson and top quark measurements would
powerfully distinguish the competing hypotheses~\cite{Fujii:2016raq}.
The ILC will give us the ability to study the heavy particles of the
Standard Model with this level of precision.  This is the principle
argument for the importance of that 
machine to the future of high energy physics.

There is still another route to the discovery of new particles beyond
those of the Standard Model.    Though the LHC experiments search for
a large range of new particles, their search is not exhaustive, even
in the range of masses that is energetically accessible.  Typical LHC
new particle searches look for reactions predicted to occur at the
rate of 1 per trillion proton-proton collisions.   These are rates at
which the Standard Model interactions are capable of producing very
complex reactions, with multiple $W$ and $Z$ bosons, top quarks, and
invisible neutrinos.    The search for new particles then depends on
the recognition of very specific predicted properties, which are used
both to select events for close analysis and to distinguish new
interactions from similar Standard Model background processes.  As a
wide variety of LHC searches have been carried out and their reach
analyzed, significant blind spots have been recognized.   The LHC
experiments have special difficulties with particles that are produced by
weak and electromagnetic rather than strong interactions, decay
primarily to $\tau$ leptons or to lighter quarks and gluons, or decay
with very small energy release.   A strength of the ILC experiments is
that they are sensitive to new particles with these hard-to-recognize
decay schemes.

Given the large number of searches carried out at the LHC, one might
be tempted to dismiss these cases as rare exceptions.   However, these
cases include some of the most important examples of new particles
that we seek to discover. The most obvious example is given by the
particle that makes up the cosmic  dark matter.  If this particle is
produced at the LHC, it will be invisible to the experimental
detectors.  The presence of dark matter particles can only be detected
by observing visible particles like e.g.\ quarks, gluons, or photons that balance the
momentum carried off by these invisible particles.  Such events are
difficult to select and are easily imitated by Standard Model
reactions with neutrino emission.  The most promising strategies for
observing dark matter involve producing a heavier particle that decays
to the dark matter particle. However, first, this particle may not
exist or be too heavy to be produced at LHC, or, second, this particle may be very close in mass to the
dark matter particle, giving too small an energy release in the decay
to be visible to the LHC experiments. Many models of dark matter
require  such close-mass partners as a part of their mechanism for
obtaining the correct observed dark 
matter density in the universe. 

New direct partners of the Higgs boson suffer from the same
difficulties.  For example, in supersymmetry, an important class of
models of new interactions,  the structure of the 
model implies that the partners of
the Higgs boson are closely spaced in mass, with the lightest one
stable and invisible to LHC detectors.   Within this model, these
particles are expected to have masses close to the mass of the
$W$, $Z$ and Higgs bosons; thus, they are expected to be found in the
energy range of the ILC.    Other models with partners of
the Higgs boson also  predict low production rates at the LHC and
decay into modes that are difficult to 
observe, making these particles targets for ILC discovery.

One of the great advantages of using electron-positron annihilation,
rather than proton-proton collisions, is that this reaction makes it
possible to search {\it comprehensively} for new reactions outside the
Standard Model.   This is the result of well-known advantages of
$e^+e^-$ reactions --- the low rate of Standard Model background processes
relative to new particle production and the high level of detail that
is visible in the final state.  In $e^+e^-$ reactions, the tight
pre-selection of events required at the LHC is not needed.  This makes
it possible not only to observe a broader range of new processes but
also to make serendipitous discoveries.   Historically, the most
unanticipated particle of the Standard Model, the $\tau$ lepton, was
discovered at an $e^+e^-$ collider.  Even the gluon, the quantum of
the strong interactions, was discovered at an $e^+e^-$ collider before
it could be recognized in proton (anti-)proton  collisions.

However it is made, whether by observation through precision measurements or
by  direct discovery of new particles, the first breakthrough to the
new particles underlying the Higgs field and the Standard Model
will be a historic event.   Not only will this be a remarkable
discovery worthy of a Nobel Prize, but also it will reveal a new stratum
of the fundamental interactions.  Its  full exploration will bring
further discoveries essential to particle physics.   Perhaps, as the
understanding of the Standard Model did, it will bring to light new possibilities
for physical laws with deep  implications throughout the physical sciences.

In this paper, we illustrate these ideas through a review of specific
studies of new particle observation that have been carried out for the
ILC. 
Throughout the document, we refer to the ILC with a center-of-mass energy of 500\,GeV 
and its 20-year operation scenario presented in~\cite{Barklow:2015tja}.
Section~\ref{sec:BSM} introduces the theoretical models that provide the
basis for the types of new particles to be considered.   Section~\ref{sec:ILC},
included for completeness, briefly reviews the capability of the ILC
to discover new interactions indirectly through precision measurements
of the Higgs boson, the top quark, and the $W$ and $Z$ bosons.
Section~\ref{sec:ILC_NP} reviews ILC studies of new particle production in the LHC
blind spots, including the production of dark matter particles,
supersymmetric particles with only electroweak interactions, and Higgs
boson partners.  Finally, section~\ref{sec:LHC_scen} maps this information onto the
grid of possibilities provided by the ILC Advisory Panel, giving
concrete examples of the conclusions that we have
 reached for each scenario.


\section{Overview on BSM Scenarios}
\label{sec:BSM}
%
%



Along with the cosmological issues mentioned earlier --- dark matter, 
dark energy, baryogenesis --- there are two troubling problems with the 
theoretical structure of the SM. The first one is closely related to the 
questions concerning the origin of the Higgs boson $h$ and its properties introduced 
above, which arise from the verified 
existence of this seemingly bonafide scalar particle. More technically speaking, 
quantum mechanical contributions to the Higgs mass diverge
quadratically with energy scale such that quantum corrections to the
Higgs mass $m_h^2$ soon exceed $m_h=125$ GeV at energy scales
$\Lambda >1$ TeV. 
While such huge mass corrections can always be cancelled off by 
adjusting some free parameters of the theory, 
such fine-tunings --- which may be as small as one part in $10^{28}$ if 
$\Lambda$ reaches as high as the grand unification scale 
$m_{GUT}\sim 2\times 10^{16}$ GeV --- 
are considered pathological and indicative of some missing ingredients 
in the theory. 
Indeed, from an historical perspective, requiring for instance not-too-large 
contributions to the $K_L-K_S$ mass difference led to the prediction of the 
charm quark mass at its measured value: 
see Ref.~\cite{Giudice:2008bi} for various other examples.

Two simple paths to addressing the Higgs mass issue include 
1. postulating a new spacetime symmetry, supersymmetry or SUSY for short, 
which guarantees cancellation of the offending quantum corrections 
to all orders in perturbation theory and 
2. understanding the Higgs field as not being fundamental but instead 
as some composite, bound state built out of new fundamental fermions.
In recent years, more intricate alternative approaches involving 
extra dimensions of spacetime or the presence of 
exotic hidden sector states 
(which may be arranged to also cancel many of the offending 
quantum corrections) have been invoked as possible solutions to the 
Higgs mass problem.

A second fine-tuning issue arises in the QCD sector in that the theory seems
to require a strongly $CP$-violating contribution to the Lagrangian whereas 
none is observed in nature. So far, the introduction of the axion field
seems the only plausible means to address the strong $CP$ problem.
As we will see, ILC may be able to shed considerable light on 
the Higgs mass problem and also some indirect light on the strong $CP$ problem.

\subsection{Dark matter}
\label{sec:DM}

There is overwhelming evidence from a large variety of observations for the existence of
dark matter (DM) in the universe.\footnote{For a recent review, 
see {\rm e.g.} \cite{Baer:2014eja}.} 
The DM content should be about five times more abundant than
baryonic matter and it must be electric (and likely color) neutral and non-relativistic in order to
seed structure formation. No particle within the SM has such properties and hence the presence of
dark matter requires new matter states to exist.

One very popular DM candidate is the weakly-interacting massive particle or WIMP, labeled as $\chi$. 
WIMPs occur in a wide variety of beyond the SM theories, including SUSY, Little Higgs
models, models with extra spacetime dimensions, and models with dark sectors among others.
A compelling feature of WIMP dark matter is that at temperatures $T\gtrsim m_\chi$, then WIMPs
are thermally produced and present in the early universe. As the universe expands and 
temperatures drop below $m_\chi$, then WIMPs can no longer be thermally produced and their
number density freezes out, locking in a relic abundance of dark matter which depends 
on the WIMP mass and its annihilation cross section. For WIMP masses in the vicinity of 
$m_{weak}\sim 100$ GeV, then roughly the measured amount of dark matter will be produced, 
a situation known as the WIMP miracle. 

Since WIMPs are weakly interacting, they should necessarily be produced at ILC, as will be discussed in section~\ref{subsec:directNP_WIMP},
while the analoguous processes might be hidden in enormous backgrounds at hadron colliders. 
It should be noted that if WIMPs exist, they should 
ultimately also be detected by underground direct detection experiments and by indirect searches 
for WIMP-WIMP annihilation within the cosmos into gamma rays or antimatter. 
Detailed measurements of WIMP properties (such as mass, spin and which particle species they annihilate into)
are required to verify or falsify the simple assumptions 
associated with thermal DM production within the WIMP miracle scenario.

It is also possible, and perhaps theoretically likely, 
that the WIMP content of the universe would be augmented by
non-thermal production of exotic particles (e.g.\ gravitinos, axinos or saxions)
followed by decay to WIMPs. It is also possible that the WIMP content might be
diminished from its thermal value for instance by decay to a lighter superWIMP
state (such as gravitino, axino or KK-graviton) 
or by significant late-time entropy injection e.g.\ by
saxion or moduli-fields, which would dilute all relics present at the 
time of decay. Late-time moduli field decay seems ubiquitous in string theory models
while early universe saxion production and decay seems required in SUSY axion models 
which address both the gauge hierarchy and strong $CP$ problems.

If indeed DM consists of super-WIMPs, then no detection of DM is expected at
either direct or indirect dark matter detection experiments. 
However, superWIMPs may leave tell-tale 
signatures at colliding beam experiments via missing energy produced as the 
end-product of cascade decays or via long-lived next-to-lightest particles 
which have suppressed decay rates into the superWIMP. 
Thus, whether or not a direct or indirect WIMP 
detection signal is found, there will remain considerable details 
regarding dark matter production in the early universe to be sorted out.

A final mention is that DM could consist of axion particles which are required by 
the Peccei-Quinn (PQ) solution to the strong $CP$ problem. 
Axions may ultimately be detected via ongoing microwave cavity or other experiments. 
At first sight, one might not expect a connection between axion physics and ILC/LHC physics. 
However, it has been recently emphasized that there is a deep connection between
PQ symmetry and supersymmetry via the superpotential $\mu$ term which is required by
naturalness to be not too far from the weak scale\cite{Bae:2013bva}. In such a case, then
DM would consist of a WIMP-axion mixture, {\it i.e.} two dark matter species. 

\subsection{Supersymmetry}
\label{subsec:SUSY}

The minimal supersymmetrized version of the Standard Model, the MSSM or Minimal Supersymmetric Standard Model, 
contains a scalar field for each chiral fermion and a gaugino for each gauge boson, a second Higgs doublet as well as fermionic partners of the Higgs bosons, the {\em higgsinos}.  
The Lagrangian for the MSSM is derived in part from 
the {\it superpotential} which contains a mass term $\mu \hat{H}_u\hat{H}_d + {\mathrm{Yukawas}}$ 
where, very importantly, $\mu$ feeds mass to both higgsinos and Higgs bosons;
$\mu$ also feeds mass to the gauge bosons $W^\pm$ and $Z$ via the vacuum expectation values of the scalar potential.
From this, absent large unnatural cancellations in the masses of these particles, 
it may already be expected that the higgsinos should have masses of 
order the $W,\ Z$ and $h$ bosons, and thus within the discovery reach of ILC.

Since the mechanism for SUSY breaking is unknown, 
we may {\it parametrize our ignorance} via the introduction of various (independent)
soft SUSY breaking terms which include mass terms for scalars and gauginos 
along with bilinear and trilinear scalar couplings ($B$ and $A$ terms).
In more complete models, the soft terms may be calculated in terms of the
more fundamental gravitino mass $m_{3/2}$ (in the case of gravity-mediated
or anomaly-mediated SUSY breaking models) or in terms of the messenger scale
$\Lambda$ (in gauge-mediated SUSY breaking models). 
From the softly broken SUSY Lagrangian of the 
Minimal Supersymmetric Standard Model (MSSM), one may compute the various mass
eigenstates and their mixings, and consequently superparticle production and
decay rates. 

A corner-stone of SUSY is that {\it sparticles couple as particles}\cite{Dimopoulos:1981zb}. 
This is independent of the mechanism responsible for SUSY breaking. 
Thus, the couplings of particles to sparticles is completely determined in SUSY theories 
and can be {\it tested} at linear $e^+e^-$ colliders in many cases\cite{Feng:1995zd}.

The MSSM receives some impressive indirect support from experiment in that
1. the measured values of the weak scale gauge couplings from LEP 
unify under MSSM renormalization group evolution, 
2. the measured value of $m_t$ is in the range needed to trigger the 
required breakdown of electroweak symmetry and 
3. the measured value of the Higgs mass falls well within 
the window required by the MSSM, namely that $m_h<135$ GeV. 
Finally, it is perhaps under-appreciated that SUSY stabilizes the 
weak/GUT gauge hierachy in grand unified theories (GUTs). GUTs
not only unify the three forces of the SM, but also unify matter fermions
(completely for each generation in $SO(10)$) 
and explain the seemingly ad-hoc quantum numbers of the Standard Model 
fermions. 

\subsubsection{SUSY dark matter}
\label{subsubsec:SUSYDM}

Under $R$-parity conservation --- which is well-motivated by 
unified theories and the need to suppress proton decay --- 
then the lightest SUSY particle (LSP) is absolutely stable 
and may serve as a dark matter candidate. 
If the LSP is a neutralino, then it is a candidate weakly interacting 
massive particle, or WIMP. 
The SUSY WIMP may be mainly bino-, wino- or higgsino-like, 
or a comparable mixture of these states.
For the popular case of a bino-like LSP, then a low bino annihilation rate
in the early universe leads to too much dark matter. 
One special enhancement mechanism for bino annihilation is called 
co-annihilation wherein the bino
may co-exist and annihilate with states nearby in mass, such as 
the lightest stau, stop or chargino. 
Another mechanism is resonance annihilation
wherein the bino could have enhanced annihilation through some
resonance such as heavy Higgs states $A$ or $H$. 
Well-tempered neutralinos consisted of comparable bino-higgsino mixtures
and could also yield the measured abundance of dark matter; this case now seems
ruled out by direct detection experiments. 
SUSY winos as comprising all of dark matter also seem ruled out 
since they would lead to large rates for gamma ray 
production in galactic annihilations. 
Light wino dark matter as just a portion of the dark matter is also allowed
since then their galactic abundance is suppressed, leading also to
suppression of their indirect detection rates.

LSPs that are mainly higgsino-like tend to give not enough dark matter unless
their mass reaches into the (unnatural) TeV regime. 
However, low mass higgsino-like WIMPs could still exist as just a 
portion of the dark matter, wherein the remainder might be made up of axions. 
Such a scenario is appealing in models of natural supersymmetry since both
the EW and QCD (strong $CP$) naturalness problems are then solved.
 
\subsubsection{Baryogenesis in SUSY}
\label{subsubsec:SUSYbaryo}

While EW baryogenesis is not viable in the SM and barely viable in
the MSSM, SUSY provides several compelling alternative possibilities
for baryogenesis. Thermal leptogenesis (and subsequent baryogenesis
via sphaleron effects) via asymmetric decays of heavy
right-hand-neutrino states in the early universe seems to require
re-heat temperatures $T_R>\sim 10^9$ GeV. Such high re-heat brings with it
possible conflict with over production of gravitinos in the early universe.
Alternatively, Affleck-Dine leptogenesis proceeds via a lepton-number
violating condensate which forms along flat directions in the SUSY scalar
potential. Affleck-Dine leptogenesis seems viable for $T_R$ as low as $10^5$ GeV.
Other possibilities include non-thermal leptogenesis where right-handed neutrinos are 
produced via inflaton decay or oscillating sneutrino leptogenesis 
which can also occur at lower $T_R$ values. 
The capacity of ILC to perform detailed measurements on SUSY particles offers
hope to sort out amongst these promising possibilities.



\subsubsection{Naturalness and upper bounds on sparticle masses}
\label{sec:BSM-SUSY-Naturalness}

An absolutely vital question --- which addresses the issue of {\it falsifiability} of
the weak scale SUSY hypothesis --- 
arises as to: how massive can the various SUSY particles be while maintaining naturalness?
In fact, this {\it naturalness} issue has led some physicists to question whether present LHC
sparticle search bounds are already putting intense pressure on the SUSY hypothesis.

Some guidance can be obtained by connecting observed particle masses to SUSY Lagrangian
parameters. For instance, the Higgs boson squared mass arises in the MSSM as
\begin{equation}
m_h^2= \mu^2+m_{H_u}^2+{\rm mixing\ terms}+{\rm radiative\ corrections},
\label{eq:mhs}
\end{equation}
where each contribution is evaluated at the weak scale and the 
latter two contributions are $\lesssim m_h^2$. 
If either $\mu^2$ or $-m_{H_u}^2$ were far larger than $m_h^2$, then the other term would have 
to be tuned to high accuracy to maintain $m_h\simeq 125$ GeV. On the contrary, to avoid
large fine-tuning in $m_h^2$, then each contribution ought to be comparable to $m_h^2$. 
This implies that $m_{H_u}^2$ should be driven to small negative values 
$\sim -(100)^2$ GeV$^2$ and also that  $\mu\sim 100-200$ GeV (the lower bound arises
from LEP2 limits on chargino production). 

Alternatively, naturalness can be expressed
in a related form in terms of $m_Z^2$ from minimizing the scalar potential:
\begin{equation}
\frac{m_Z^2}{2} =\frac{m_{H_d}^2+\Sigma_d^d-(m_{H_u}^2+\Sigma_u^u)\tan^2\beta}{\tan^2\beta -1}-\mu^2\sim -m_{H_u}^2-\Sigma_u^u-\mu^2 
\label{eq:mzs}
\end{equation} 
where the last partial equality obtains for moderate-to-large $\tan\beta$.
To avoid large unnatural cancellations on the right-hand-side, we again 
see that the weak scale magnitudes of $m_{H_u}$ and $\mu$ must be comparable to $m_Z$. 
The {\it electroweak} naturalness measure $\Delta_{EW}$\cite{Baer:2012up} 
compares the largest term on the right-hand-side of Eq.~(\ref{eq:mzs}) to $m_Z^2/2$.
A low (natural) value\footnote{The onset of fine-tuning for $\Delta_{EW}>30$ is visually displayed in Fig. 1 of Ref. \cite{Baer:2015rja}.} of $\Delta_{EW}<30$ 
implies {\it light higgsinos} $\tilde{\chi}_1^\pm$ and $\tilde{\chi}_{1,2}^0$ 
with mass $\sim 100-300$ GeV, the closer to $m_Z$ the better.
The radiative corrections $\Sigma_u^u$ and $\Sigma_d^d$ contain over 40 one-loop contributions. 
The largest of these usually arise from the top-squark sector and implies for $\Delta_{EW}<30$ 
that $m_{\tilde{t}_1}<3$ TeV as long as the stop is highly mixed 
(which helps lift $m_h$ up to $\sim 125$ GeV). 
The gluino feeds into $m_Z^2$ at two-loop order and calculations show $m_{\tilde{g}}\lesssim 4$ TeV:
well-beyond present LHC bounds and perhaps beyond the reach of HL-LHC. A more detailed discussion 
about the relations between different fine-tuning measures can be found e.g.\ in~\cite{Baer:2014ica}.


\subsubsection{Overview of natural SUSY parameter space}

\begin{figure}[htbp]
\begin{center}
\hspace{3mm} \includegraphics[width=8cm]{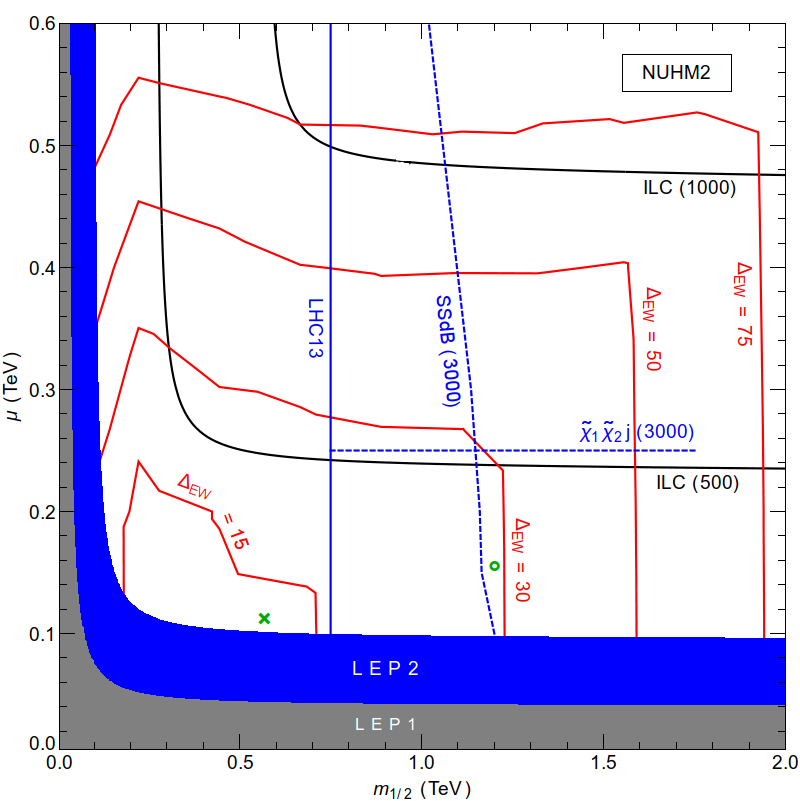}
\end{center}
\caption{The $m_{1/2}$ vs. $\mu$ parameter space of the NUHM2 SUSY model with $m_0=5$ TeV, 
$A_0=-8$ TeV, $\tan\beta =15$ and $m_A=1$ TeV.
Naturalness contours of $\Delta_{EW}=15$, 30 50 and 75 are shown in red.
We also show the current exclusion limit of LHC13 (solid-blue) 
and future reach of HL-LHC (dashed-blue). The reach of ILC with $\sqrt{s}=500$ and 1000 GeV is also shown.
We show the locus of two ILC benchmark points which have been studied in green.
}
\label{fig:mhfvsmu}
\end{figure}

A grand overview of natural SUSY parameter space for the case of gravity-mediated SUSY breaking 
in the two-extra-parameter non-universal Higgs model (NUHM2) is shown in Fig.~\ref{fig:mhfvsmu} 
where we plot the unified GUT scale gaugino mass $m_{1/2}$ vs. superpotential $\mu$ parameter 
with GUT scale scalar masses $m_0=5$ TeV, $A_0=-8$ TeV, $\tan\beta =15$ and $m_A=1$ TeV\cite{Baer:2016usl}. 
The grey and blue-shaded regions were long ago excluded by chargino pair searches at LEP and LEP2. 
Current LHC13 searches require $m_{\tilde{g}}>1.9$ TeV
which excludes the region to the left of the blue line labeled LHC13.
Contours of naturalness $\Delta_{EW}$ values are indicated in red. 
Fine-tuning already sets in for $\Delta_{EW}>30$.
It can be seen that a large region of natural SUSY remains with $\Delta_{EW}<30$ 
but which is well beyond current LHC bounds. 
The future reach of HL-LHC, estimated at the generator-level with $\sqrt{s}=14$ TeV and 3000 fb$^{-1}$ of integrated luminosity is shown
by the dashed lines for the same-sign-diboson channel (SSdB) 
and the monojet plus soft dilepton channel ($\tilde{\chi}_1^0\tilde{\chi}_2^0j$). 
These reach contours can be compared to the reach of ILC with $\sqrt{s}=500$ GeV and $\sqrt{s}=1000$ GeV
(black contours). 
The location of two ILC benchmark points are denoted by green tags.

In models with gaugino mass unification such as NUHM2, then the mass hierarchy $\mu\ll M_1<M_2<M_3$ is expected
and HL-LHC should be able to identify same-sign diboson (SSdB) production signals arising from
wino pair production, or else higgsino pair production $\tilde{\chi}_1^0\tilde{\chi}_2^0 j$ followed
by $\tilde{\chi}_2^0\rightarrow \ell^+\ell^-\tilde{\chi}_1^0$ decay.

\begin{figure}[tbp]
\begin{center}
\includegraphics[height=0.2\textheight]{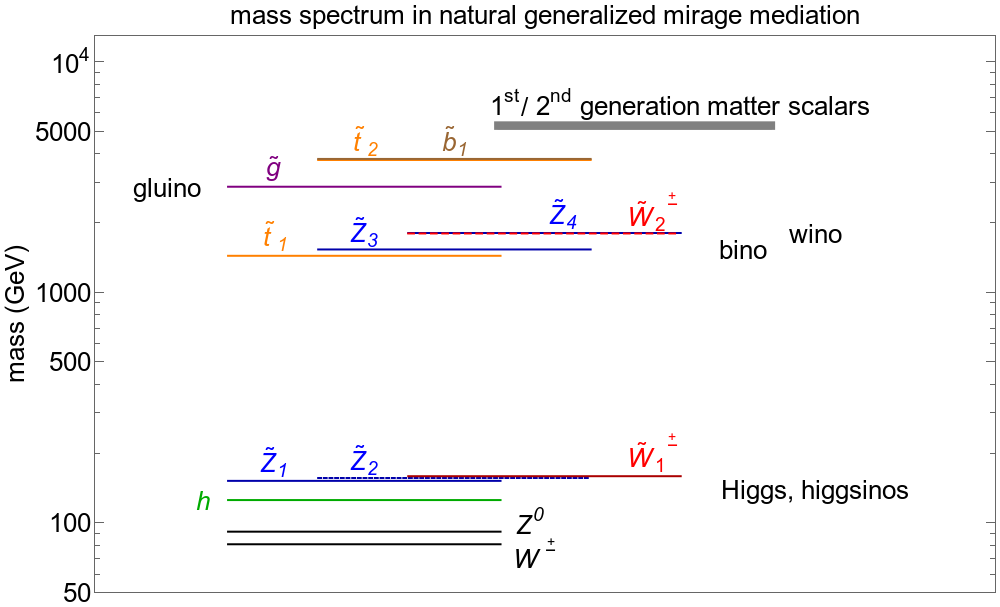}
\includegraphics[height=0.2\textheight]{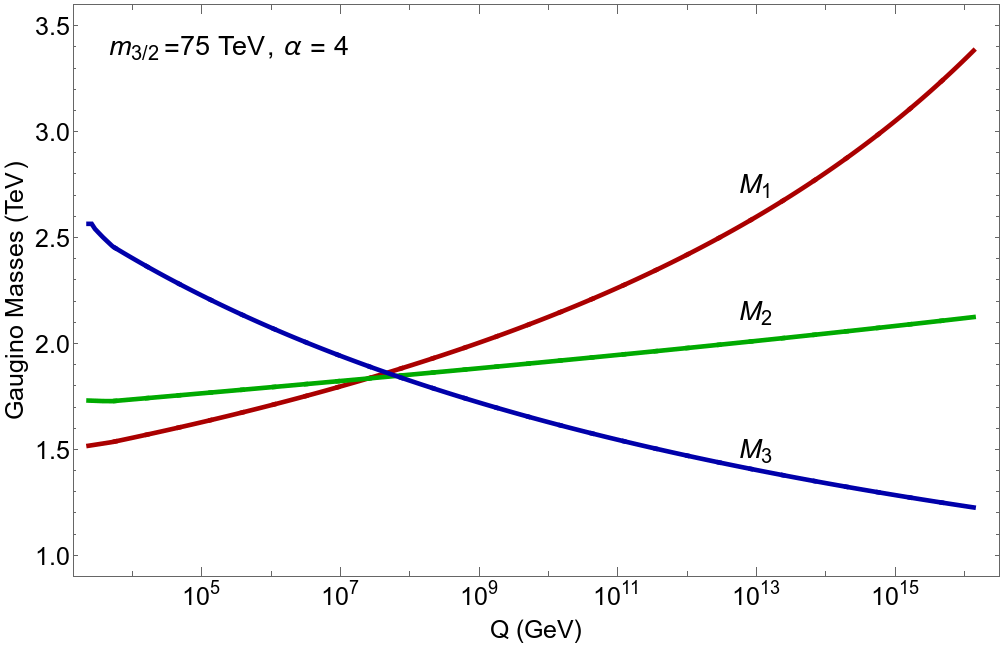}
\caption{{\it a}) A typical superparticle mass spectrum generated from 
natural generalized mirage mediation (nGMM)~\cite{Baer:2016hfa}.
{\it b}) Evolution of gaugino masses from the nGMM benchmark point with 
$m_{3/2}=75$ TeV, $\alpha =4$.
\label{fig:spect}}
\end{center}
\end{figure}

However, several compelling SUSY models which are {\it also natural} can defy this mass pattern. 
For instance, in {\it mirage mediation} SUSY models, the gauginos gain comparable mass 
contributions from moduli (gravity)-mediation and anomaly mediation. 
The GUT scale splitting of the gaugino masses, 
which is proportional to the gauge group beta functions, is compensated
by the gaugino mass RG running so that gaugino masses apparently unify at some
intermediate scale (which is determined by the relative amounts of anomaly- and moduli-mediated
mass contributions). In such mirage mediation models, the gaugino spectrum is more compressed: if the
gluino lies in the 3 TeV range, then winos and binos may also lie in the few TeV range so that
wino pair production is suppressed and the $\tilde{\chi}_2^0-\tilde{\chi}_1^0$ mass gap is
compressed to the few GeV level. In such a case, SUSY would be completely natural but largely 
inaccessible to HL-LHC searches~\cite{Baer:2016hfa}. 
However, the required light higgsinos  would still be accessible to ILC with $\sqrt{s}>2m(higgsino)$.
A concrete example is shown in Fig. \ref{fig:spect} where {\it a}) presents the mass spectrum and
{\it b}) illustrates the characteristic mirage unification of gaugino masses for a sample benchmark point.

\subsubsection{Neutrino Physics at the ILC}

One outstanding question the Standard Model cannot explain is the smallness of neutrino masses.
Starting with the compelling data on atmospheric neutrinos in 1998, it has now been firmly established
that neutrinos are massive, and the different flavors are strongly
mixed. However, neutrino masses are much smaller than the masses of the electrically charged
fermions and flavor mixing in the leptonic sector differs considerably from that in the
quark sector. These experimental findings, along with the observed hierarchies in
the masses of the charged leptons and quarks form the so-called flavor puzzle.
Even though we can parametrize our understanding, 
we do not know the underlying principles leading to the experimentally observed patterns.

Several models have been proposed to explain the smallness of neutrino masses.
At low energies they usually lead to the so-called Weinberg operator $(LH)^2$
giving Majorana masses to light neutrinos after electroweak symmetry breaking.
The most compelling among them is the {\it seesaw mechanism} which comes in three varieties
depending on the details of how this operator is generated: type I in case of gauge singlet
fermions (usually the right-handed neutrinos), type II in case of an $SU(2)_L$ triplet
Higgs boson and type III in case of $SU(2)_L$ triplet fermions. 
The type-I see-saw easily melds with $SO(10)$ SUSY GUTs which require a right-handed neutrino
to fill out the 16-dimensional spinor representation. It should be remarked that SUSY
is highly desirable in this case to stabilize the Higgs mass from blowing
up to the see-saw scale $\sim 10^{11}-10^{13}$ GeV.

These additional particles associated with the see-saw mechanism 
are usually too heavy to be produced directly in collider experiments.
In supersymmetric models, however,  they may leave imprints on the RG evolution of
the mass parameters~\cite{Baer:2000hx} 
and give rise to additional flavor structures which are linked
to the underlying mechanism for generating neutrino masses. 
These flavor structures induce flavor violating decays of sleptons 
which can potentially be studied at the ILC\cite{Deppisch:2003wt}.

{\it See-saw scale, axions and the origin of the SUSY $\mu$-parameter:}

The hidden sector SUSY breaking scale $m_{hidden}$, the Peccei-Quinn (PQ) scale $f_a$ 
and the see-saw neutrino 
scale $m_N$ are all expected to lie in the intermediate range: $\sim 10^{10}-10^{13}$ GeV.
Are these scales all related? One class of models begins with PQ symmetry which forbids the SUSY
superpotential $\mu$ term, but then both $\mu$ and the see-saw scale $m_N$ are generated due to
radiatively-broken PQ symmetry which is triggered by SUSY breaking\cite{Murayama:1992dj,Bae:2014yta}. In such models, 
the $\mu$ term is generated as $\mu=\lambda_\mu f_a^2/M_P$ leading to a Little Hierarchy 
where $\mu\ll m_{SUSY}$ arising as a consequence of $f_a<m_{hidden}$. 
A Majorana neutrino scale $m_N\sim f_a$ is also generated. In the case where an axion is detected by
microwave cavity or other experiments, where $m_a\simeq m_\pi f_\pi/f_a\cdot \sqrt{Z}/(1+Z)$ with 
$Z=m_u/m_d\simeq 0.56$, then the axion mass determines $f_a$ while
an ILC measurement of light higgsinos of mass $\sim \mu$ determines the PQ coupling $\lambda_{\mu}$.

{\it Bilinear R-parity violation}
\label{subsubsec:bRPV_theo}

Beside the usual mechanisms to generate neutrino masses, supersymmetry offers an 
additional possibility: breaking of $R$-parity in the lepton sector. The simplest
model is the one where only bilinear terms are present in the superpotential as well
as the corresponding terms in the soft SUSY sector. As $R$-parity is broken,
the lightest supersymmetric particle (LSP) is not stable anymore but decays. 
The six parameters explaining neutrino data are then also those responsible for the decay
properties of the LSP: ratios of decay branching ratios are proportional
to neutrino mixing angles, e.g.\ 
$BR(\tilde \chi^0_1 \to W \mu)/BR(\tilde \chi^0_1 \to W \tau) \simeq \tan^2\theta_\mathrm{atm}$ or
$BR(\tilde \chi^0_1 \to \nu \mu \tau)/BR(\tilde \chi^0_1 \to \nu e \tau) \simeq \tan^2\theta_\mathrm{sol}$
\cite{Porod:2000hv} where
$\theta_\mathrm{atm}$ and $\theta_\mathrm{sol}$ are the atmospheric and solar neutrino mixing angles. 
We will illustrate this possibility with a dedicated simulation study in section~\ref{subsubsec:directNP_bRPV}.
Moreover, the smallness of the neutrino masses
also implies that the lifetime of the LSP is measurable at the ILC in a large part of the parameter
space. For completeness we note that the existence of such correlations does not depend
on the nature of the LSP --- e.g.\ whether it is a neutralino or a chargino or a slepton --- but only
the concrete form of these correlations \cite{Hirsch:2003fe}.

\subsection{Two-Higgs-Doublet Models}
\label{sec:2HDM}

A minimal extension of the SM Higgs sector consists of adding a second Higgs doublet to the one present in the SM.
In these Two Higgs Doublet Models (2HDMs), the scalar potential will contain mixing mass parameters and both doublets will acquire vacuum expectation values, $v_1/\sqrt{2}$ and $v_2/\sqrt{2}$, respectively,  and the gauge boson masses will keep their SM 
expressions with the Higgs vacuum expectation value $v$ replaced by $v = \sqrt{v_1^2 + v_2^2}$.  The complete 2HDM  is defined only after considering the interactions of the Higgs fields to fermions.
Yukawa couplings of the generic form
\begin{equation}
-h_{ij}^a  \bar{\Psi}^i_L H_a \Psi^j_R + h.c.
\label{eq:2HDM Yukawa}
\end{equation}
may be added to the renormalizable Lagrangian of the theory. Contrary to the SM, the two Higgs doublet structure does not ensure the alignment of the fermion mass terms  with the Yukawa couplings, and the neutral Higgses can mediate flavor changing interactions. Such flavor changing interactions should be suppressed in order to comply with flavor data constraints.  
Based on the Glashow--Weinberg criterion, it is clear that 
the simplest way of avoiding such transitions is to assume the existence of a symmetry that ensures the couplings of the fermions of each given quantum number (up-type and  down-type quarks, charged
and neutral leptons) to only one of the two Higgs doublets.  Different models may be defined depending on which of these fermion fields couple to a given Higgs boson. Models of type-I are those
in which all SM fermions couple to a single Higgs field.  In type-II models, of which the MSSM is a prime example, down-type quarks and charged leptons couple to a common Higgs field, while the up-type quarks and neutral leptons couple to
the other.  In models of type-III (lepton-specific) quarks couple to one of the Higgs bosons, while leptons couple to the other. Finally, in models of type-IV (flipped), up-type quarks and charged leptons couple to one of the
Higgs fields while down-quarks and neutral leptons couple to the other. 

In 2HDMs, the electroweak phase transition can be first order, opening the possibility for a successful baryogenesis at the weak scale. Regions of the parameter space leading to a strong first order phase transition also exhibit large deviations of the Higgs self-coupling~\cite{Kanemura:2004ch}. This is also valid in the context of various other Higgs sectors beyond the 2HDM~\cite{Noble:2007kk}.

Other extensions of the Higgs sector can include multiple copies of
SU(2)$_L$ doublets, additional Higgs singlets, triplets or more
complicated combinations of Higgs multiplets.  It is also possible to
enlarge the gauge symmetry beyond SU(2)$_L\times$U(1)$_Y$
along with the necessary Higgs structure to generate gauge boson
and fermion masses. Direct searches for additional Higgs bosons with masses below or above 125\,GeV, deviations in the 125-GeV Higgs boson couplings to fermions and gauge bosons and deviations of the its self-coupling are ways to probe these extended Higgs sectors.

\subsection{Randall-Sundrum Models}
\label{sec:RS}
Randall-Sundrum models~\cite{Randall:1999ee} feature extra-dimensional spaces strongly curved by the gravitational effect of a bulk large vacuum energy. The 5D geometry corresponds to an Anti-de-Sitter space with two boundaries: 
the Planck brane and the TeV brane. 
The effect of the Planck brane is to render the graviton zero mode normalizable. The TeV brane provides a mass gap  into the Kaluza--Klein expansion of all types of fields.
The distance between the Planck and TeV branes is arbitrary, corresponding to the massless radion. In fact, the size of the extra dimension is unstable to small perturbations and must be stabilized~\cite{Goldberger:1999uk}, and this stabilization results in a mass for the radion. A light radion decays predominantly to gluons due to the 
trace anomaly.

 In realistic Randall-Sundrum models, the SM fermions propagate in the bulk and the support of zero-mode wavefunctions on the TeV brane, where the Higgs vev is localized, sets the size of the Yukawa interactions.  The wave functions of the third generation quarks and in particular that of the $t$ quark extends to the TeV
side of the 5th dimension, while the
wavefunctions of light fermions peak close to the Planck brane. This  constitutes an
elegant explanation of the striking mass hierarchy in the fermion
sector.   A consequence of this effect is that the couplings of the
$t$ quark to the $Z$ boson are expected to have large shifts due to
mixing with KK states, of independent size for $t_L$ and $t_R$.
Models of this effect are described, for example, in Refs.~\cite{Djouadi:2006rk,Carena:2006bn}.

\subsection{Composite Higgs}
\label{sec:comphiggs}

Within the SM, electroweak symmetry breaking is posited but has no dynamical origin. Furthermore, the Higgs boson appears to be unnaturally light. 
A scenario that remedies these two catches is to consider the Higgs boson as a bound state of new  dynamics becoming strong around the weak scale. Strongly coupled models of electroweak symmetry breaking have been recently and concisely reviewed in Ref.~\cite{Csaki:2015hcd},
that has been used as a source for the presentation below. For an exhaustive review on composite Higgs models, see Ref.~\cite{Panico:2015jxa}. 

The Higgs boson can be made significantly lighter than the other resonances of the strong sector if it appears as a pseudo-Nambu--Goldstone boson. More precisely, the  strongly coupled sector is supposed to be invariant under a global symmetry $G$ spontaneously broken to a subgroup $H$ at the  scale~$f$ and the Higgs boson belongs to the coset space $G/H$. 
To avoid conflict with EW precision measurements, it is better if the strong interactions themselves do not break the EW symmetry, hence the SM gauge symmetry itself should be contained in $H$.  A canonical and minimal example (called MCHM) is based on $SO(5)/SO(4)$.

The SM (light) fermions and gauge bosons cannot be part of the strong sector itself since LEP data have already put stringent bounds on the compositeness scale of these particles far above the TeV~scale. 
The gauge bosons couple to the strong sector by a weak gauging of a SU(2)$\times$U(1) subgroup of the global symmetry $G$. 
The couplings of the SM fermions to the strong sector could a priori take two different forms:
(i) a bilinear coupling of two SM fermions to a composite scalar operator, ${\cal O}$, of the form $ {\cal L}=y\, \bar q_L u_R {\cal O} +{\rm h.c.}$ in simple analogy with the SM Yukawa interactions. This is the way fermion masses were introduced in Technicolor theories and it generically comes with severe flavor problems and calls for extended model building gymnastics to circumvent them; 
(ii) a linear mass mixing with fermionic vector-like operators and the physical states are a linear combination of elementatry and composite fields. 
The SM fermion mass hierarchy  emerges from the dynamics controlling the mixing, $\theta_i$, between the elementary and composite sectors: the light fermions are mostly elementary states ($\sin \theta_i \ll 1$), while the third generation quarks need to have a sizable degree of compositeness.
While the introduction of partial compositeness greatly ameliorated the flavor problem of the original composite Higgs models, nevertheless it did not solve the issue completely, at least in the case where the strong sector is assumed to be flavor-anarchic~\cite{Csaki:2008zd}. 

Another nice aspect of the partial compositeness structure is the dynamical generation of the Higgs potential. The Higgs being a pseudo-Nambu--Goldstone boson, its mass does not receive any contribution from the strong sector itself but it is generated at the one-loop level via the couplings of the SM particles to the strong sector.
The leading contribution to the potential arises from top loops and it takes the form
\begin{eqnarray}
V(H) =&  m_\rho^4 \frac{\sin \theta_{t_L} \sin \theta_{t_R} }{16 \pi^2} \left(
\alpha \cos (H/f) +
 \beta \sin^2 (H/f)
+\gamma \sin^4 (H/f) \right),
\label{eq:CH_potential}
\end{eqnarray}
where $\alpha, \beta, \gamma$ are numbers of order 1 and $\theta_{t_L}$ and $\theta_{t_R}$ are the mixing controlling the compositeness of left-handed and right-handed top quarks respectively. 
The gauge contribution to the potential takes the form ($g$ denotes the SU(2) gauge coupling, $g_\rho f$ is the typical coupling of the strong sector and $m_\rho \approx g_\rho f$ is the typical mass scale of the strong sector resonances)
\begin{equation}
m_\rho^4 \frac{g^ 2 /g_\rho^2}{16\pi^2} \sin^2 (H/f),
\label{eq:gauge_PC}
\end{equation}
which is parametrically suppressed with respect to the top contribution  by $g^2/(g_\rho y_t)$. The gauge term is always positive, and cannot trigger electroweak symmetry breaking by itself. When $\alpha=0$, the minimization condition of the potential simply reads
\begin{equation}
\sin^2 \frac{\langle H\rangle}{f} = - \frac{\beta}{2\gamma},
\label{eq:VEV_PC}
\end{equation}
which implies that the natural expectation is that the scale $f$ is generically of the order of the weak scale. Obtaining $v \ll f$, as required phenomenologically, requires some degree of tuning, which scales like $\xi \equiv v^2/f^2$. A mild tuning of the order of 10\% ($\xi\approx 0.1$) is typically enough to comply with electroweak precision constraints. This is an important point: in partial compositeness models, the entire Higgs potential is generated at one loop, therefore the separation between $v$ and $f$ can only be obtained at a price of a tuning. 

After minimization, the potential~(\ref{eq:CH_potential}) leads to an estimate of the Higgs mass as
\begin{equation}
m_H^2 \approx {g_\rho^3\, y_t}{2 \pi^2} v^2.
\label{mH_PC}
\end{equation}
It follows that the limit $f\to \infty$, i.e. $\xi \to 0$, is a true decoupling limit: all the resonances of the strong sector become heavy but the Higgs whose mass is protected by the symmetries of the coset $G/H$. When compared to the experimentally measured Higgs mass, this estimate puts an upper bound on the strength of the strong interactions: $g_\rho \ltsim 2$. In this limit of not so large coupling, the Higgs potential receives additional contributions. In particular, the fermionic resonances in the top sector which follow from the global symmetry structure of the new physics sector can 
help raise the Higgs mass. 
For instance in the minimal SO(5)/SO(4) model, using some dispersion relation techniques, one obtains~\cite{Matsedonskyi:2012ym}
\begin{equation}
m_H^2 \approx \frac{6}{\pi^2} \frac{m_t^2}{f^2} \frac{m_{Q_4}^2 m_{Q_1}^2}{m_{Q_1}^2-m_{Q_4}^2} \log \left( \frac{m_{Q_1}}{m_{Q_4}} \right)
\label{eq:light_partners}
\end{equation}
where $Q_4$ and $Q_1$ are fermionic color resonances transforming respectively as a weak bidoublet and a weak singlet.
Therefore a 125\,GeV mass can be obtained if at least one of the fermionic resonances is lighter than $\sim 1.5\, f$. As in supersymmetric scenarios, the top sector is playing a crucial role in the dynamics of electroweak symmetry breaking and can provide the first direct signs of new physics. The direct searches for these top partners, in particular the ones with exotic electric charges 5/3 appearing in minimal models, are already exploring the natural parameter spaces of these models.

The main physics properties of a pseudo Nambu--Goldstone Higgs boson can be captured in a model-independent way by a few higher-dimensional operators, namely the ones that involve extra powers of the Higgs doublets and that are therefore generically suppressed by a factor $1/f^2$ as opposed to the operators that involve extra derivatives or gauge bosons and are suppressed by a factor $1/(g_\rho^2 f^2)$. The relevant effective Lagrangian describing a strongly interacting light Higgs is:
\begin{eqnarray}
{\cal L}_{\rm SILH} = & 
\frac{c_H}{2f^2} \left( \partial_\mu \left( \Phi^\dagger \Phi \right) \right)^2
- \frac{c_6\lambda}{f^2}\left( \Phi^\dagger \Phi \right)^3 
+ \left( \sum_f \frac{c_f \, y_f}{f^2}\Phi^\dagger \Phi  {\bar f}_L \Phi f_R +{\rm h.c.}\right). 
\label{eq:silh}
\end{eqnarray}
Typically, these new interactions induce deviations in the Higgs couplings that scale like ${\cal O}(v^2/f^2)$, hence the measurements of the Higgs couplings can be translated into some constraints on the compositeness scale, $4\pi f$, of the Higgs boson. The peculiarity of these composite models is that, due to the Goldstone nature of the Higgs boson, the direct couplings to photons and gluons are further suppressed and generically the coupling modifiers~\cite{LHCHiggsCrossSectionWorkingGroup:2012nn} scale like
\begin{eqnarray}
\kappa_{W, Z, f} \sim 1 + {\cal O}\left( \frac{v^2}{f^2} \right) \, , \quad
\kappa_{Z\gamma} \sim {\cal O}\left( \frac{v^2}{f^2} \right) \, , \quad
\kappa_{\gamma, g}  \sim {\cal O}\left( \frac{v^2}{f^2} \times \frac{y_t^2}{g_\rho^2} \right)\, ,
\label{eq:kappa_SILH}
\end{eqnarray}
where 
$y_t$ is the top Yukawa coupling that is assumed to be the largest interaction that breaks the Goldstone symmetry. 

In composite models, large corrections to the top couplings are naturally predicted and can be of two origins: (i) a strong mixing  with the composite dynamics  in scenarios with partial compositeness, (ii) the existence of top partners with exotic charges  and with a relatively low mass, as required to generate the correct Higgs mass 
(see Eq.~(\ref{eq:light_partners})). While internal left-right symmetry in the strong sector can be invoked to suppress the corrections to the $Z \bar{b}_L b_L$ vertex \cite{Agashe:2006at}, the top quark sector is not immuned and receives potentially big tree-level corrections. Two dimension-6 operators are responsible for correcting the $Z \bar{t}_L t_L$ vertex:
\begin{equation}\label{eq:operators_tL}
{\cal L} = i \frac{c_{Hq}}{f^2} (\overline q_L \gamma^\mu q_L)
\left(H^\dagger \overleftrightarrow{D_\mu} H\right)
+ i \frac{c'_{Hq}}{f^2} (\overline q_L \sigma^i \gamma^\mu q_L)
\left(H^\dagger \sigma^i \overleftrightarrow{D_\mu} H\right)\,.
\end{equation}
But the left-right symmetry of the strong sector forces $c'_{Hq} = - c_{Hq}$ and as a consequence, the leading correction to the $Z \bar{t}_L t_L$ vertex is equal to the correction to the $V_{tb}$ matrix element. The $Z \bar{t}_R b_R$
 vertex can receive tree-level corrections through the mixing, induced by the non-zero top mass, between $t_R$ and composite resonances with different quantum numbers. And these corrections are obviously enhanced if the top partners are light, see Ref.~\cite{Grojean:2013qca} for an extended discussion.

\subsection{Little Higgs}

Little Higgs models originally were introduced as minimal cases of
models with deconstructed
dimensions~\cite{ArkaniHamed:2001nc, ArkaniHamed:2002qx}. The main
motivation behind those models is to eliminate traces of nearby new
strong interactions in electroweak precision data, and to eliminate
the sensitivity to new physics in the Higgs sector perturbatively at
the one-loop level. Generally speaking, Little Higgs models are
composite models with an additional (global and local) symmetry
structure which allows to greatly reduce the fine tuning compared to
plain composite models. Paradigm implementations are the Littlest
Higgs~\cite{ArkaniHamed:2002qy} and the Simplest Little
Higgs~\cite{Schmaltz:2004de}. The general feature of both types of
models (with global symmetries of either simple Lie algebras or direct
sums) are heavy vector-like fermions, new heavy vector bosons and
modifications of the Higgs couplings of the order $v/f$, the ratio of
the vacuum expectation values of the electroweak symmetries and that
of the Little Higgs symmetries. In many, but not all, models an
extended Higgs sector is predicted. In general, these additional heavy
Higgs bosons are even more difficult to detect at the LHC than those
of the generic, supersymmetric or not, Two-Higgs-Doublett models. 
The mixing with the heavy vector-like fermions leads
to deviations of the electroweak and Yukawa couplings of the SM
fermions, with the effect generically being proportional to the mass
of the SM fermion. This can be seen in a benchmark model in
section~\ref{subsec:ILC_top}.

In many setups of Little Higgs models (as well as composite Higgs
models, too), there are additional $U(1)$ factors in the global
symmetry structure there are similar to the $U(1)$ that gives rise to
the $\eta$ and $\eta^\prime$ masses in the SM. These global symmetries
lead to (sometimes very) light pseudoscalar particles that decay
into the heaviest accessible fermions of the SM (mostly $b$s), two
gluons or two photons~\cite{Kilian:2004pp}.

\subsection{Twin Higgs or Neutral Naturalness models}


In all composite models presented above, the particles responsible for canceling the quadratic divergences in the Higgs mass are charged under the SM gauge symmetries. In particular, the top partner carries color charge, implying a reasonably large minimal production cross section at the LHC. An alternative scenario, which is experimentally quite challenging and might explain the null result in various new physics searches, is the case nowadays referred to as ``neutral naturalness"~\cite{Chacko:2005pe, Chacko:2005un,Craig:2015pha}, where the particles canceling the 1-loop quadratic divergences are neutral under the SM. The canonical example for such theories is the Twin Higgs model of Ref.~\cite{Chacko:2005pe, Chacko:2005un}. This is an example of a pseudo-Goldstone boson Higgs theory, with an approximate global $SU(4)$ symmetry broken to $SU(3)$.  The Twin Higgs model is obtained by gauging the $SU(2)_A\times SU(2)_B$ subgroup of $SU(4)$, where $SU(2)_A$ is identified with the SM $SU(2)_L$, while $SU(2)_B$ is the twin $SU(2)$ group. Gauging this subgroup breaks the $SU(4)$ symmetry explicitly, but   quadratically divergent corrections given do not involve the Higgs boson when the gauge couplings of the two SU(2) subgroups are equal, $g_A=g_B$. The $SU(4)\to SU(3)$ breaking will also result in the breaking of the twin $SU(2)_B$ group and as a result three of the seven Goldstone bosons will be eaten, leaving 4 Goldstone bosons corresponding to the SM Higgs doublet $h$.  In fact imposing the $Z_2$ symmetry on the full model will ensure the cancellation of all 1-loop quadratic divergences to the Higgs mass. Logarithmically divergent terms can however arise for example from gauge loops, leading to a Higgs mass of order $g^2 f /4\pi$, which is of the order of the physical Higgs mass for $f \sim 1$\,TeV. The quadratic divergences from the top sector can be eliminated if the $Z_2$ protecting the Higgs mass remains unbroken by the couplings that result in the top Yukawa coupling. This can be achieved by introducing top partners charged under a twin $SU(3)_c$. In this case the quadratic divergences are cancelled by top partners that are neutral under the SM gauge symmetries. 

Neutral naturalness models are the best motivated scenarios featuring exotic Higgs decays, $h \to XX \to \textrm{SM}$, with displaced vertices, the intermediate decay products being a hidden glueball or a hidden quarkonia that both decay back to the SM via an off-shell Higgs boson~\cite{Craig:2015pha}.


\section{ILC Capabilities for Precision Measurements}
\label{sec:ILC}
%
The ILC physics programme and its added value with respect to the (HL-)LHC has been
studied extensively in the past years and is documented in detailed reports~\cite{Baer:2013cma, Moortgat-Picka:2015yla, Djouadi:2007ik, Asner:2013psa, Baer:2013vqa, Baak:2013fwa, Agashe:2013hma, Vos:2016til},
including a recent update~\cite{Fujii:2015jha} according to a 20-year running strategy~\cite{Barklow:2015tja}.
In this section, we will briefly recall the most important precision measurements and highlight their
implications for physics beyond the Standard Model. We will furthermore summarize the potential to
discover new particles by direct production and present some new results in this area. 

%
\subsection{Higgs Measurements at the ILC}
\label{subsec:ILC_Higgs}
%
%
\paragraph{Higgs Couplings to Fermions and Gauge Bosons:}

With the complete ILC running scenario~\cite{Barklow:2015tja}, the couplings of the H-125 to SM fermions and gauge bosons can be
precisely determined in a model-independent way, most of them to the level of at least $1\%$ or better~\cite{Fujii:2015jha}, thus
typically gaining an order of magnitude compared to HL-LHC projections~\cite{CMS:2013xfa, bib:ATLAS_HLLHC_Higgs}. Notably
the couplings to the $Z$ and the $W$ boson are expected to be measured to better than $0.5$\% at the ILC, while invisible decay modes will be constrained to be less than $0.3$\% at $95$\%\,CL.
Beyond measuring the absolute size of the couplings, also their $CP$ properties can be probed with high precision at the ILC.
A recent full simulation study showed that 
at the ILC $\sqrt{s}=250$ GeV the $CP$ phase $\phi$ can be measured with a precision of
$3.8^\circ$ using $e^+e^-\to Zh$ and 2 $\rm{ab}^{-1}$ data \cite{Jeans:2016}, which is already a factor 2-3 better than 
expected at the HL-LHC~\cite{Harnik:2013aja}.
Better sensitivities are expected in the full H20 ILC run scenario by further including the data at $\sqrt{s}=500$ GeV. 
It is also possible to probe the Higgs $CP$ mixing in $htt$ coupling
at the ILC using the channel $e^+e^-\to t\bar{t}h$~\cite{BhupalDev:2007ftb,Godbole:2011hw,Hagiwara:2016rdv}, for which
a sensitivity for $\phi$ of $\sim15^\circ$ has been estimated on generator level assuming $2.5$\,ab$^{-1}$ of data at 1 TeV \cite{Asner:2013psa}.

\begin{figure}[ht]
  \centering
    \subfigure[]{\includegraphics[width=0.44\linewidth]{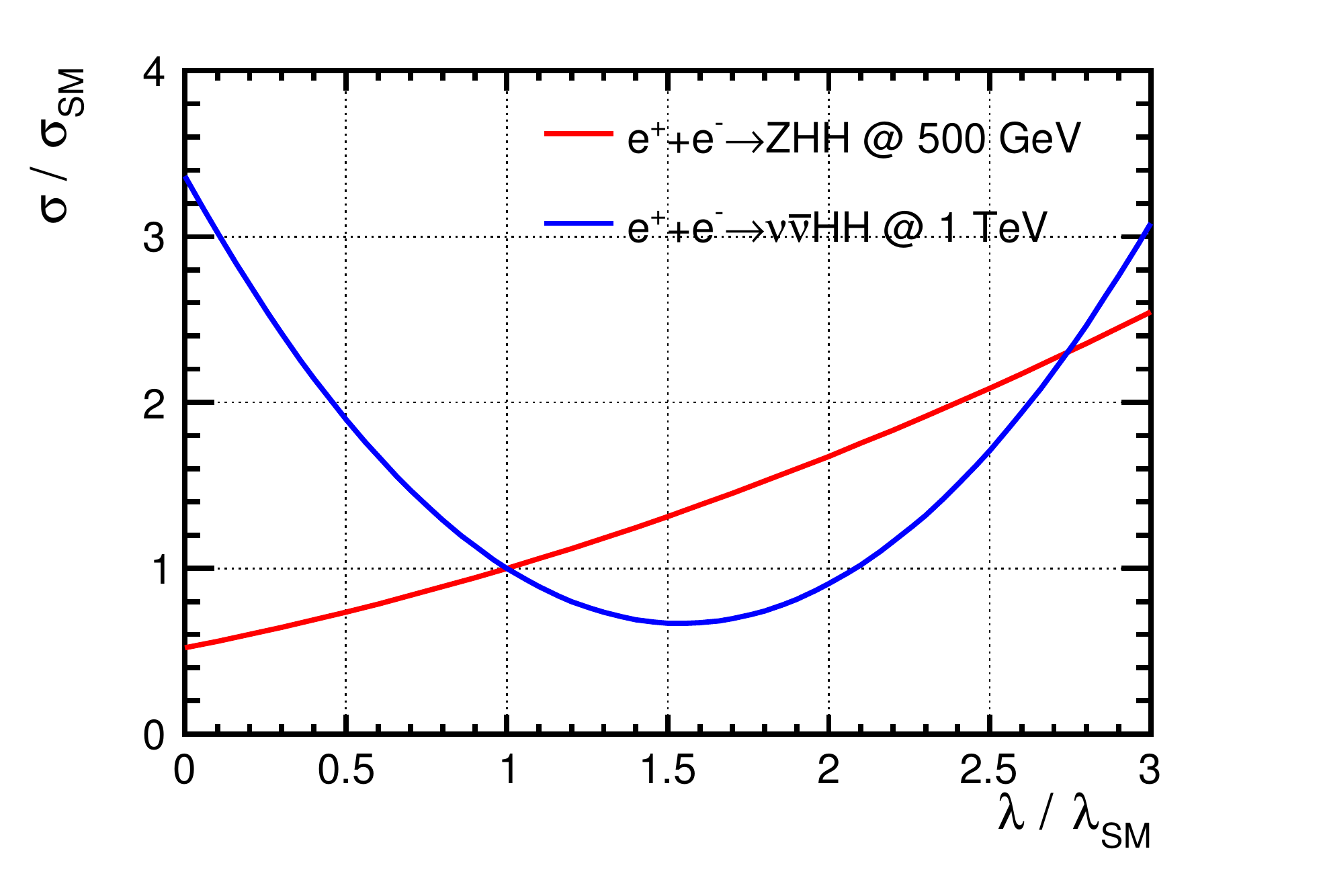}}
    \hspace{0.1\linewidth}
    \subfigure[]{\includegraphics[width=0.44\linewidth]{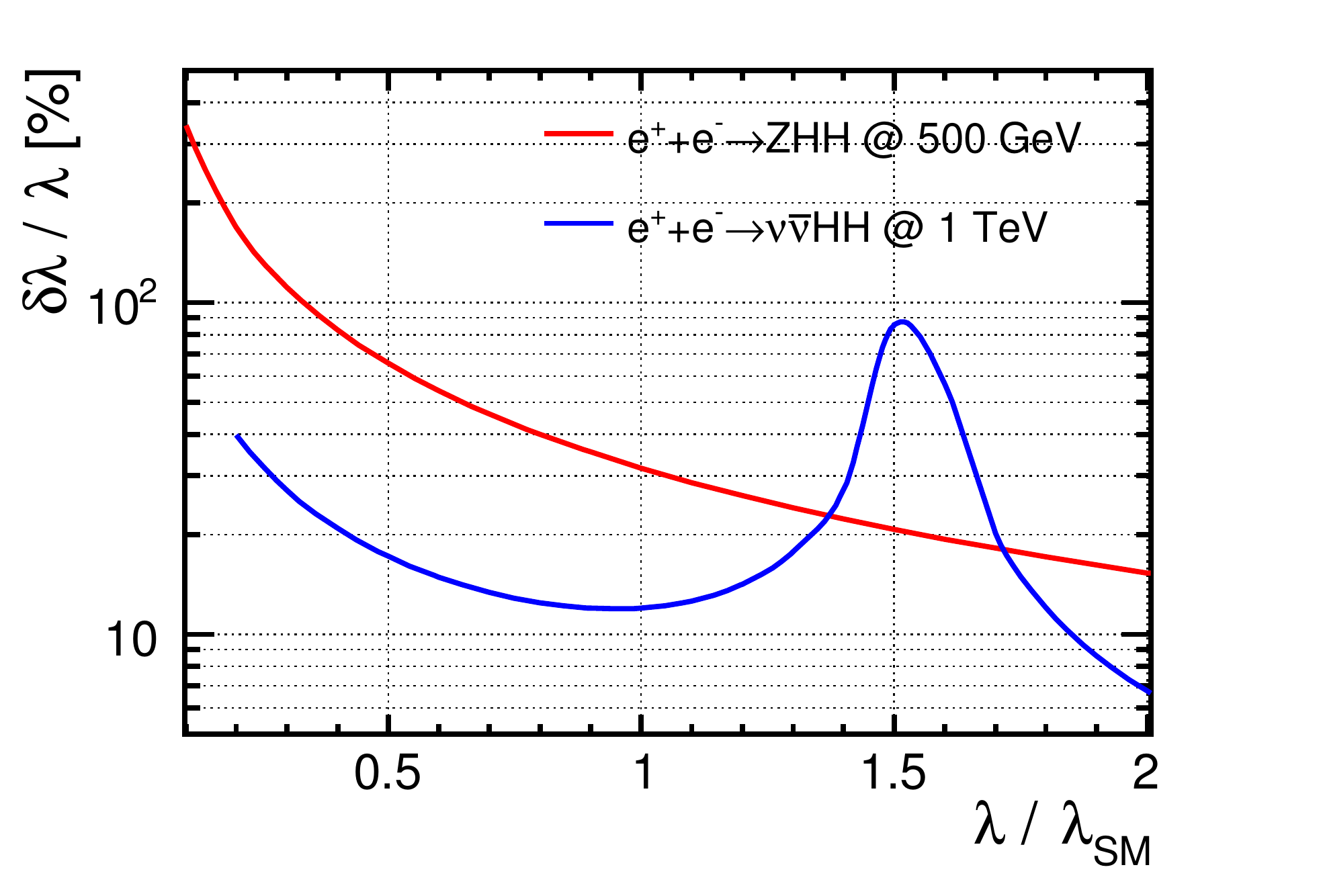}}   
  \caption{(a) Cross sections of the two major double Higgs production processes at the ILC normalized to the SM values
%
as a function of $\lambda / \lambda_{SM}$, 
  (b) Expected relative precisions on $\lambda$ as a function of $\lambda/ \lambda_{SM}$ for those two processes.}
  \label{fig:HHHBSM}
\end{figure}

\paragraph{Higgs Self-Coupling:}
At the ILC, current projections based on the full simulation studies suggest that $\lambda_{SM}$ 
can be measured with a precision of 26\% at $\sqrt{s}=500$ GeV for the H20 scenario~\cite{Duerig:2016dvi}, and with a 
precision of 10\% at $\sqrt{s}=1$ TeV with 8 $\rm{ab}^{-1}$ data~\cite{Tian:2013,Kurata:2013}. 
Recent studies \cite{Tian:2015} pointed out that if $\lambda$ deviates from its SM prediction, the projections can be dramatically modified as shown in 
Fig.~\ref{fig:HHHBSM} which shows the cross sections (left) and the expected precisions of $\lambda$ (right)
 as a function of the value of $\lambda$. In the interesting case $\lambda=2\lambda_{SM}$, 
 the cross section of $e^+e^-\to Zhh$ at $\sqrt{s}=500$ GeV will be enhanced by 60\%, and the expected
 precision on $\lambda$ will be a factor of 2 better, $\delta\lambda/\lambda\sim15\%$,
 which indicates that it is possible to not only discover the trilinear Higgs self-coupling by $7\sigma$
 significance, but also to see deviation with respect to its SM expectation with $>3\sigma$ significance, at the $500$\,GeV ILC.

\subsection{Top Quark Measurements at the ILC}
\label{subsec:ILC_top}
%
%
\newcommand{\eplus}{e^+}
\newcommand{\eminus}{e^-}
\newcommand{\epem}{\eplus\eminus}
\newcommand{\ttbar}{ \tQ \bar{\tQ}}
\def\roots{ \sqrt{s} }
\newcommand{\pem} { {\cal P}_{\e-} } 
\newcommand{\pep} {{\cal P}_{\e+} }
\definecolor{orange}{rgb}{1,0.5,0}
\definecolor{gray}{rgb}{0.5,0.5,0.5}
\definecolor{roetlich}{rgb}{1, .7, .7}
\definecolor{camel}{rgb}{0.76, 0.6, 0.42}
\definecolor{bronze}{rgb}{0.8, 0.5, 0.2}
\definecolor{britishracinggreen}{rgb}{0.0, 0.26, 0.15}

The ILC would allow to study the $t$ quark using a precisely defined leptonic initial state. Therefore individual events can be analysed in more detail. It also changes the production mechanism for $\tQ$ quark pairs from the strong to the electro-weak interactions, which are a step closer to the phenomena of electro-weak symmetry breaking. Finally, this change brings into play new experimental observables -- weak interaction polarisation and parity asymmetries -- that are very sensitive to the coupling of the $t$ quark to possible new interactions. It is very possible that, while the $t$ quark might respect Standard Model expectations at the LHC, it will break those expectations when studied with higher precision at the ILC. 
%
\paragraph{Top Quark Mass:} 
\label{subsubsec:top_mass}
One of the unique capabilities of an $\epem$ collider is the ability
to carry out scans of particle production thresholds.  The $\ttbar$ pair production threshold around a centre-of-
mass energy $\roots \approx 2m_t$ enables a precise measurements of the $t$ quark mass
$m_t$ in theoretically well-defined mass schemes. This is in contrast to the mass measurements at the LHC, where the highest precision is obtained in measurements relying on the use of event generators, resulting in additional, currently not well understood, uncertainties when translating the experimental result to mass definitions used in theoretical frameworks.




Using the methodology described in \cite{Seidel:2013sqa} with state-of-the-art NNNLO QCD calculations of $\ttbar$ production  \cite{Beneke:2015kwa} as input, the effects of the ILC luminosity spectrum and initial state radiation as well as signal efficiencies and background contributions are taken into account. From the simulated data points, the statistical precision as well as theoretical uncertainties based on NNNLO scale uncertainties are extracted following the techniques developed in \cite{Simon:2016htt, Simon:2016pwp}, resulting in a statistical precision of $\delta m_t \approx 13 \MeV$. At present, the scale uncertainties result in a theory systematic of $\sim$40 MeV, which is comparable to the expected experimental and parametric systematics. Table \ref{tab:Top:Systematics} summarizes the current status of the estimated uncertainties of the top quark mass measurement in a threshold scan.  These studies are performed assuming unpolarized beams. In the combined uncertainties, the lower end of the given range illustrates the effect of some mild improvements assumed on $\alpha_s$ and theory uncertainties expected by the time of ILC data taking, as well as a better suppression of non-$\ttbar$ background due to the capability for high beam polarization of electrons and positrons at the ILC.

\renewcommand{\arraystretch}{1.2}
\begin{table}
\centering
\begin{tabular}{l|c|l}
\hline
\hline 
error source & $\Delta m_t^{\mathrm{PS}}$ [MeV] & references\\
\hline
stat. error (200 fb$^{-1}$) & 13 & \cite{Seidel:2013sqa, Simon:2016pwp}\\
\hline
theory (NNNLO scale variations, PS scheme) & 40 & \cite{Simon:2016htt, Simon:2016pwp}\\
parametric ($\alpha_s$, current WA) & 35 & \cite{Simon:2016htt}\\
\hline
non-resonant contributions (such as single top) & $< 40$ & \cite{Fuster:2015jva}\\
residual background / selection efficiency & 10 -- 20 & \cite{Seidel:2013sqa}\\
luminosity spectrum uncertainty & $< 10$ & \cite{Simon:2014hna}\\
beam energy uncertainty & $< 17$ & \cite{Seidel:2013sqa}\\
\hline
\hline
combined theory \& parametric & 30 -- 50 & \\
combined experimental \& backgrounds & 25 - 50 &\\
\hline
total (stat. + syst.) & 40 -- 75 & \\
\hline
\hline
\end{tabular}
\caption{\label{tab:Top:Systematics} Summary of the estimated uncertainties of top mass measurements at threshold. The upper parts of the table reflect the current understanding, based on the references given. In the bottom part of the table, the lower end of the given ranges corresponds to moderate assumptions on improvements expected by the time of ILC data taking, see text.}
\end{table}


The dependence of the $t$ quark cross section shape on the $t$ quark mass and interactions is computable to high precision with full control over the renormalisation scheme dependence of the $t$ quark mass parameter. The authors of~\cite{Marquard:2015qpa} show that the PS or 1S masses as resulting from the described analysis can be translated to e.g. the $\overline{MS}$ mass, typically used in theoretical calculations to a precision of about $10 \MeV$.

\paragraph{Top Quark Electroweak Couplings:}
The unique feature of linear colliders to provide polarised beams allow for a largely unbiased disentangling of the individual left- and right-handed couplings of the $t$ quark to the $Z^0$ boson and the photon, $ g_{L,R}^{\gamma, Z}$ or equivalently of the form factors $F^{\gamma, Z}_{(1,2),(V,A)}$. These quantities can be measured at the sub-percent level at the ILC~\cite{Amjad:2013tlv,Amjad:2015mma}, as indicated by the red ellipse in Figure~\ref{fig:models-rp}. This is-- when referring to the results in~\cite{Baur:2004uw,Baur:2005wi}-- considerably better than will be possible at the LHC\footnote{The improving analyses of the LHC experiments, as e.g.~\cite{Khachatryan:2015sha}, will however be observed with great interest.} even with an integrated luminosity of ${\cal L} = 3000\,\ifb$. The expected precision at the ILC would allow for the verification of a great number of models for physics beyond the Standard Model, for which representative examples are given in the figure.

\begin{figure}[ht]
  \centering
\setlength{\unitlength}{1.3mm}
\begin{picture}(150,80)
\linethickness{0.3mm}
  \put(9,33){\line(0,1){2}}
  \put(5,34){\line(1,0){8}}  
  \put(15,34){...} 
  \put(21,34){\vector(1,0){60}} 
  \put(82,33){\Large{$\delta g^Z_R / g^Z_R$}}
  \multiput(31,33)(10,0){5}{\line(0,1){2}} 
  \put(6.5,31){\footnotesize{-330\%}} 
  \put(29,31){\footnotesize{-20\%}} 
  \put(39,31){\footnotesize{-10\%}}
  \put(60,31){\footnotesize{10\%}}
  \put(70,31){\footnotesize{20\%}}
  \put(51,4){\vector(0,1){60}} 
  \put(45,66){\Large{$\delta g^Z_L / g^Z_L$}}
  \multiput(50,14)(0,10){5}{\line(1,0){2}} 
  \put(44,13.5){\footnotesize{-20\%}} 
  \put(44,23.5){\footnotesize{-10\%}} 
  \put(44.5,43.5){\footnotesize{10\%}} 
  \put(44.5,53.5){\footnotesize{20\%}}
  \put(51,34){\color{red}\circle*{2}} 
  \put(53,36){\color{red}SM}
  \put(51,24){\color{britishracinggreen}\circle*{1.5}}
  \put(53,23){\color{britishracinggreen}Light top partners~\cite{Grojean:2013qca}}
  \put(41,24){\color{britishracinggreen}\circle*{1.5}} 
  \put(15,26){\color{britishracinggreen} Light top partners}
 \put(15,23){\color{britishracinggreen} Alternative 1~\cite{bib:panico-priv}}
  \put(76,59){\color{britishracinggreen}\circle*{1.5}} 
  \put(54,61.5){\color{britishracinggreen} Light top partners Alternative 2~\cite{bib:panico-priv}}
   \put(51,19){\color{cyan}\circle*{1.5}} 
   \put(53,18){\color{cyan}Little Higgs~\cite{Berger:2005ht}}
  \put(51,14){\color{gray}\circle*{1.5}} 
  \put(53,13){\color{gray}RS with Custodial SU(2)~\cite{Carena:2006bn}}
  \put(51,9){\color{orange}\circle*{1.5}} 
  \put(53,8){\color{orange}Composite Top~\cite{Pomarol:2008bh}}
  \put(31,14){\color{magenta}\circle*{1.5}} 
  \put(20,16){\color{magenta}5D Emergent~\cite{Cui:2010ds}}
  \put(56,29){\color{camel}\circle*{1.5}} 
  \put(57,28){\color{camel}4D Composite Higgs Models~\cite{Barducci:2015aoa}}
  \put(9,34){\color{blue}\circle*{1.5}} 
  \put(1,36){\color{blue}RS with Z-Z' Mixing~\cite{Djouadi:2006rk}}
\multiput(56,40)(30,0){2}{\line(0,1){12}} 
\multiput(56,40)(0,12){2}{\line(1,0){30}}
\put(61,47){\large{ILC Precision}}
\color{red}
\qbezier(73.3463, 43.7167)(73.2699, 43.9671)(72.5286, 43.9342)
\qbezier(72.5286, 43.9342)(71.7873, 43.9013)(70.8154, 43.6044)
\qbezier(70.8154, 43.6044)(69.8435, 43.3075)(69.2103, 42.9205)
\qbezier(69.2103, 42.9205)(68.5772, 42.5336)(68.6537, 42.2833)
\qbezier(68.6537, 42.2833)(68.7301, 42.0329)(69.4714, 42.0658)
\qbezier(69.4714, 42.0658)(70.2127, 42.0987)(71.1846, 42.3956)
\qbezier(71.1846, 42.3956)(72.1565, 42.6925)(72.7897, 43.0795)
\qbezier(72.7897, 43.0795)(73.4228, 43.4664)(73.3463, 43.7167)
\end{picture}
\caption{\sl Predictions of several Randall-Sundrum (RS) models and/or compositeness or Little Higgs models on the deviations of the left- and right-handed couplings of the $t$~quark to the $Z^0$ boson. The ellipse in the frame in the upper right corner indicates the precision that can be expected for the ILC at $\roots = 500\,\GeV$ with ${\mathcal L}=500\ifb$ of integrated luminosity shared equally between the beam polarisations $\pem,\,\pep =\pm0.8,\mp0.3$~\cite{Richard:2014upa}. 
}
\label{fig:models-rp}
\end{figure}
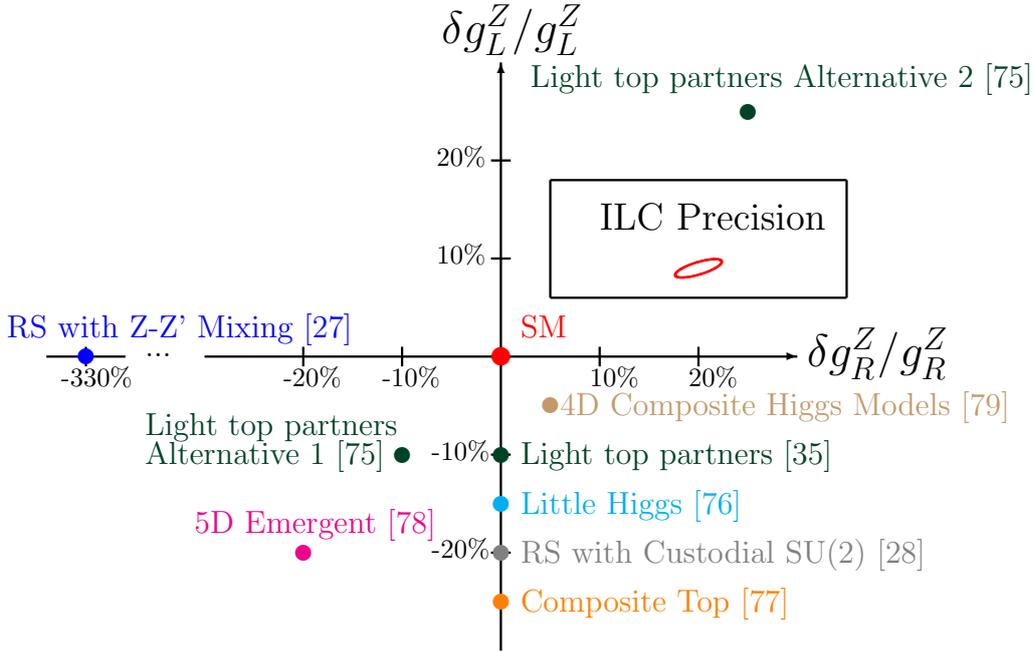


Beam polarisation is a critical asset for the high precision measurements of the electroweak $t$ quark couplings. Experimental and theoretical effects manifest themselves differently for different beam polarisations. It seems to be that the configuration positive electron beam polarisation is more benign in both, experimental aspects due to the suppression of migrations in the polar angle spectrum of the final state $t$ quark, see e.g.~\cite{Amjad:2013tlv,Amjad:2015mma}  and theoretical aspects due to the somewhat simpler structure of higher order electroweak corrections~\cite{Khiem:2015ofa}. 
\subsection{Electroweak Precision Measurements at the ILC}
\label{subsec:ILC_ew}
%
%
\paragraph{Two-fermion production:}
Precise measurements of $e^+e^- \to f \bar f$ for all types of fermions, 
making use of polarized beams, are a stringent test
on the existence of possible forces between matter particles beyond
those in the SM. Those comprise gauge interactions (generically labelled $W'$ and $Z'$), scalar resonances
like the Higgs or also tensor resonances as in gravity models or
models with new strong interactions. Also compositeness or partial
compositeness of fermions could lead to deviations of two-fermion
processes from their SM values due to mixing effects. 
In $e^+e^-$ collisions (far) below the production threshold for the extra gauge
bosons, they manifest themselves as deviations from SM predictions due to
interference between the new physics and the SM $\gamma/Z$ contributions,
see~\cite{Freitas:2013xga} for a recent review. 
This is similar to indirect observations of the presence of the $Z$ boson
in $e^+e^-\rightarrow \mu^+\mu^-$ scattering at the PETRA and TRISTAN 
colliders well before the turn-on of LEP. 

Studies for the ILC~\cite{Osland:2009dp} have shown that
already with $500$\,fb$^{-1}$ at 500\,GeV or with $1$\,ab$^{-1}$ at 1~TeV ILC 
evidence for a $Z'$ with mass exceeding $\sim 7$\,TeV and $\sim 12$\,TeV can be obtained in 
many models, respectively. The measurements will allow to distinguish between
a variety of different $Z'$ models by measuring their vector and axial-vector couplings, starting already at 
$\sqrt{s} = 500$~GeV~\cite{Godfrey:2005pm}. Similar numbers hold also for
$W'$~bosons~\cite{Godfrey:2000hc}. 


\paragraph{$W$ Mass:}
At the ILC, the $W$ mass can be measured by three different methods:
(i) by a polarized threshold scan in different final states, (ii) by
measuring the $\ell\nu jj$ and partially also the $jjjj$ process at
the design energies of 250, 350 and 500\,GeV as well as (iii) from
single-$W$ production ($e^+e^- \to W \ell\nu$) favorably at the
highest possible energies. Both methods (ii) and (iii) rely on
kinematic reconstruction of the $W$ system similar to the top-quark
mass measurement at hadron colliders, while (i) uses in principle
template fits for the $WW$ threshold curve. These methods have been
shown to be able to reach precisions of $\delta M_W = 3$\,MeV~\cite{Baak:2013fwa},
which presents an improvement of a factor $5$ w.r.t.\ current precisions.

\paragraph{Weak mixing angle $\sweff$:}
When exploiting the option to operate the ILC at the $Z$ pole with polarised beams, collecting
about 1000 times more events than at LEP, the effective weak mixing angle can be measured
about a factor~10 better than today~\cite{Baak:2013fwa}. 

\section{Direct Production of New Particles at the ILC}
\label{sec:ILC_NP}
%
%
In this section, we summarize the potential of ILC for directly 
producing new particles, thus updating an earlier 
Snowmass white paper~\cite{Baer:2013vqa}. 
In particular, we highlight cases where the discovery potential of ILC 
is complementary to that of the LHC.

\subsection{Generic WIMPs}
\label{subsec:directNP_WIMP}

The prospects to detect WIMPs at the ILC and to determine their properties have been studied at a theoretical level~\cite{Baer:2001ia,Birkedal:2004xn,Dreiner:2012xm,Chae:2012bq} and specifically for a center-of-mass energy of $500$\,GeV in detailed detector simulation~\cite{Bartels:2012ex}. The experimental prospects have been interpreted both in the framework of effective
operators\footnote{Note that under ILC conditions the effective field theory approximation is accurate, while it is questionable
in similar analyses at hadron colliders. Reinterpretations of ILC projections in simplified models are in preparation for comparison with upcoming LHC results.} as well as in a cosmological approach, where the annihilation fraction $\kappa_e$ of WIMPs
into electron-positron pairs is a free parameter, assuming that the WIMP makes up all the dark matter in the universe.
As can be seen in the left panel of figure~\ref{fig:WIMPs}, evidence for WIMP production could be obtained over a wide range
of masses already with an initial ILC dataset of $500$\,fb$^{-1}$, even if the annihilation fraction into $e^+e^-$ pairs provides only a few percent of the total annihilation rate~\cite{bib:chaus}. The right panel shows the extrapolation of these results to a wide range of integrated luminosities and center-of-mass energies based on an effective operator approach~\cite{bib:habermehl}. For the
full $500$\,GeV-program of the ILC, scales of new physics ($\Lambda$) of up to $3$\,TeV can be probed, while the $1$\,TeV-energy-upgrade of the ILC would extend this even to $4.5$\,TeV or more, depending on the integrated luminosity.
Once a WIMP would be discovered, its properties could be determined precisely due to the known initial state of a lepton collider~\cite{Bartels:2012ex}. 
In particular, its mass could be determined with a precision of about 1\%, and the type of operator (or the angular momentum of dominant partial wave) of the WIMP pair production process can be determined. 
By such detailed measurements of WIMP properties as offered at the ILC, it is often possible to constrain WIMP production rates in the early universe along with WIMP scattering or annihilation
rates and the local WIMP abundance~\cite{Baltz:2006fm}. Such checks could verify or falsify the simple assumptions 
associated with thermal DM production within the WIMP miracle scenario, thus giving
important insights into the nature of dark matter, as explained in section~\ref{sec:DM}.

\begin{figure}[htb]
\setlength{\unitlength}{1.0cm}
\includegraphics[width=0.55\linewidth]{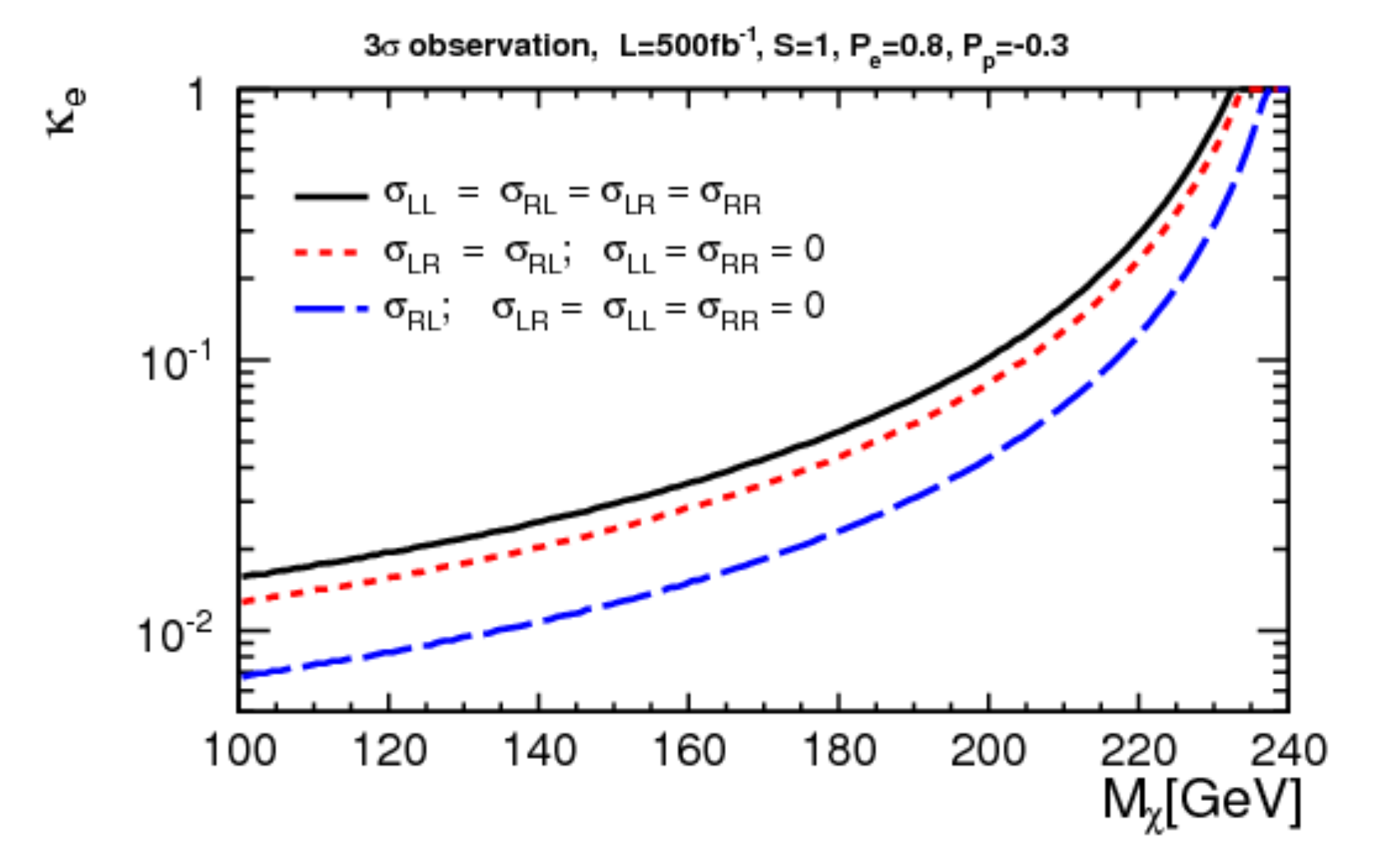}
\hspace{0.1cm}
\includegraphics[width=0.45\textwidth]{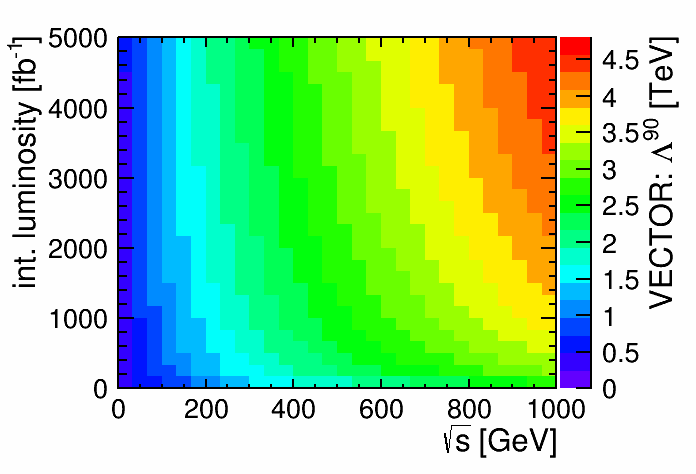}
  \caption{\label{fig:WIMPs} Left: Observational reach ($3\sigma$) of the ILC for a Spin-1 WIMP in terms of WIMP mass and $\kappa_e$ for three different chiralities of the WIMP-fermion couplings~\cite{bib:chaus}.
  Right: Expected sensitivity for a vector operator in an EFT-based interpretation as a function of integrated luminosity and center-of-mass energy~\cite{bib:habermehl}.}
\end{figure}

\subsection{SUSY with no loopholes}
\label{subsec:directNP_NoLoophole}

The ILC will be able to detect new particles with electroweak interactions nearly up to the kinematic
limit of $\sqrt{s}/2$. In particular,  in SUSY 
(where, as discussed in section~\ref{subsec:SUSY}, the couplings
cannot become arbitrarily small but are given by the couplings of the corresponding SM partners and their mixings) systematic, loophole-free searches can be performed for production of pairs of NLSPs. In the $R$-parity conserving
case they have to decay into the LSP and their SM partner (either on-shell or virtual), and in the clean environment
of the ILC, these decays can be detected even for extremely small mass difference. Figure~\ref{fig:noloophole1}
shows as an example the experimentally most challenging case of a $\stone$-NLSP~\cite{Berggren:2013vna}; in other cases,
the discovery reach approaches even closer to $\sqrt{s}/2$. 

\begin{figure}[tb]
   \centering

     \subfigure[]{\includegraphics[width=0.35\linewidth]{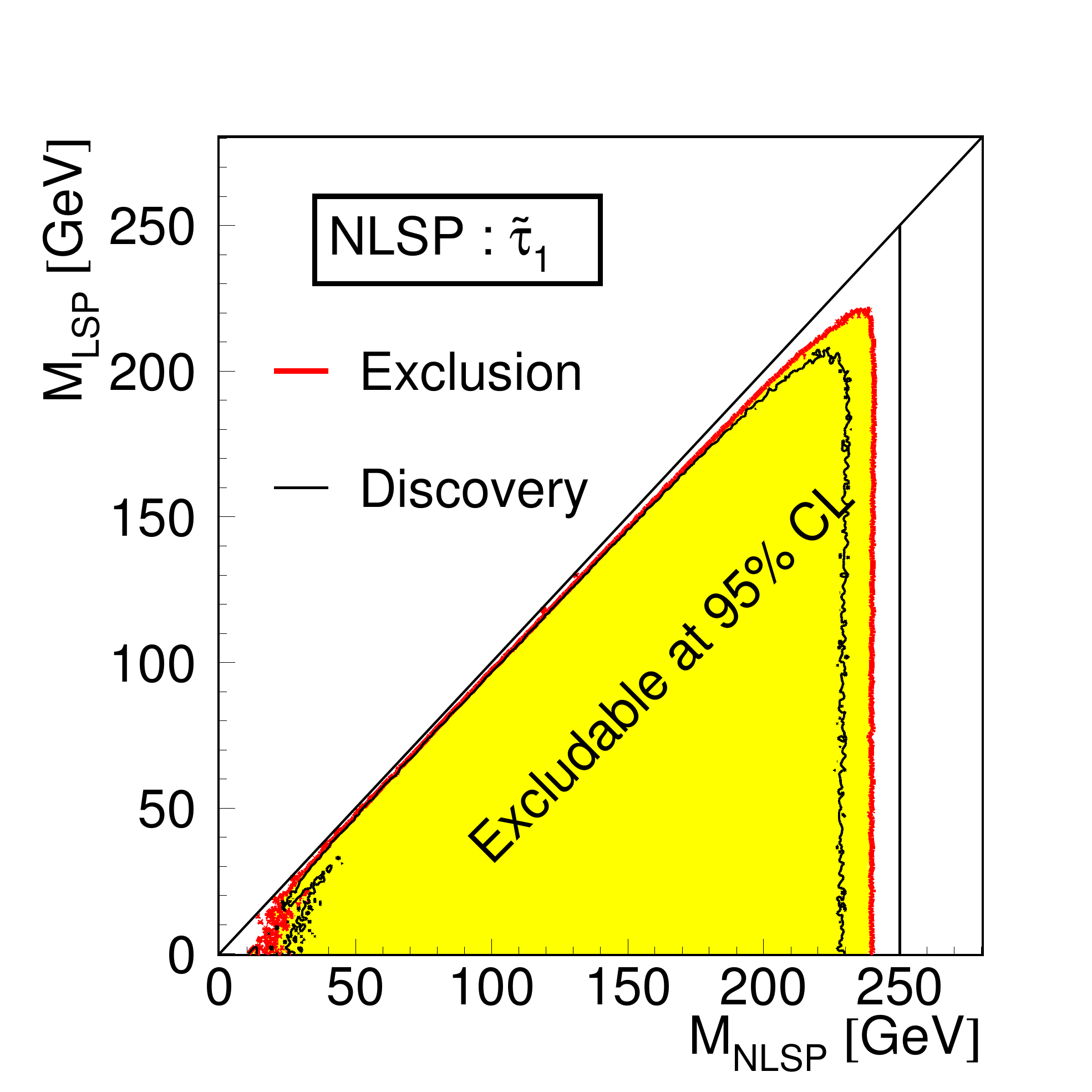}}
     \hspace{0.1\linewidth}
     \subfigure[]{\includegraphics[width=0.35\linewidth]{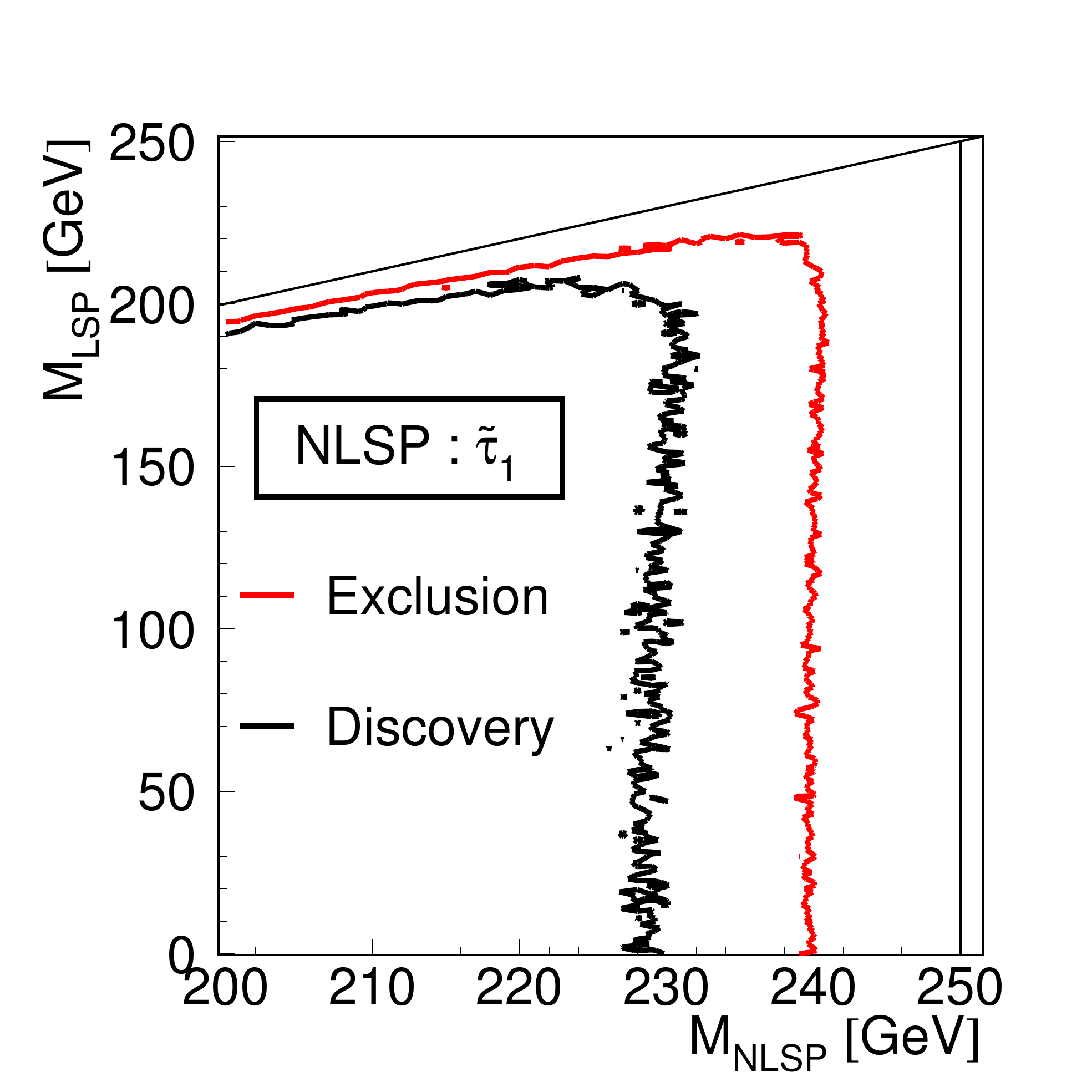}}
\caption{ILC discovery reach for a $\stone$ (bottom) NLSP for $\int \mathcal{L} \, \mathrm{dt}$ = 
500 fb$^{-1}$ at $\sqrt{s}$ = 500 GeV. The mixing angle was chosen to give the lowest
possible production cross-section. (a) full scale, (b) zoom to last few GeV before the
kinematic limit~\cite{Berggren:2013vna}. \label{fig:noloophole1}}
\end{figure}

\subsection{SUSY Dark Matter}
\label{subsec:directNP_SUSYDM}

Over a large region of SUSY parameter space, co-annihilation with the NLSP
is an attractive mechanism which acts to reduce the relic density of the LSP to its
cosmologically observed value~\cite{deVries:2015hva}. Co-annihilation requires
a small mass difference between the NLSP and the LSP in order to be effective,
and thus the expected value of the relic density depends strongly on the exact
masses and mixings of the involved particles, requiring measurements at the permille and
percent-level, respectively~\cite{Lehtinen:2016qis}. 
Here, we present as an example the case of $\stau$ co-annihilation, for which
the ILC prospects have been studied for different benchmark scenarios~\cite{Bechtle:2009em, Berggren:2015qua}. 
In these studies it has been shown that the relevant masses of the
NLSP and the LSP can be measured at the permille-level from kinematic edges,
but also from measuring cross sections near production thresholds. Both methods are illustrated
in Fig.~\ref{fig:sleptons}. The threshold scan in addition determines the spin of the particle
unambiguously, in this case as a scalar.
The production cross sections and
the polarisation of the $\tau$ from $\stone$ decays can be determined at the percent level~\cite{Bechtle:2009em}, 
giving access to the mixing of the $\stone$ and the bino content of the \XN{1}.

\begin{figure}[htb]
  \begin{center}
    \subfigure[]{\includegraphics[width=0.3\linewidth]{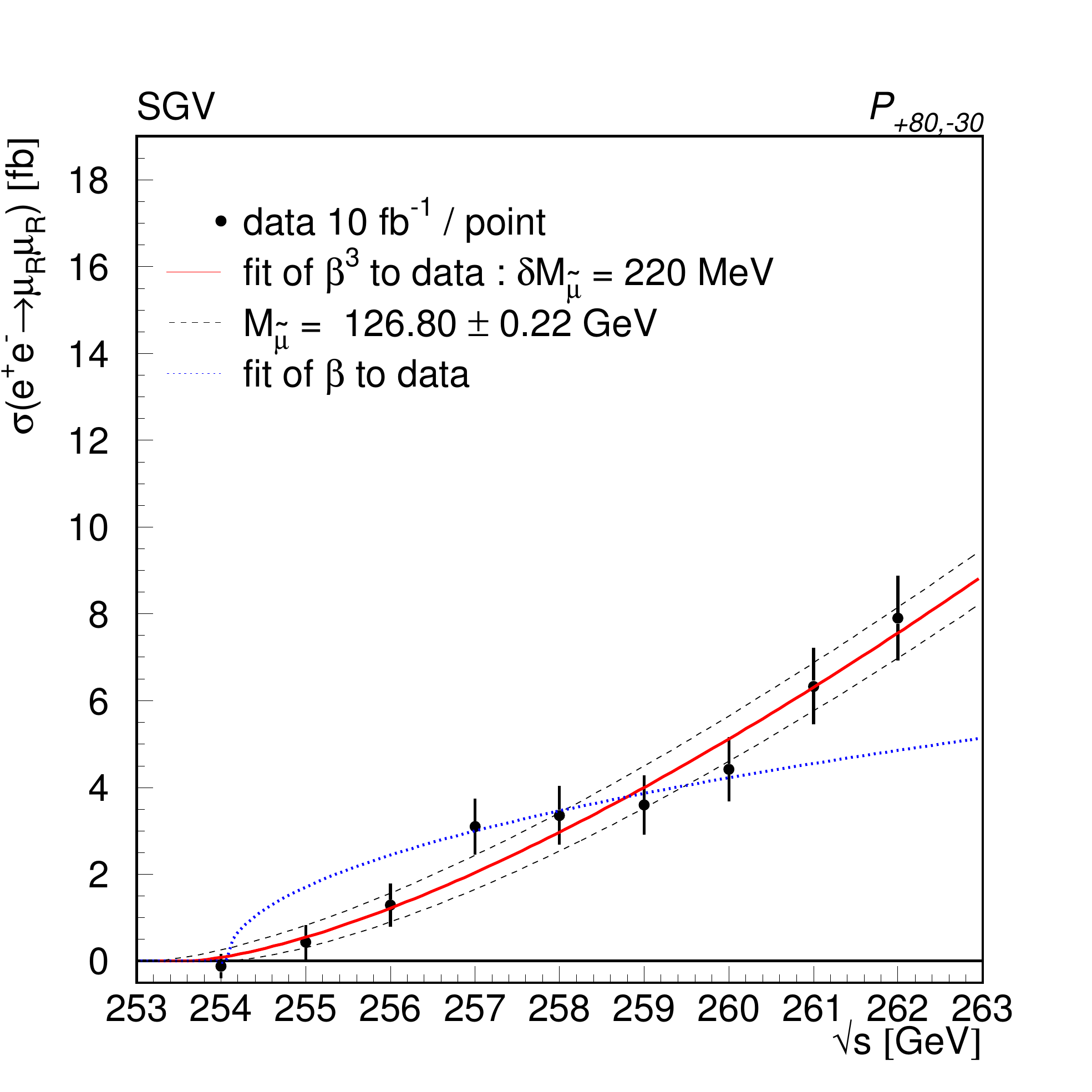}}
    \hspace{0.01\linewidth}
    \subfigure[]{\includegraphics[width=0.3\linewidth] {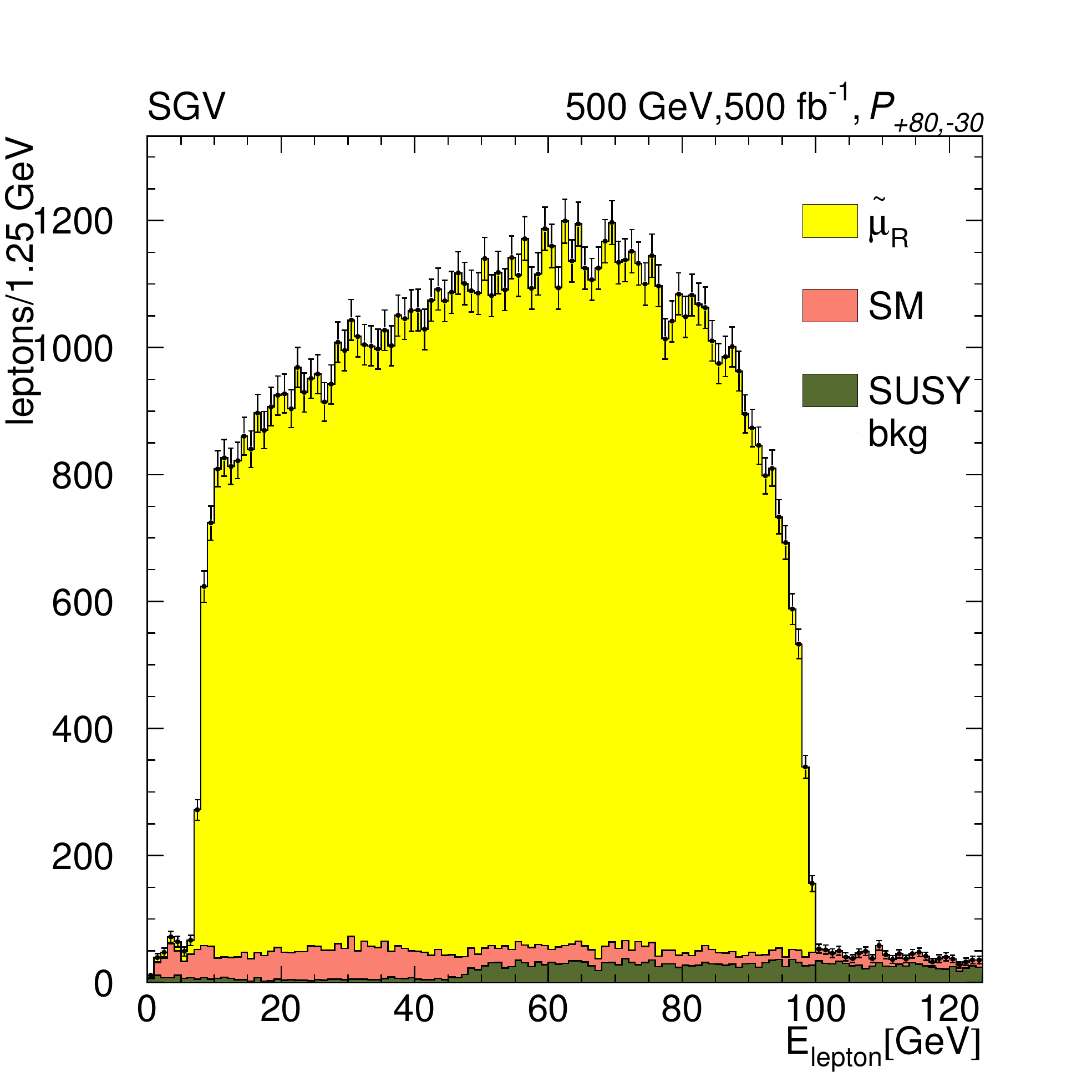}}
    \hspace{0.01\linewidth}
    \subfigure[]{\includegraphics [width=0.3\linewidth]{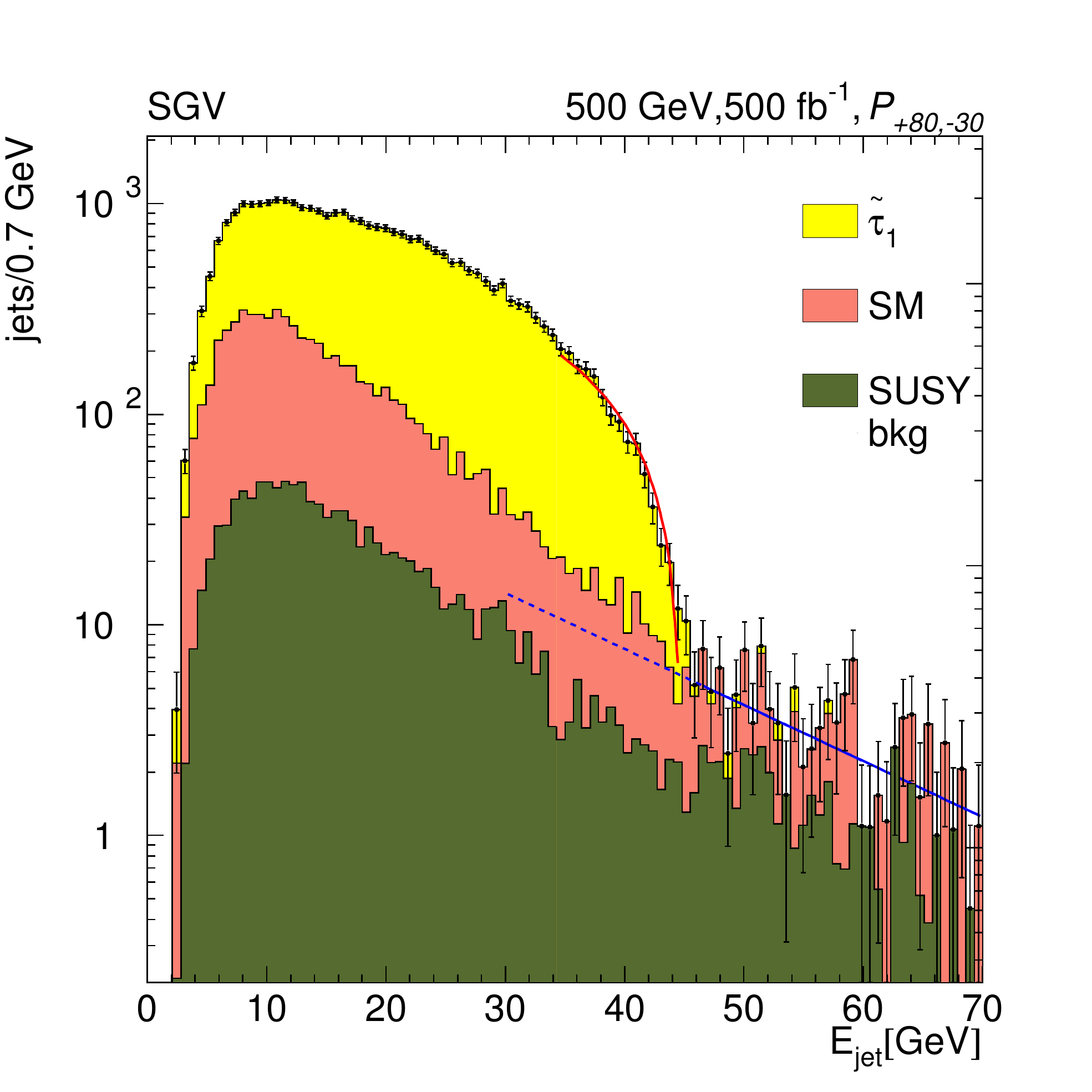}}
  \end{center}
  \caption{\label{fig:sleptons} Property determination of SUSY Dark Matter: (a) scan of the $\eeto \smur \smur$ threshold, (b) muon and (c) $\tau$-jet energies in selected di-leptons events
after collecting
500 fb$^{-1}$ of data for beam-polarisation  $\mathcal{P}_{-80,+30}$~\cite{Berggren:2015qua}. }
\end{figure}

With such precision, ILC has the capacity to cross check DM properties~\cite{Baltz:2006fm} 
such as the thermally-produced relic abundance with a precision
close to the cosmological observations on CMB from the Planck collaboration,
and hence to determine if indeed the lightest neutralino constitutes the bulk of 
dark matter in the universe --- 
or whether other non-thermal and/or non-WIMP processes play an important role in determining the
ultimate dark matter relic density, as discussed in section~\ref{subsubsec:SUSYDM}.

\subsection{Light Higgsinos}
\label{subsec:directNP_higgsinos}
As discussed in section~\ref{sec:BSM-SUSY-Naturalness}, light higgsinos are a fundamental requirement 
of natural SUSY models, while the other SUSY particles can be more heavy.
The ILC with center-of-mass energy of $\sqrt{s}=500$~GeV will be able to 
cover the most natural portion of this parameter space while the remainder
would be fully covered with an energy upgrade.
Mass differences within the higgsino sector are small, typically  below $20$\,GeV,
depending on the values of the other SUSY parameters, in particular the Bino and Wino
mass parameters $M_1$ and $M_2$. In the clean environment of the ILC, their soft visible
decay products can be easily detected --- without any need to rely on large-mass-gap decays
of heavier particles. The ILC capabilities 
have been studied in detector simulations performed for different benchmark points with mass 
differences ranging from $770$\,MeV~\cite{Berggren:2013vfa}
to $20$\,GeV~\cite{Baer:2016new}. 
Two examples of the striking signals and the extraction of 
kinematical endpoints are given in Fig.~\ref{fig:higgsinos}. 
The resulting precisions on masses and polarised 
cross sections reach the percent level even in the experimentally most difficult 
cases and allow to determine other SUSY parameters, as will be discussed in secttion~\ref{sec:LHC_scen}. 
They will also play an important role in unveiling the nature of dark matter: 
in this case with the result that the LSP only contributes a small fraction of the 
total abundance. Such a situation might call for additional, non-WIMP constituents of 
dark matter such as axions: c.f.\ section~\ref{subsubsec:SUSYDM}.

\begin{figure}[htb]
  \begin{center}
    \subfigure[]{\includegraphics[width=0.44\linewidth] {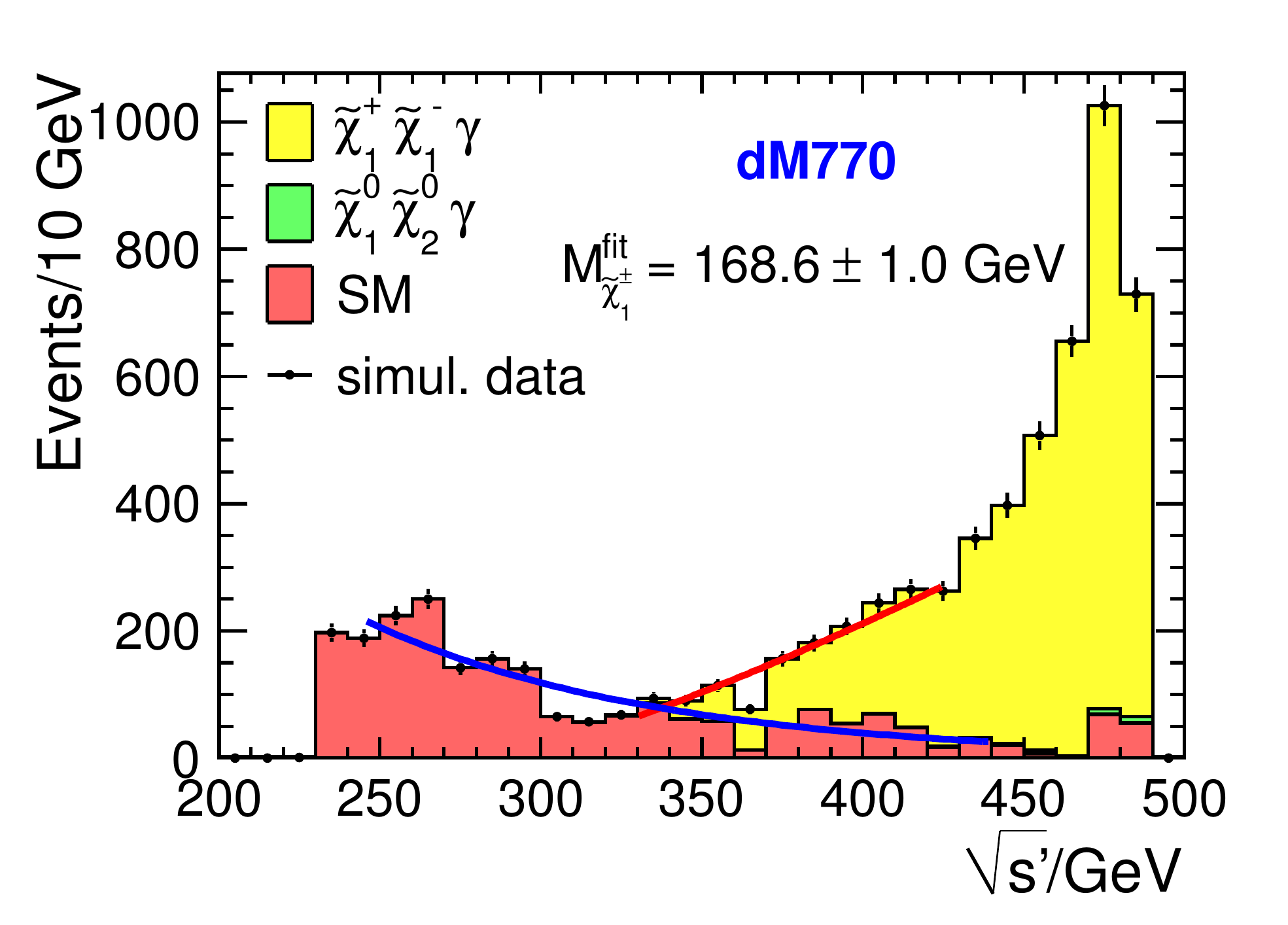}}
    \hspace{0.1\linewidth}
    \subfigure[]{\includegraphics [width=0.44\linewidth]{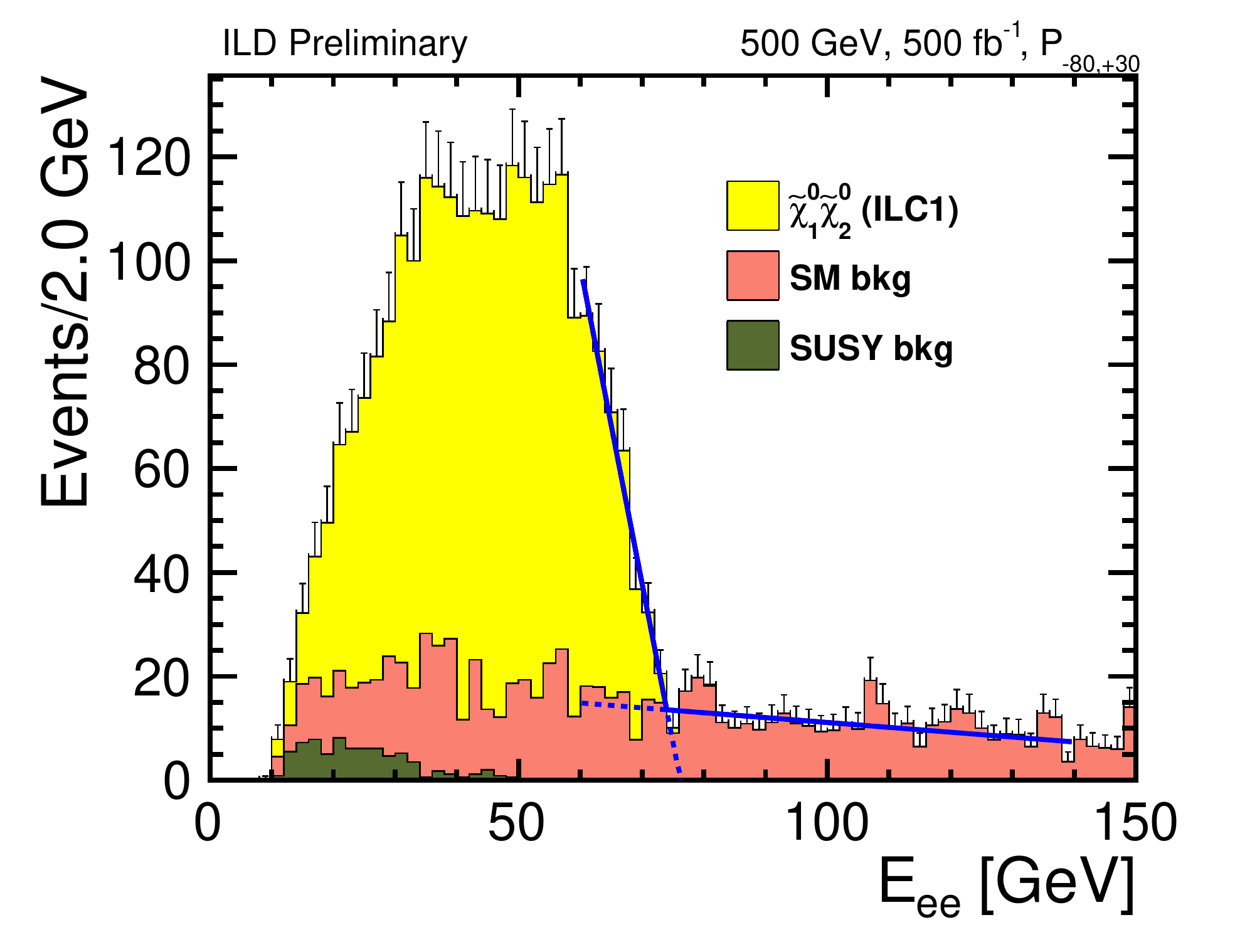}}
  \end{center}
  \caption{\label{fig:higgsinos} Higgsino mass determination for (a) the charged higgsino from the recoil against an ISR photon in a scenario with a mass splitting of $770$\,MeV~\cite{Berggren:2013vfa}, (b) the neutral higgsino from the energy of its visible decay products in a scenario with a mass splitting of  $20$\,GeV~\cite{Yan:2016LCWS, Baer:2016new}. }
\end{figure}

\subsection{Additional Higgs Bosons}
\label{subsubsec:directNP_SUSYHiggs}
The ILC will be able search for additional Higgs bosons of extended Higgs sectors, for instance 2HDMs (c.f.\  section~\ref{sec:2HDM}) or additional singlets as in the NMSSM. A loophole-free search for heavier Higgs bosons can probe masses up to about $\sqrt{s}/2$. Even more interesting, however, is the ILC's capability to detect
lighter Higgs bosons, even if their couplings to the $Z$ boson are strongly reduced. Figure~\ref{fig:h60_plots} shows an example from the NMSSM featuring a light scalar and a light pseudo-scalar Higgs boson with masses of $60$ and $10$\,GeV, respectively~\cite{Potter:2015wsa}.

\begin{figure}[htb]
  \begin{center}
    \subfigure[]{\includegraphics[width=0.4\linewidth] {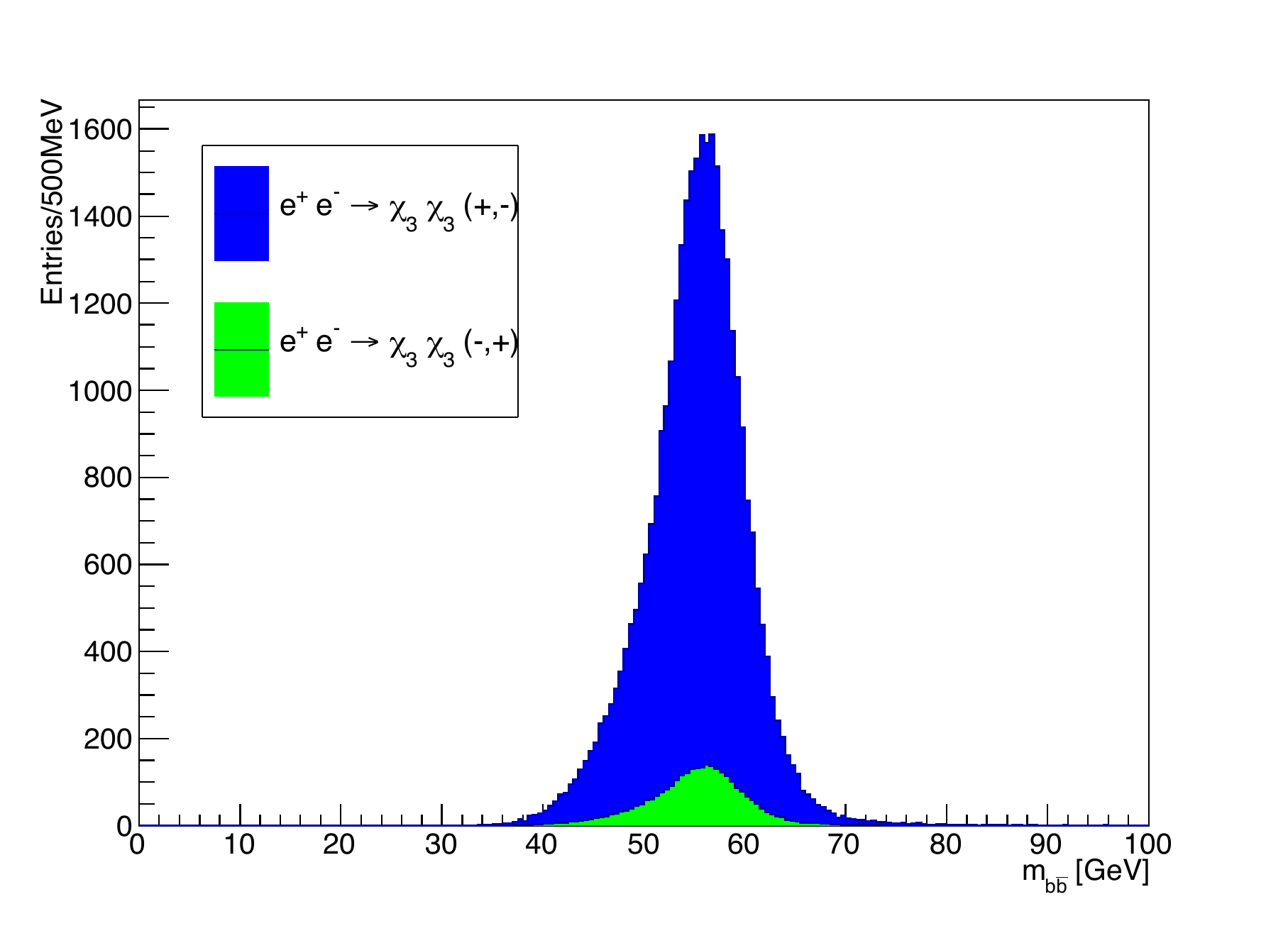}}
    \subfigure[]{\includegraphics[width=0.4\linewidth] {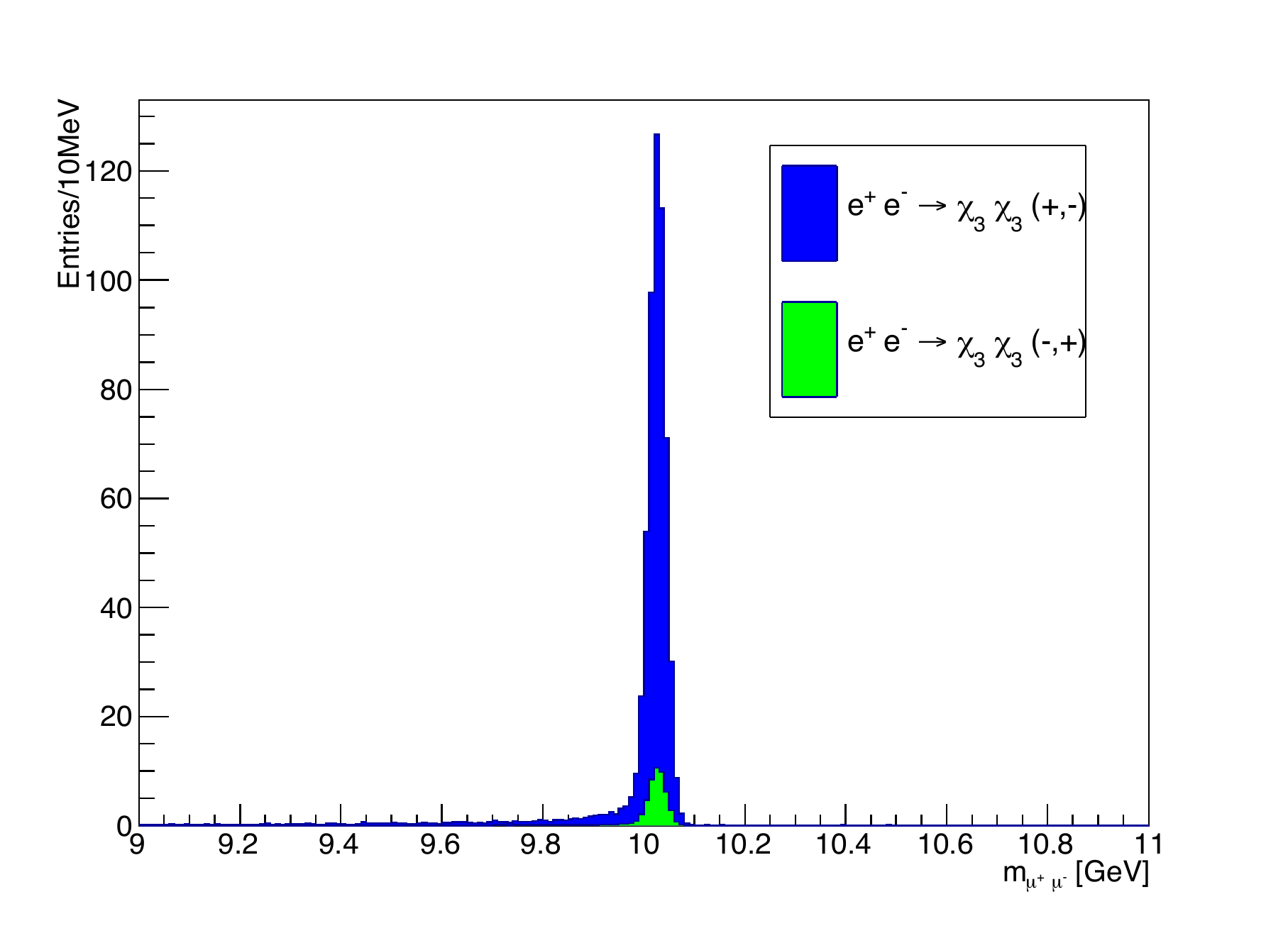}}
  \end{center}
  \caption{\label{fig:h60_plots} Reconstruction of $e^+e^-\rightarrow \chi_3\chi_3$ events with decays to light Higgs bosons in a NMSSM benchmark~\cite{Potter:ECFALC2016} 
%
  (a) $h_{1}\rightarrow b\overline{b}$ ,  (b) $a_1\rightarrow \mu^+\mu^-$.
   An integrated luminosity of 2000~fb$^{-1}$ at $\sqrt{s}=500$~GeV is assumed for each beam polarization configuration. Background is not yet evaluated.
}
\end{figure}

\subsection{$R$-Parity Violating SUSY}
\label{subsubsec:directNP_bRPV}


As an example for the ILC capabilities in the case of $R$-parity violating SUSY (c.f.\ section~\ref{subsubsec:bRPV_theo}),
Fig.~\ref{fig:bRPV_reach} shows the $5\sigma$ discovery reach for selectron-mediated Bino pair production of an initial ILC run~\cite{List:2013dga}. In this scenario, the Bino-LSP decays into a $W$ boson and a lepton, where the
relative rates for the different lepton flavours are related to neutrino mixing. Fig.~\ref{fig:final_prec} compares
the resulting extraction of $\sin^2\theta_\mathrm{atm}$ from LSP branching ratios as could be measured at the ILC with the known value from neutrino oscillations.


%
\begin{figure}[htb]
  \begin{center}
\subfigure[]{\includegraphics[width=0.35\linewidth]{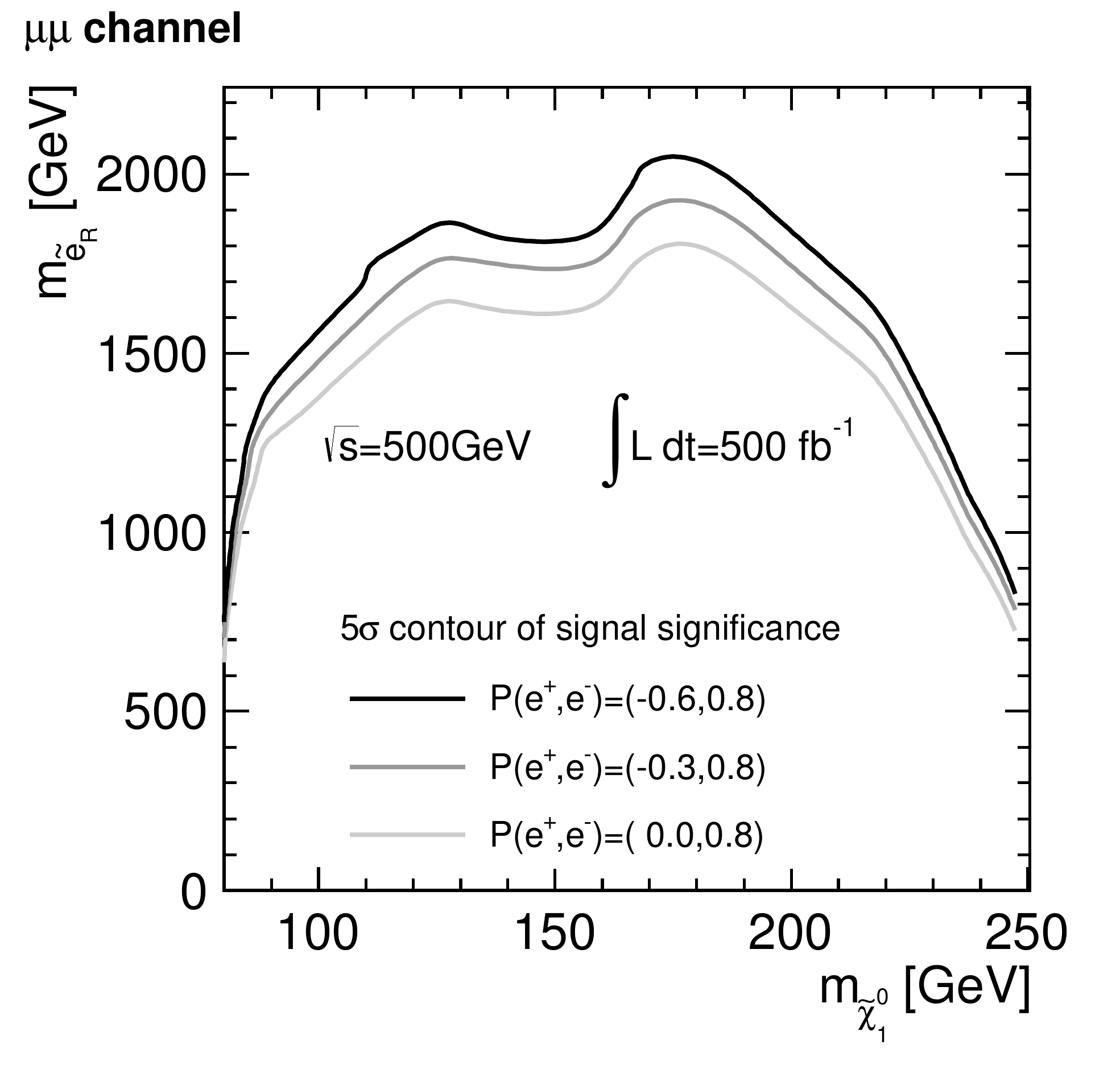}\label{fig:bRPV_reach}}
\hspace{0.1\linewidth}
\subfigure[]{\includegraphics[width=0.35\linewidth]{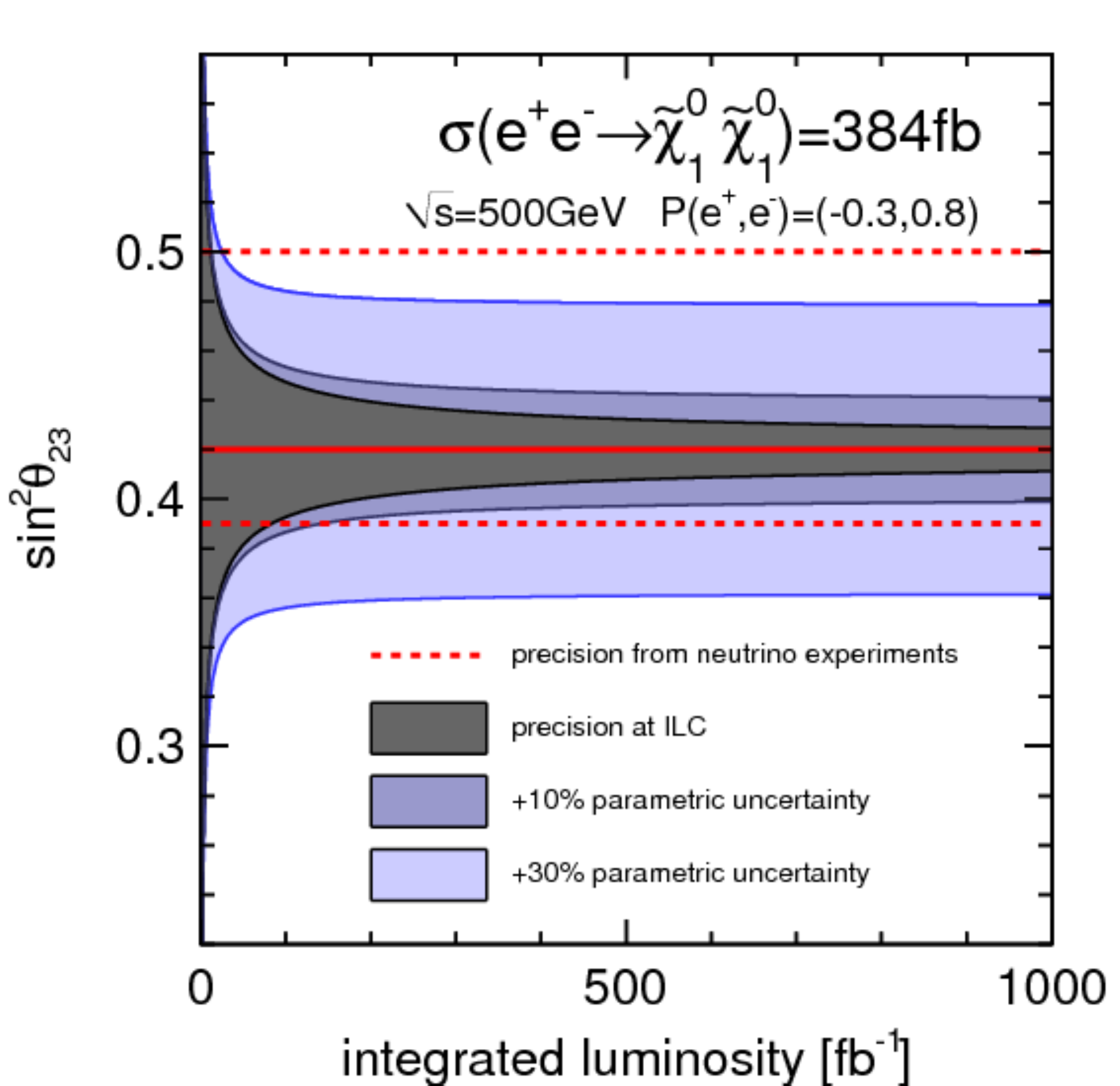}\label{fig:final_prec}}
 \end{center}
  \caption{\label{fig:bRPV} (a) Discovery reach for an initial ILC run. (b) Derived uncertainty on the atmospheric neutrino mixing angle $\sin^2\theta_{23}$, 
  including the additional parametric uncertainty due to limited knowledge of other SUSY parameters. Both from~\cite{List:2013dga}.}
\end{figure}
%
%

%

\section{LHC Discovery Scenarios}
\label{sec:LHC_scen}
%
In this section, we will discuss the ILC discovery potential in the light of the three LHC scenarios presented in~\cite{ICFAtoMEXT}: 
In section~\ref{subsec:noNP}, we will address the case that the LHC experiments will not discover
any further new particles, neither in the near future nor in the long run. 
In sections~\ref{subsec:lightNP} and~\ref{subsec:heavyNP}, we will discuss
examples for the cases that the LHC experiments will discover relatively light or relatively heavy
new particles, respectively. We will show that the ILC has significant potential to 
discover phenomena beyond the Standard Model in each of the cases.

%
\subsection{Scenario 1: LHC Does Not Discover New Particles}
\label{subsec:noNP}
%
%

In this section, we assume that the LHC does not discover new particles
or other significant deviations from the SM.
Concretely, we assume that even after Run~III (i.e.\ around the year
2023) with more than
${\cal O}$(300 fb$^{-1}$) and both experiments combined no evidence for new
particles will have been seen. 
The picture does not change in a relevant way if the same is assumed for the
HL-LHC, i.e.\ no evidence for new particles with ${\cal O}$(3 ab$^{-1}$).
Similarly, besides the current status
of precision measurements, we assume that neither at the LHC, nor at
another (future) experiment any significant deviation from the SM
predictions will be observed.

We will review that even in this case the physics potential of the ILC
is very rich. It will be reviewed that the ILC can discover new
physics via direct production of new particles and
via the observation of deviations in precision measurements that
cannot be done at other experiments. This ensures a strong physics
case even in the quite pessimistic scenario of no new discoveries at
the LHC.


\subsubsection{Discovery via Precise Measurements of Standard Model Parameters}
\label{subsubsec:noNP_indirect}



In this section we concentrate on the ILC discovery potential via
precise measurements of SM quantities. This includes electroweak
precision observables (EWPO), top quark measurements, the production
of electroweak gauge bosons and Higgs-boson mass measurements. 
The corresponding Higgs-boson coupling
measurements will be discussed in the next subsection.

\paragraph{Electroweak Precision Observables and the Mass of the Top Quark:}
As discussed in sections~\ref{subsec:ILC_top} and~\ref{subsec:ILC_ew}, 
the ILC will improve substantially the main electroweak precision measurements,
including the masses of the
$W$~boson, \MW, and of the top quark, \mt, as
well as the effective weak mixing angle, typically
by factors $\sim 5$ to $10$ with respect
to current precisions (which will not be substantially improved at the
LHC)\footnote{For $\mt$, the improvement at the ILC in the precision of the
strong coupling constant \als by a factor of $\sim 5$ will be crucial.}.
With these results, the SM
Higgs boson mass will be predicted indirectly by
a factor of $\sim 5$ better than today~\cite{Moortgat-Picka:2015yla}, as shown
in Fig.~\ref{fig:blueband}.
Figures~\ref{fig:mwmt} 
and~\ref{fig:mwsw}~\cite{Heinemeyer:2007bw, Heinemeyer:2013dia} illustrate that 
the high precision measurements of \MW, \mt\ and \sweff\ will have the power
to exclude the SM experimentally and reveal new, unknown physics scales
up to energy regimes far beyond the direct ILC reach.
\begin{figure}[htb]
\centering
\subfigure[]{\includegraphics[trim=0 130 0 140,  clip, width=0.3\linewidth]{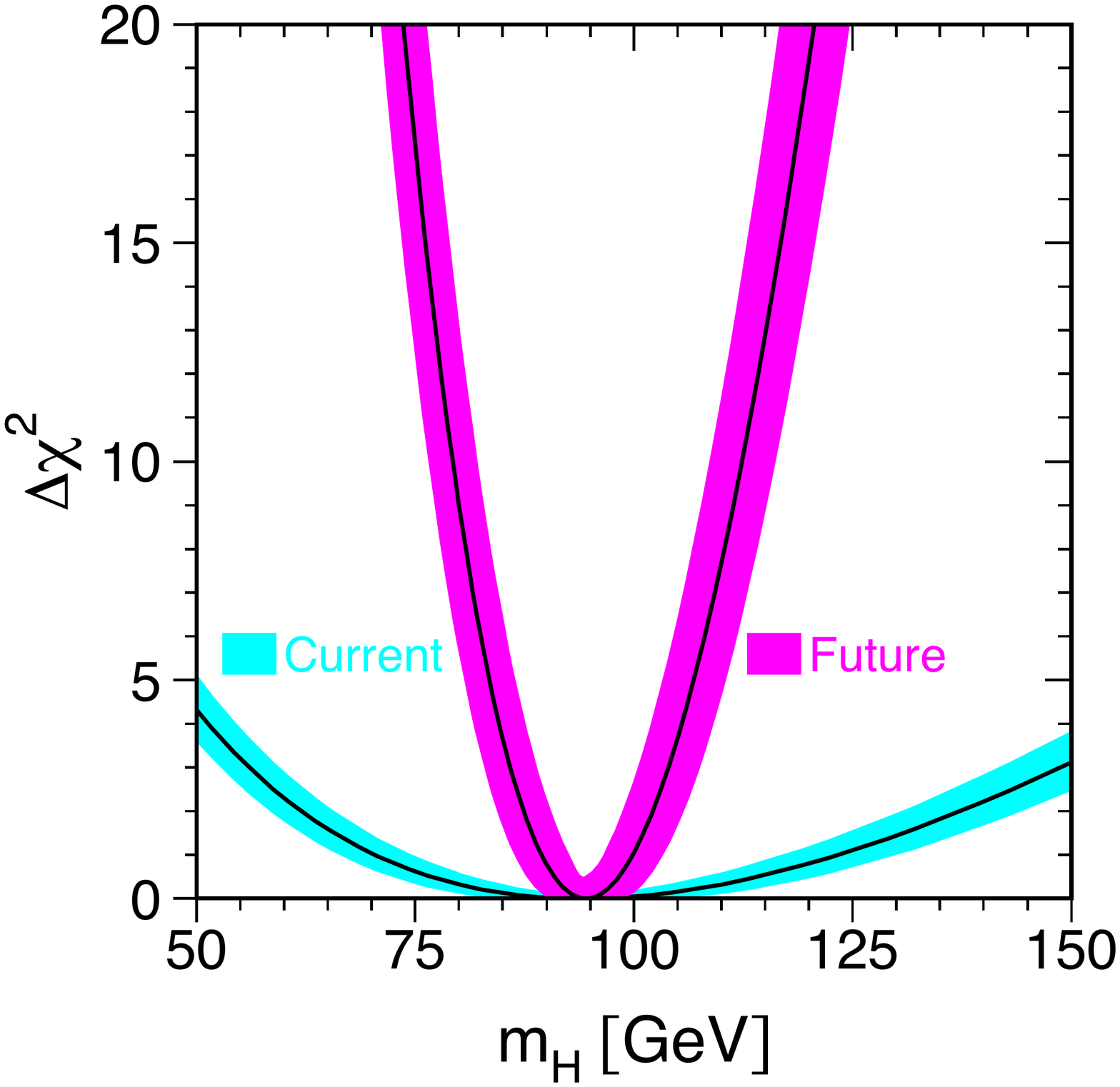}\label{fig:blueband}}
\hspace{0.01\linewidth}
\subfigure[]{\includegraphics[trim=0 0 50 200,  clip, width=0.275\linewidth]{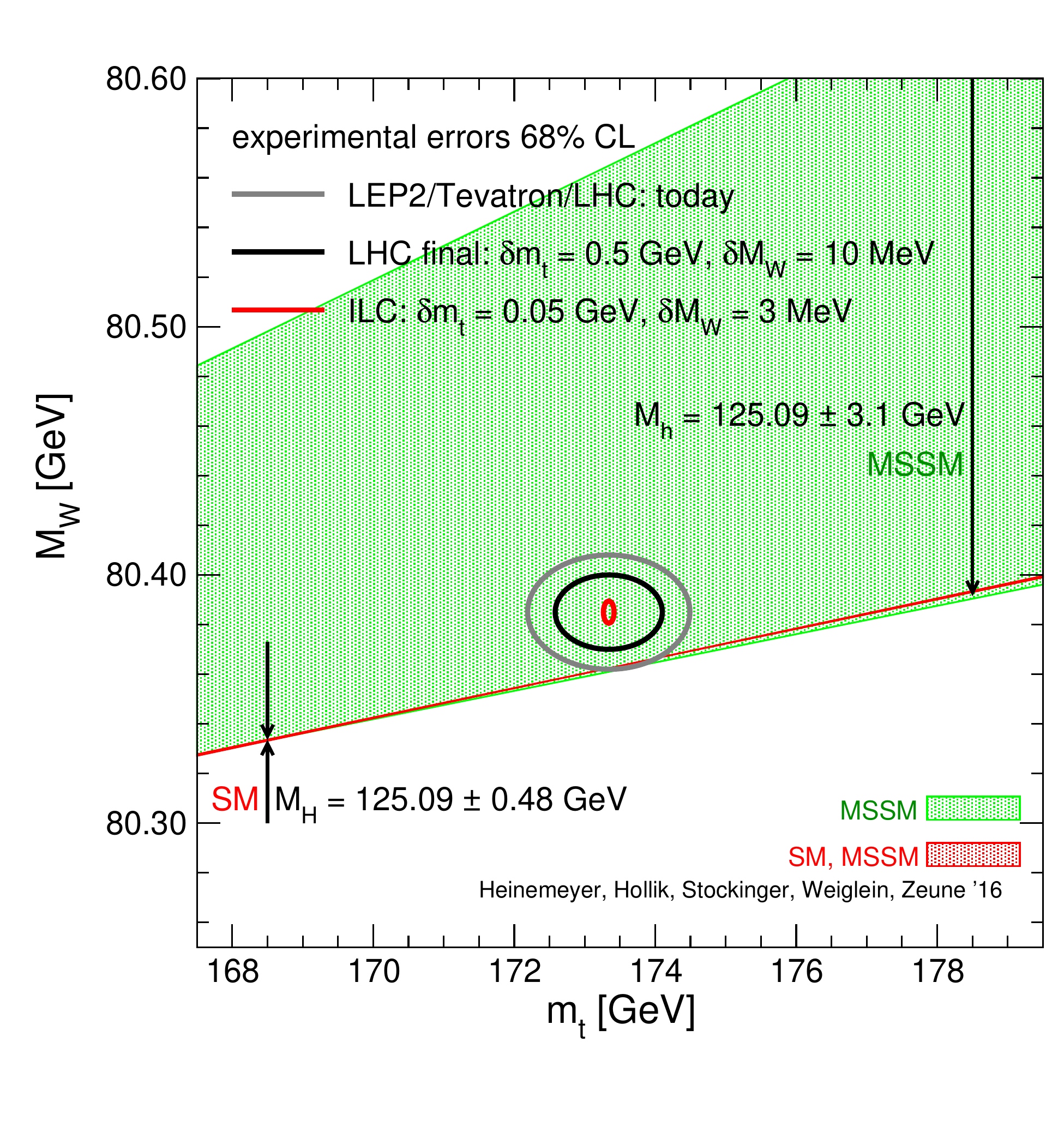}\label{fig:mwmt} }
\hspace{0.01\linewidth}
\subfigure[]{\includegraphics[trim=0 0 50 200,  clip, width=0.275\linewidth]{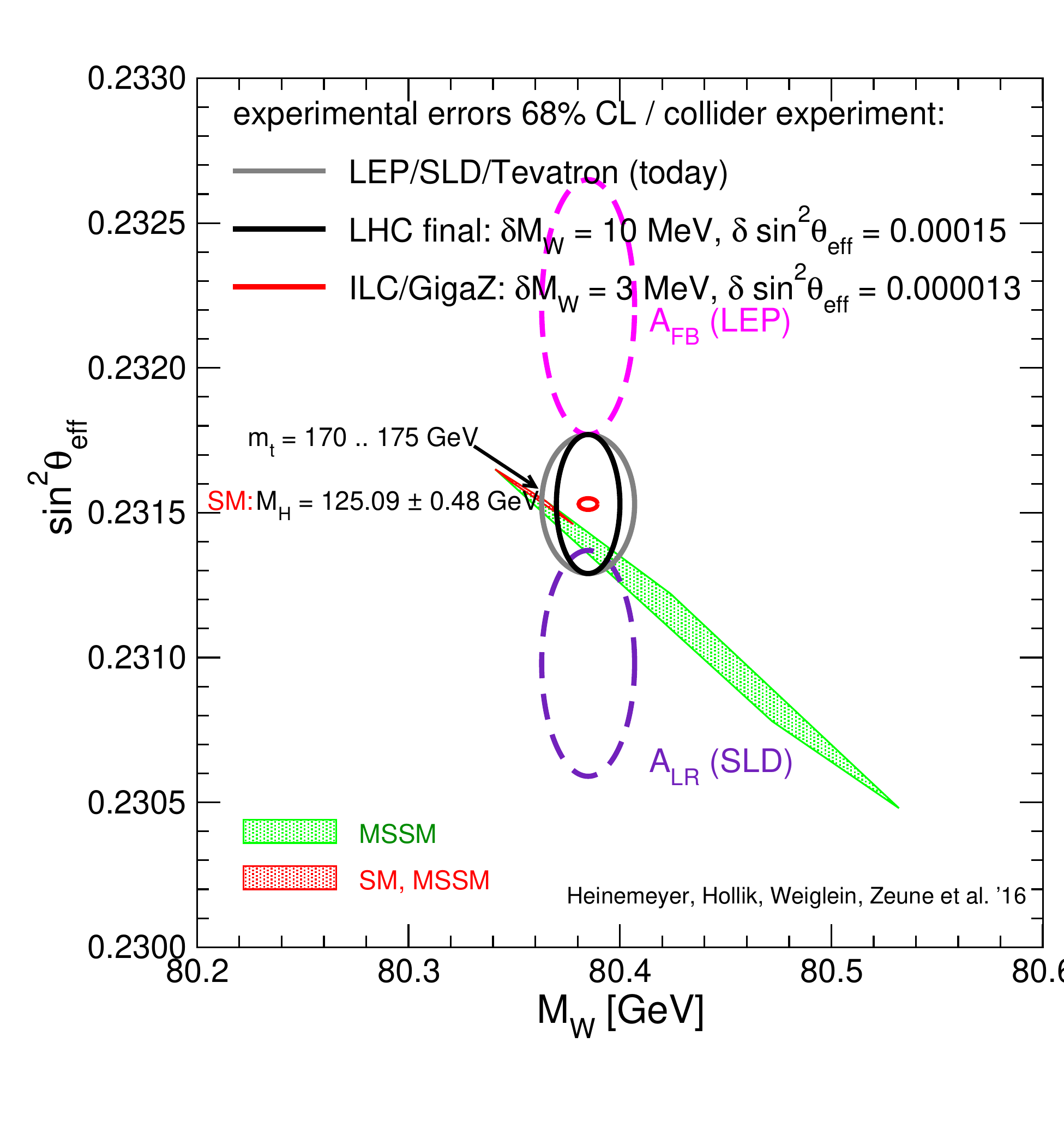}\label{fig:mwsw} }\\
\caption{ILC discovery potential via electroweak precision observables: (a) The prediction of the SM Higgs boson mass
from EWPOs will improve by a factor 5 compared to today due to ILC measurements~\cite{Moortgat-Picka:2015yla}. (b),(c) With  precision offered by the ILC (red ellipses), a clear conflict with the measured value of the Higgs mass
(red lines) could be discovered over a wide range of BSM parameter space (green areas)~\cite{Heinemeyer:2007bw, Heinemeyer:2013dia}. In the case of the weak mixing angle,
ILC could resolve also the long standing discrepancy between the SLD and LEP measurements.
}
\label{fig:discEWPO}
\end{figure}


\paragraph{Electroweak Couplings of the Top Quark: } Another guaranteed
part of the ILC physics program is measurement of the left- and right-handed couplings of the top quark to the $Z$~boson with an order of magnitude higher precision than possible at the LHC, c.f.\ section~\ref{subsec:ILC_top}. 
Many BSM theories predict deviations from the SM in
these couplings, which cannot be resolved at the LHC. As shown in Fig.~\ref{fig:topcoup-lumi}, the excellent ILC precision
will not only enable the discovery of new particles up to energies of several tens of TeV,
far above the kinematic reach of the LHC, but
even allow for the distinction between the various BSM theories, as illustrated in Fig.~\ref{fig:models-rp}. 

\paragraph{Two-Fermion Production and New Gauge Bosons: } As discussed in section~\ref{subsec:ILC_ew}, the ILC measurements
 of $e^+e^- \to f \bar f$ for all types of fermions, 
making use of polarized beams,
will offer indirect access to a large variety of BSM models featuring
 new gauge bosons $Z'$ or $W'$. Their masses can easily be larger than the 
 kinematic reach of the LHC, i.e.\
larger than $\sim 4-5$~TeV. In this case they would escape detection 
at the LHC, while they could still be discovered at the ILC. This is
illustrated in Fig.~\ref{fig:zprime-discovery} for a variety of 
representative models~\cite{Osland:2009dp}, which then could be distinguished
by measurements of the vector and axial-vector couplings of the new gauge boson.
Thus, the discovery reach as well as the diagnostic reach of the
ILC extends far beyond the capabilities of the LHC.  

\begin{figure}[htb]
\centering
\subfigure[]{\includegraphics[width=0.6\columnwidth]{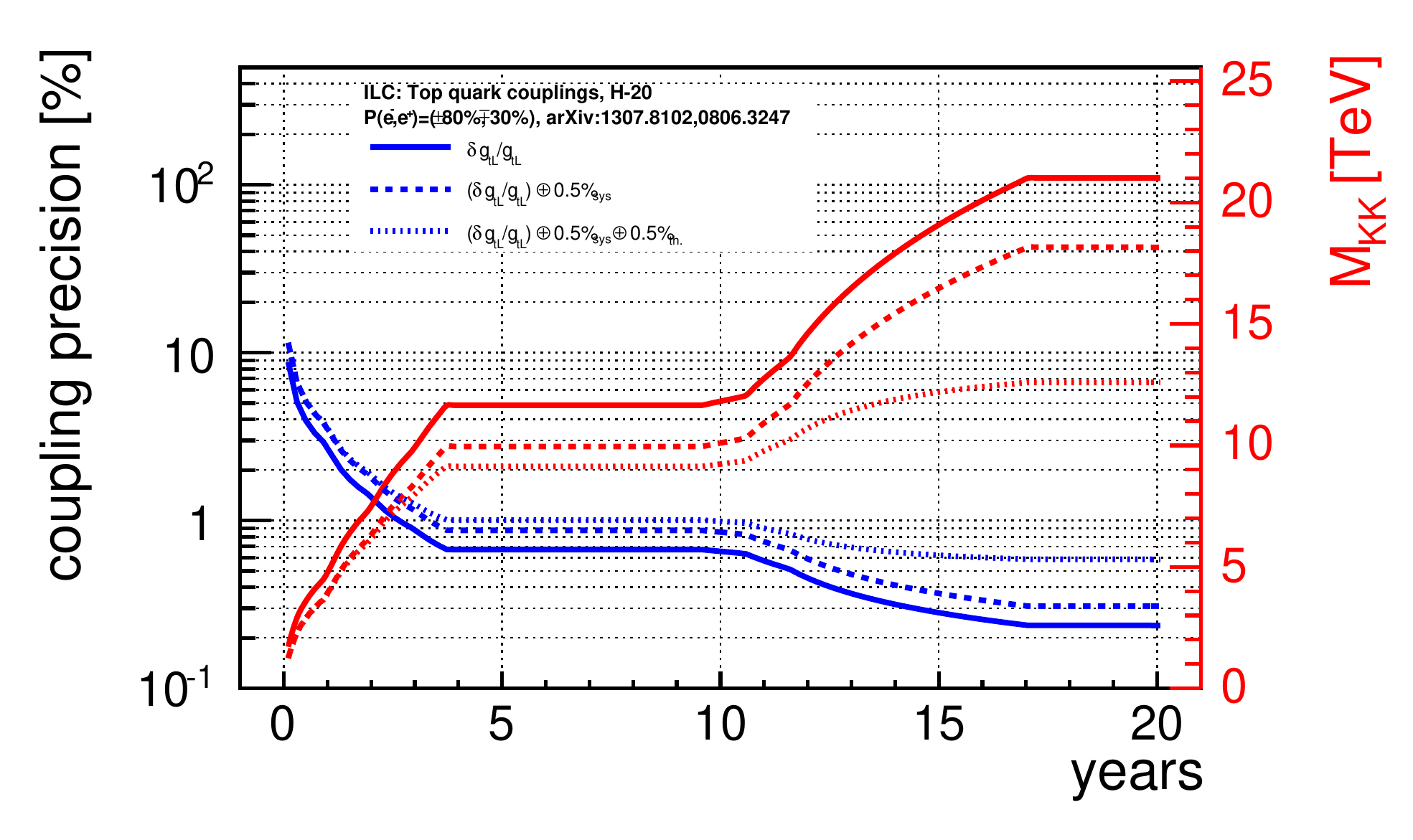}\label{fig:topcoup-lumi}}
\hspace{0.1cm}
\subfigure[]{\includegraphics[trim=10 80 0 100, clip, width=0.325\columnwidth]{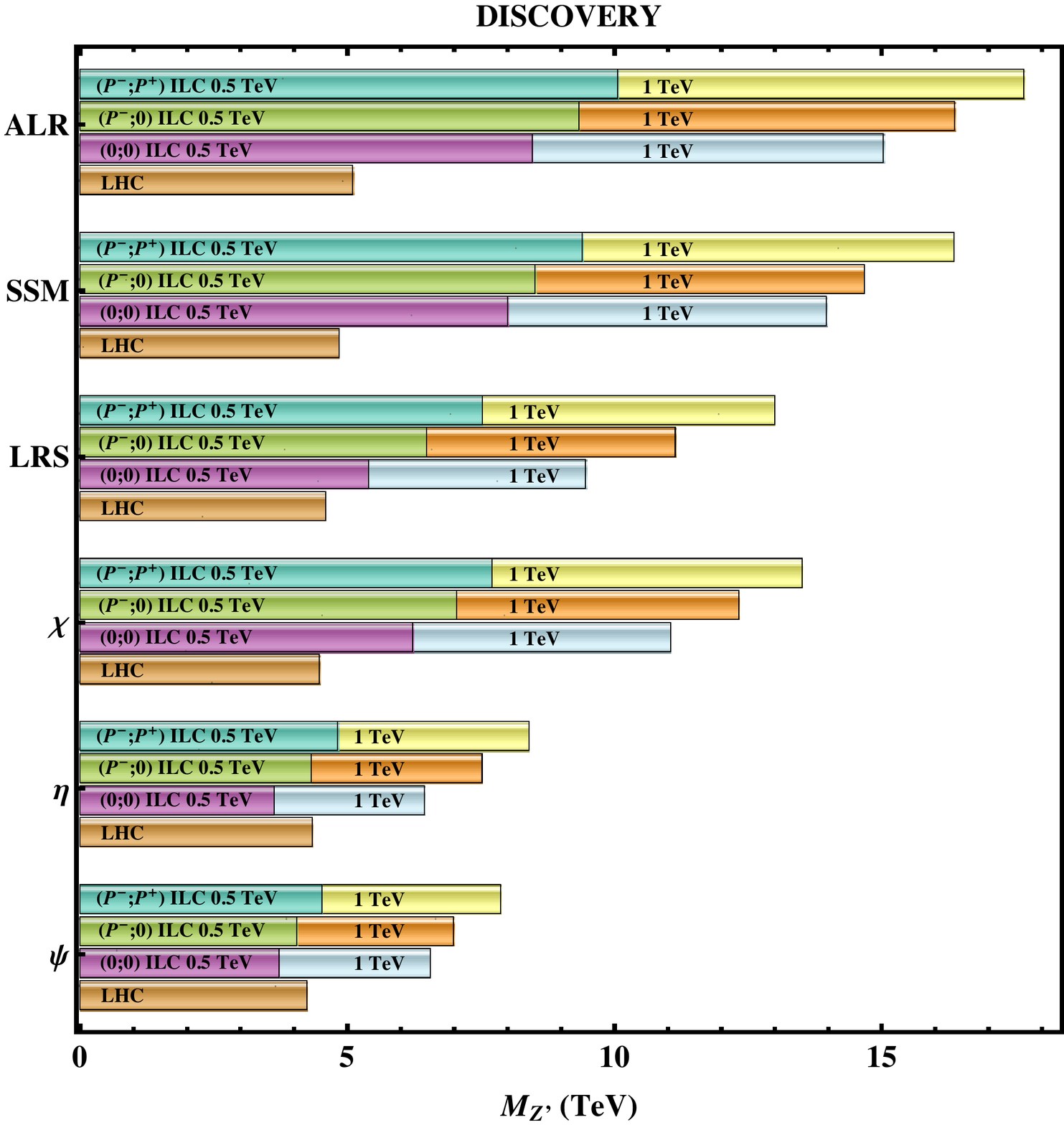}\label{fig:zprime-discovery}}
\caption{
 (a) Precision of the electroweak $t$ quark coupling $g_L$ (left scale, lower curves) and the mass reach expressed in terms of $M_{KK}$ (right scale upper curves) for a typical BSM scenario, here~\cite{Pomarol:2008bh} as a function of ILC running time~\cite{Barklow:2015tja}. 
%
 (b) The discovery reach of ILC for new $Z'$ gauge bosons exceeds that of the LHC in various BSM models~\cite{Osland:2009dp}. 
 }
\end{figure}



\paragraph{Stability of the Higgs potential:}

If no evidence for new physics is found at the LHC, the SM might be valid up
to the Planck scale. 
It was shown, see
e.g.~\cite{Degrassi:2012ry}, that the extrapolation of the Higgs self-coupling
to the Planck scale indicates at the level of $2\,\sigma$ that our vacuum state
is only meta-stable and could tunnel to the true vacuum. The life-time of our
vacuum, however, appears to be sufficiently long-lived on cosmological time
scales. 
The high precision measurement of $\mt$ that can be performed 
at the ILC (c.f.\ section~\ref{subsec:ILC_top}) is indispensable to decide 
with certainty whether or not the SM vacuum is meta-stable, and to
determine its lifetime. 
Conversely, if a clearly ``too short'' lifetime is derived, 
this would be an unambigous proof
that physics beyond the SM must exist so that our vaccum could survive until today.
Or, if the SM vacuum turns out to be right on the boundary between the meta-stable
and stable regions within the ILC precisions, this would suggest the existence of
some completely new principle that leads to the apparent finetuning.


\subsubsection{Extended Higgs Sectors}
\label{subsubsec:noNP_SUSYHiggs}



\paragraph{Additional Higgs bosons heavier than \boldmath{$125$}\,GeV:}
As discussed in section~\ref{subsec:ILC_Higgs}, pair production of heavy Higgs bosons at the ILC
will be limited to masses of about half the center-of-mass energy, but it will cover this
regime in a largely model-independent way and in particular also for strongly reduced couplings to the $Z$
boson, as they are expected to be if the H(125) continues to look SM-like.
In specific MSSM scenarios (e.g.\ $m_h^{\mathrm{mod+}}$~\cite{Carena:2013ytb}), additional Higgs bosons 
with masses above $125$~GeV are already now significantly constrained by LHC searches. In more
general cases, like e.g.\ 2HDM Type I or II, this is not the case~\cite{CMS:pas-hig-16-007}.
Thus, in absence of a future discovery at the LHC, the corresponding searches at the ILC
will provide important additional information and offer complementary discovery potential. 

Furthermore, the ILC will probe Higgs masses far beyond the kinematic limit and beyond the 
reach of the LHC by its very precise and model-independent determinations of the coulings 
of H(125) to fermions and gauge bosons (c.f.\ section~\ref{subsec:ILC_Higgs}).   
Using simplified estimates,
the ILC will be able to probe $m_A$ up to at least $1$\,TeV, independently of $\tan{\beta}$, 
and compositeness scales $f$ up to at least $3$\,TeV, thus investigating parameter space that
remains inaccessible at the LHC.


\paragraph{Additional Higgs bosons lighter than \boldmath{$125$}\,GeV:}
Also additional Higgs bosons below $125$~GeV can
escape searches at the LHC and have been undetectable at LEP due to their
reduced coupling to the $Z$ boson.
This scenario is a unique opportunity for the ILC. At the LHC large
backgrounds and soft decay products make the discovery of light Higgs
bosons particularly difficult. At the ILC, it will be easy to discovery
additional light Higgs bosons in a very large mass range up to 125~GeV. 
The ILC can thus considerably enlarge the discovery range of the LHC.




\paragraph{The Higgs self-coupling:} A special role is played by the Higgs-boson self-couping
$\lambda$, since it can deviate sizeably from the SM prediction even if
all other couplings of the Higgs boson are rather SM-like.
In particular values of $\lambda / \lambda_{SM} > 1.2$ are well motivated since they
would be a prerequisite for electroweak baryogenesis~\cite{Noble:2007kk}, e.g.\ as in the example of 2HDMs discussed in section~\ref{sec:2HDM}. 
According to current projections (based on fast detector simulations 
or four-vector smearing and assuming $100\%$ trigger efficiency), 
it seems highly unlikely that even the HL-LHC would 
discover double Higgs production: For $\lambda = \lambda_{SM}$, ATLAS~\cite{bib:atlas_higgsself_bbgg, bib:atlas_higgsself_bbtt} and CMS~\cite{bib:cms_higgsself} would only reach about $2\sigma$ significance for this process each, corresponding to less than $3\sigma$ in combination. For  $\lambda > \lambda_{SM}$, the cross section for
double Higgs production at the LHC diminishes further due to destructive
interference, thus further reducing the chances to observe this process.
At $95\%$\,CL, values of  $\lambda / \lambda_{SM}$ less than $-1.3$ and larger than $8.7$ could be excluded~\cite{bib:atlas_higgsself_bbgg}. 

As discussed in section~\ref{subsec:ILC_Higgs}, the situation is quite different at the ILC: Already at $\sqrt{s}=500$\,GeV, double Higgs production could be discovered with a significance of nearly $6\sigma$,
based on full detector simulation. As opposed to the situation at LHC and at $e^+e^-$ colliders with $\sqrt{s} \ge 1$\,TeV, the cross section 
{\itshape increases} for $\lambda > \lambda_{SM}$, making the discovery
even more significant in case of new physics leading to a larger values
of $\lambda$. Smaller values of $\lambda$ could be measured with at least $10\%$ precision at the $1$\,TeV upgrade of the ILC.

\paragraph{\boldmath{$CP$} odd admixtures:} Additional spectacular discovery potential is provided by the ILC's sensitivity to small $CP$-odd admixtures in the Higgs-to-fermion couplings, as discussed in section~\ref{subsec:ILC_Higgs}.


\subsubsection{Loophole-free Search for Light States }
\label{subsubsec:noNP_loopholefree}


As pointed out in Section~\ref{sec:BSM}, many extensions to the SM predict 
light particles with only electroweak quantum numbers.
Their production rates at a hadron collider therefore is quite
small compared to the strong interaction mediated SM processes,
and are only accessible via the decay of heavier particles produced 
via the strong interaction or if they produce sizable missing transverse 
momentum or sufficiently high-energetic leptons.
Therefore, such states can quite conceivably exist even
if there is no further discovery at the LHC.
In section~\ref{subsec:directNP_NoLoophole}, it was shown that
in the case such state are SUSY partners to the SM particles,
the ILC is guaranteed to discover them nearly up to the kinematic limit, 
even in the case of very small mass splittings. 

This is illustrated by a specific example in figure~\ref{fig:noloophole2}, 
which shows the current limits in the $\MXN{1}$ - $\MXC{1}$ 
plane from ATLAS~\cite{Aad:2014vma}, together with the
projected discovery reach at 14 TeV with $\int \mathcal{L} \, \mathrm{dt}$ = 300 or 3000 fb$^{-1}$ \cite{ATLAS:2013hta}.
Here it is assumed that  $\MXN{2}=\MXC{1}$, that $\XPM{1}$ and $\XN{2}$ are pure Winos, and that Br($\chi\rightarrow W^{(*)}/Z^{(*)}\XN{1}$)=1
\footnote{Note that the more difficult case $\chi\rightarrow h^{(*)}\XN{1}$ is not considered.}.
The brown-shaded area indicates the corresponding limit from LEP \cite{Heister:2002mn,Abdallah:2003xe,Abbiendi:2002vz},
which assumes only  $\XPM{1}$ pair production, with no assumption on the decay mode.
The expected limits for the ILC at $\sqrt{s}=500$ or $1000$\,GeV are also shown with the same 
assumptions as for the LEP exclusion.
As can be seen from the (loophole) region not covered by the LHC, there is a large
discovery potential for the ILC, even after the high luminosity LHC data has been fully exploited.

\begin{figure}[h]
   \centering
      \includegraphics[width=0.5\linewidth]{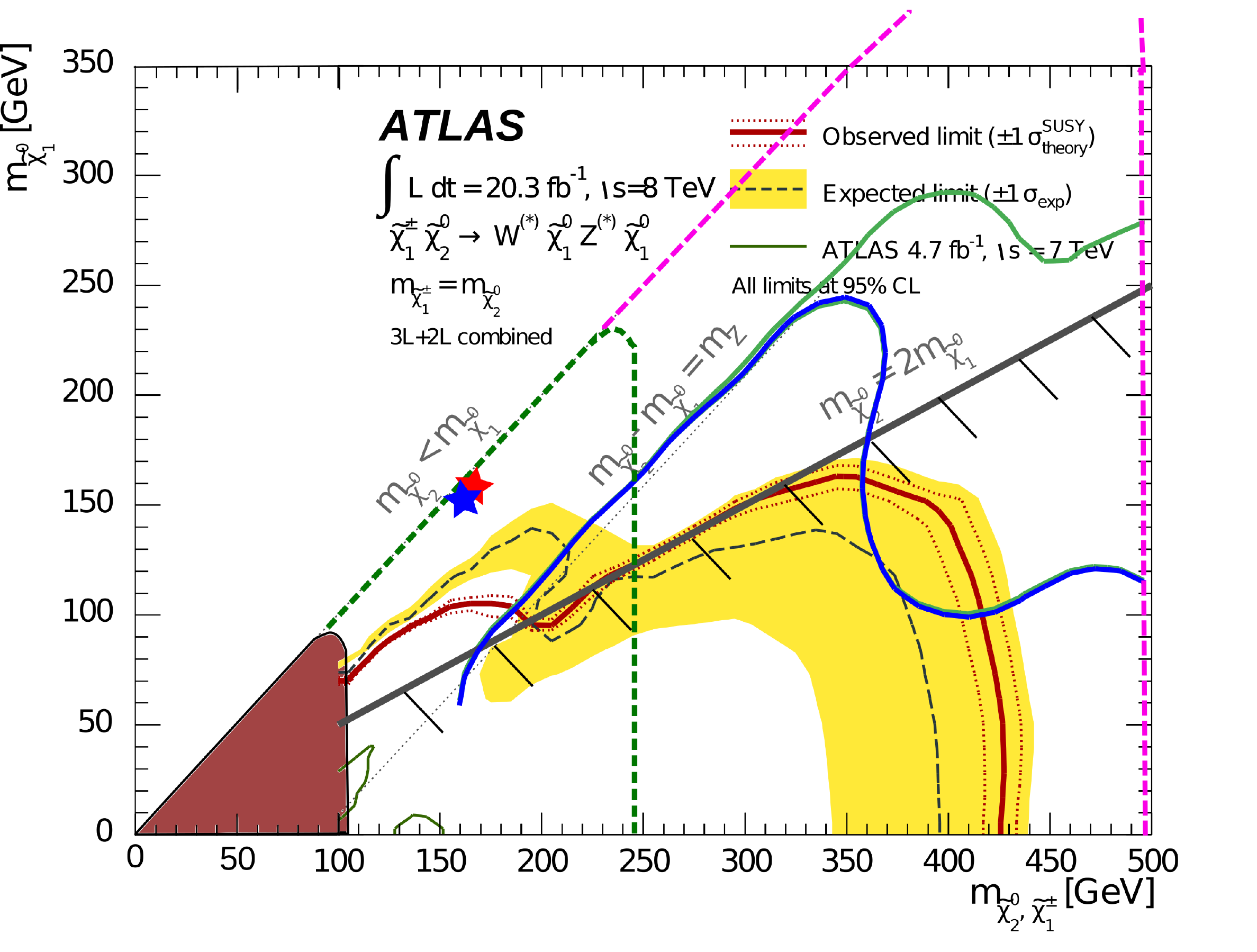}
\caption{Discovery or exclusion regions in the $M_{NLSP} - M_{LSP}$ plane
  for a $\XPM{1}$ or $\XN{2}$ NLSP. Solid brown area: LEP exclusion;
  Solid red and dashed gray lines: ATLAS exclusion (observed and expected);
  Solid blue (green) lines: ATLAS 14 TeV discovery projections for $\int \mathcal{L} \, \mathrm{dt}$ = 300 (3000) fb$^{-1}$;
  Dashed green (magenta) lines:  ILC discovery expectation for $E_{CMS}$ = 500 (1000) GeV;
  Solid black line: below line, no GUT scale gaugino mass unification.
 \label{fig:noloophole2}}
\end{figure}

Also in the case of other extensions to the SM,
such particles can be searched for over the whole mass range.
E.g.\ the tell-tale couplings to the third-generation SM particles
to new light states in composite models can be searched for
in the invariant mass spectrum of two $b$ jets in processes with two tagged
$W$s and four $b$ jets (c.f.\ e.g.\ \cite{Kilian:2004pp}). 
At LHC, on the other hand, such processes would lead to
anomalies in the top quark processes in fiducial regions that are
partially in background regions, so they could very likely be missed
even if they light enough to lie in the kinematic range of the ILC.
In the same way, decays into charm, tau's or photons can be detected.
Thus, by using its electroweak production modes and low backgrounds, 
the ILC can scan for all
existing particles in Nature with electromagnetic, hyper-charge or
electroweak quantum numbers and thus provides discovery potential
complementary to that of the LHC.  



\subsubsection{Natural SUSY}

\label{subsubsec:noNP_NatSUSY}

As discussed in section~\ref{sec:BSM-SUSY-Naturalness}, a core prediction of natural SUSY models
is the existence of a triplet of light higgsinos with masses not too far above $100$\,GeV. 
In addition, top squarks may range up to $\sim 3$ TeV and gluinos up to $\sim 4$ TeV with litle cost to naturalness. 
Such heavy top squarks and gluinos may well lie beyond the reach of even HL-LHC.
Higgsinos are challenging to detect at the LHC since a) their production cross sections are significantly smaller
even than those for wino pairs (which were assumed in Fig~\ref{fig:noloophole2}), and b) because their mass splittings
are small, ranging from $20$\,GeV down to a few GeV or even less than $1$\,GeV in extreme cases.
Thus, SUSY may well remain natural even in the case where LHC ultimately discovers no new particles.
 
The ILC on the other hand (as shown in section~\ref{subsec:directNP_higgsinos}), will be able to discover
the required light higgsinos for masses nearly up to half the center-of-mass energy even for sub-GeV mass differences --- 
{\itshape and independently of any assumption on the rest of the sparticle spectrum.} 
And proceeding beyond this unique discovery potential, 
the ILC will be able to derive information about the rest of the superparticle mass spectrum at the weak scale 
as well as on mass unification at a high scale from precision measurements of the higgsino properties.
As illustration of the power of these measurements, 
Fig.~\ref{fig:higgsino_unification} shows the RGE extrapolation of the ILC measurements 
presented in section~\ref{subsec:directNP_higgsinos} up to the GUT scale in two different
models-- the NUHM2 model with gaugino mass unification and the nGMM model with intermediate scale mirage unification
of gaugino masses-- leading to slightly different higgsino mass splittings.
As the figure shows, the two gaugino mass unification scenarios can clearly be distinguished. 
Under the assumption that all three gaugino masses unify at the same scale, 
then also the gluino mass at the weak scale can be predicted. 
Even without these model  assumptions, from the weak-scale parameter determination alone 
the masses of other sparticles (e.g.\ the stops, the heavier electroweakinos and the heavier Higgses) can be
predicted, giving important guidance for an energy upgrade of the ILC, the LHC  or for even more energetic future colliders.

\begin{figure}[htb]
\centering
\subfigure[]{\includegraphics[width=0.44\linewidth]{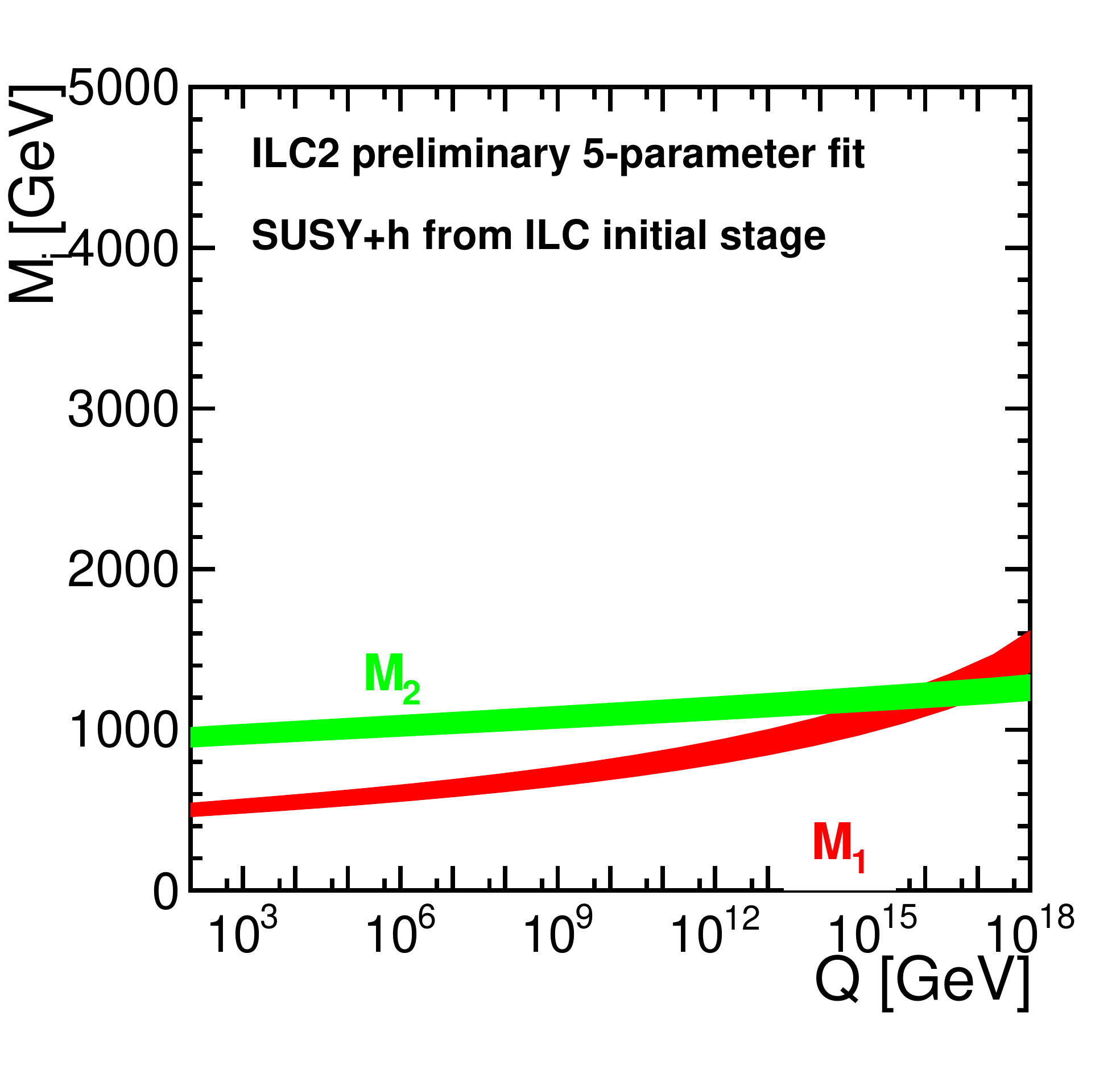}\label{fig:runnig_ILC2_noLHC}}
\hspace{0.1\linewidth}
\subfigure[]{\includegraphics[width=0.44\linewidth]{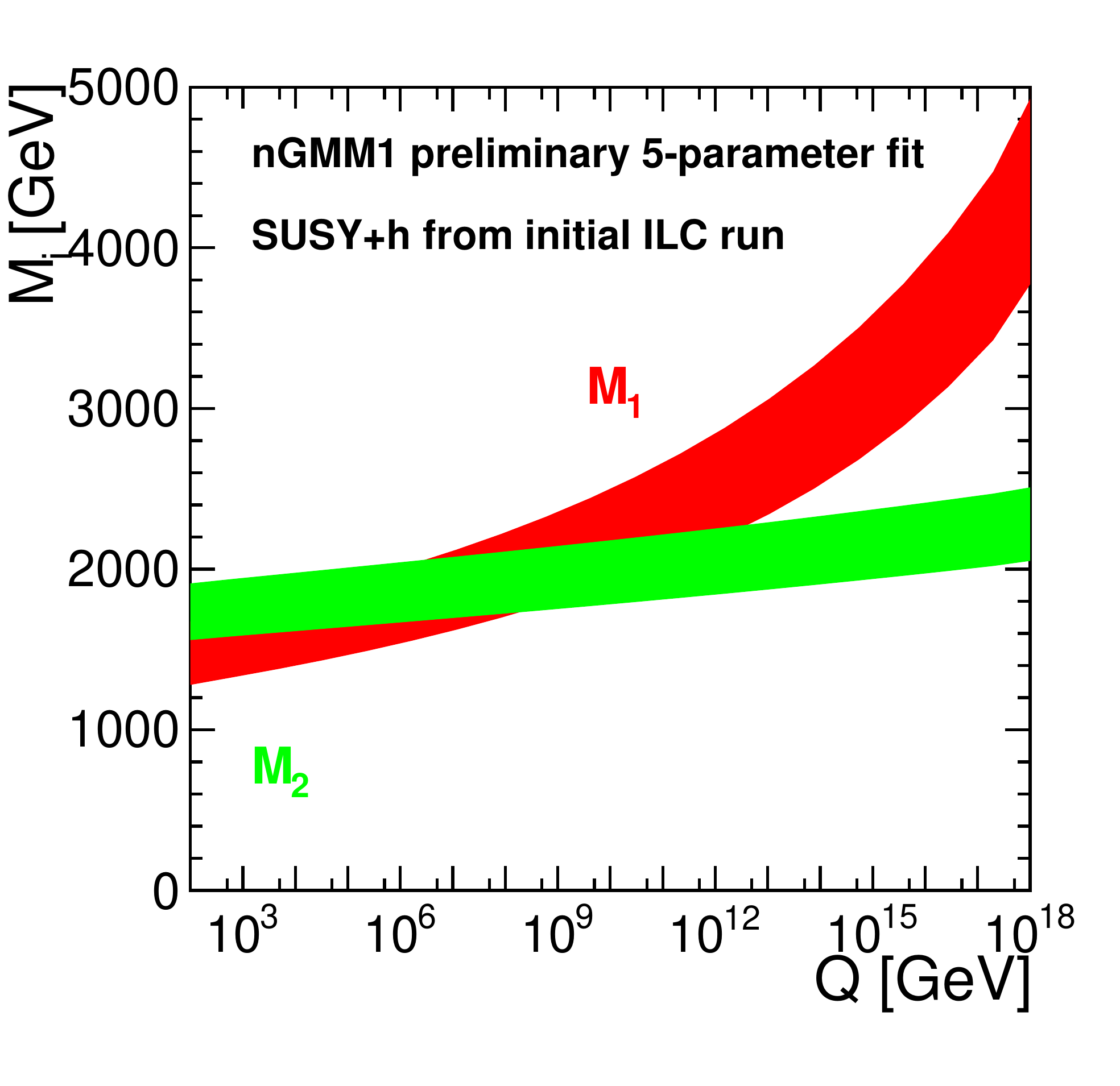}\label{fig:running_nGMM1_noLHC}}
\caption{ILC probing the GUT scale: RGE extrapolation of the gaugino mass parameters $M_1$ and $M_2$ as determined from higgsinos properties measured at the ILC (c.f.\ section~\ref{subsec:directNP_higgsinos}) in two different high-scale models. (a) An NUHM2-based model with mass unification at the GUT scale (ILC2). (b) A mirage mediation model with mass unification at an intermediate mirage scale (nGMM1). Note that both figures only assume an initial ILC data set before the luminosity upgrade, with significant improvements expected after the full H-20 running scenario.
}
\label{fig:higgsino_unification}
\end{figure}

\subsubsection{WIMP Dark Matter}
\label{subsubsec:noNP_WIMPs}

Searches for WIMP dark matter at the ILC are highly complementary to those
at hadron colliders and at direct detection experiments: as an electron-positron collider, 
ILC is sensitive to WIMP couplings to electrons, 
whereas hadron colliders and direct detection experiments are sensitive to WIMP 
couplings to quarks. Depending on the
type of particle mediating the WIMP-SM interaction, there is a priori no reason for these couplings to be of similar strength.Thus, if the LHC does not discover a deviation from the SM expectation in its ``mono-$X$'' searches, it is essential to complement the picture by probing the WIMP-lepton couplings at an electron-positron collider.

Moreover, while LHC can probe larger WIMP masses due to its higher center-of-mass energy, ILC can probe smaller couplings, thus higher energy scales for the WIMP-electron interaction due to its higher precision. As discussed in section~\ref{subsec:directNP_WIMP}, the ILC can probe new physics scales of several TeV and thus has here a unique discovery potential independent of LHC results.

\subsubsection{$R$-Parity Violating SUSY}
\label{subsubsec:noNP_bRPV}

Another example of complementary discovery potential can be found in $R$-parity violating SUSY. Here, the LSP decays
into standard model particles and thus the characteristic feature of MET for hadron colliders 
will not be present, making it quite hard to probe such signatures at the LHC. 
For instance in the case of bRPV summarized in section~\ref{subsubsec:directNP_bRPV}, 
the relevant limits from LHC~\cite{bib:ATLAS_bRPV} are much weaker than in the $R$-parity conserving case, 
and rely even more crucially on strong production mechanisms. Thus, if the coloured SUSY particles are too heavy, LHC will most likely
not be able to detect light electroweak states which decay fully into standard model particles without significant MET.
Therefore, the example presented in section~\ref{subsubsec:directNP_bRPV}, but also other cases with similar signatures, will remain
unknown territory for the ILC to explore even if LHC does not discover further new particles.


%
\subsection{Scenario 2: LHC Discovers Relatively Light New Particles}
\label{subsec:lightNP}
%
%
In this section, we consider the scenario where LHC discovers new light 
matter states during Run II or Run III or at HL-LHC, 
and what impact such discoveries might have on the ILC physics program.

\subsubsection{Discovery via Precise Measurements of Standard Model Parameters}
If the LHC discovers relatively light new particles, these particles most likely will have
an impact on precision observables. The full ILC precision program layed out in section~\ref{subsubsec:noNP_indirect}
will not loose any of its importance, since it will provide a crucial closure test and since it will be indispensable 
in order to fully identify the kind of new physics discovered and to pin down the underlying model and its parameters.
This includes explicitly the model-independent precision determination of the 
H(125) couplings as discussed in~\ref{subsubsec:noNP_SUSYHiggs}, in particular 
in the case where the discovered particle is a candidate for a heavier Higgs boson.

\subsubsection{Loophole-free Search for Lighter States }
\label{subsubsec:lightNP_simplifiedSUSY}


In case the LHC discovers new particles, these will most likely not be the lightest
states of the BSM particle spectrum, but those which either have strong interactions 
and/or sufficiently large mass gaps towards lighter new particles in the decay chain.
Therefore the complementary capabilities of the ILC to search in a loophole-free manner
for lighter new states with electroweak quantum numbers, as discussed in section~\ref{subsubsec:noNP_loopholefree}. 
This includes in particular also searches for lighter Higgs bosons as discussion in section~\ref{subsubsec:noNP_SUSYHiggs}.  
In any case, a discovery of light new particles will lead to a strong interplay of LHC and ILC measurements.
In the following we will give two explicit examples for this interplay and the joint power of the two complementary colliders.

\subsubsection{LHC-ILC Interplay - a SUSY Dark Matter example}
\label{subsubsec:lightNP_SUSY}

In this section, we take as an example a Dark Matter motivated benchmark scenario~\cite{Baer:2013ula} featuring 
$\tilde{\tau}$-coannihilation, which has been studied with respect to future LHC and ILC 
capabilities~\cite{Berggren:2015qua}, with some of
the latter also being summarized in~\ref{subsec:directNP_SUSYDM}.
In such scenarios, the LHC would discover some of the heavier SUSY particles, in particular the lighter stop and
sbottom, as well as some hints of the heavier neutralinos and charginos. However a full determination
of the mass spectrum would be extremely challenging based on LHC data alone. In particular the discovery of the 
$\tilde{\tau}_1$ and a determination of its properties, which are crucial to test whether coannihilation is indeed
the explanation for the cosmologically observed Dark Matter relic density~\cite{Lehtinen:2016qis}, will have to 
wait until the ILC turns on.

On the other hand, the detailed information from the
ILC can be used in the analysis of the LHC data in order to disentangle the contributions
of different production modes and to reconstruct quantities which are sensitive to masses of the
heavier sparticles with more complex decay chains. An example is the  case
of the \XPM{2} shown in Fig.~\ref{fig:chargino2_mass}, which is out of reach to the ILC, 
but which can --- with the knowledge of the masses of the \XPM{1}~ and the \snu\ from the ILC --- 
be isolated from the electroweakino mix at the HL-LHC. Its mass can then be reconstructed on an
event-by-event basis with a resolution of about $50$\,GeV at the HL-LHC. 
\begin{figure}[htb]
\centering
\subfigure[]{\includegraphics[width=0.44\linewidth]{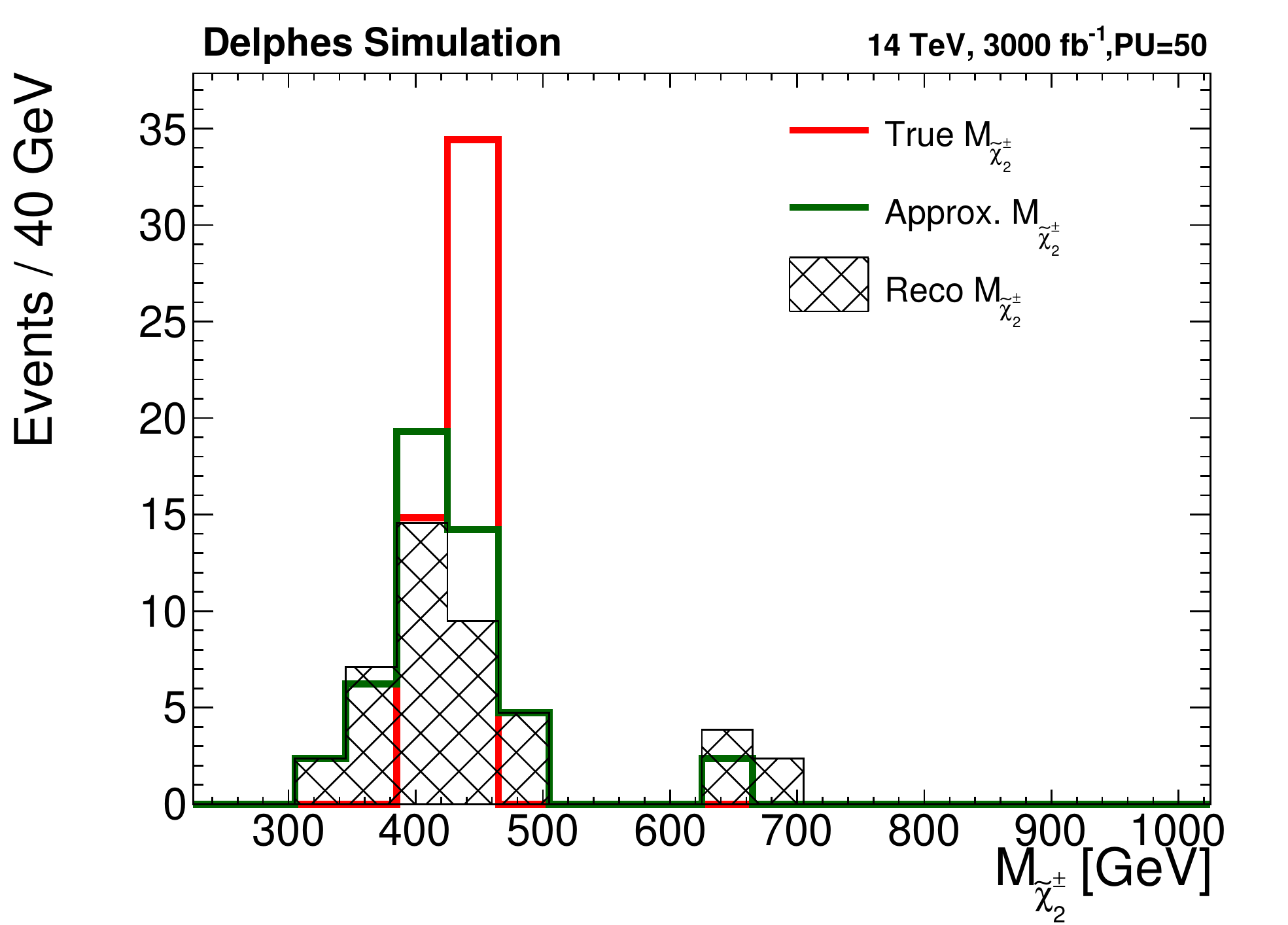} }
\subfigure[]{\includegraphics[width=0.44\linewidth]{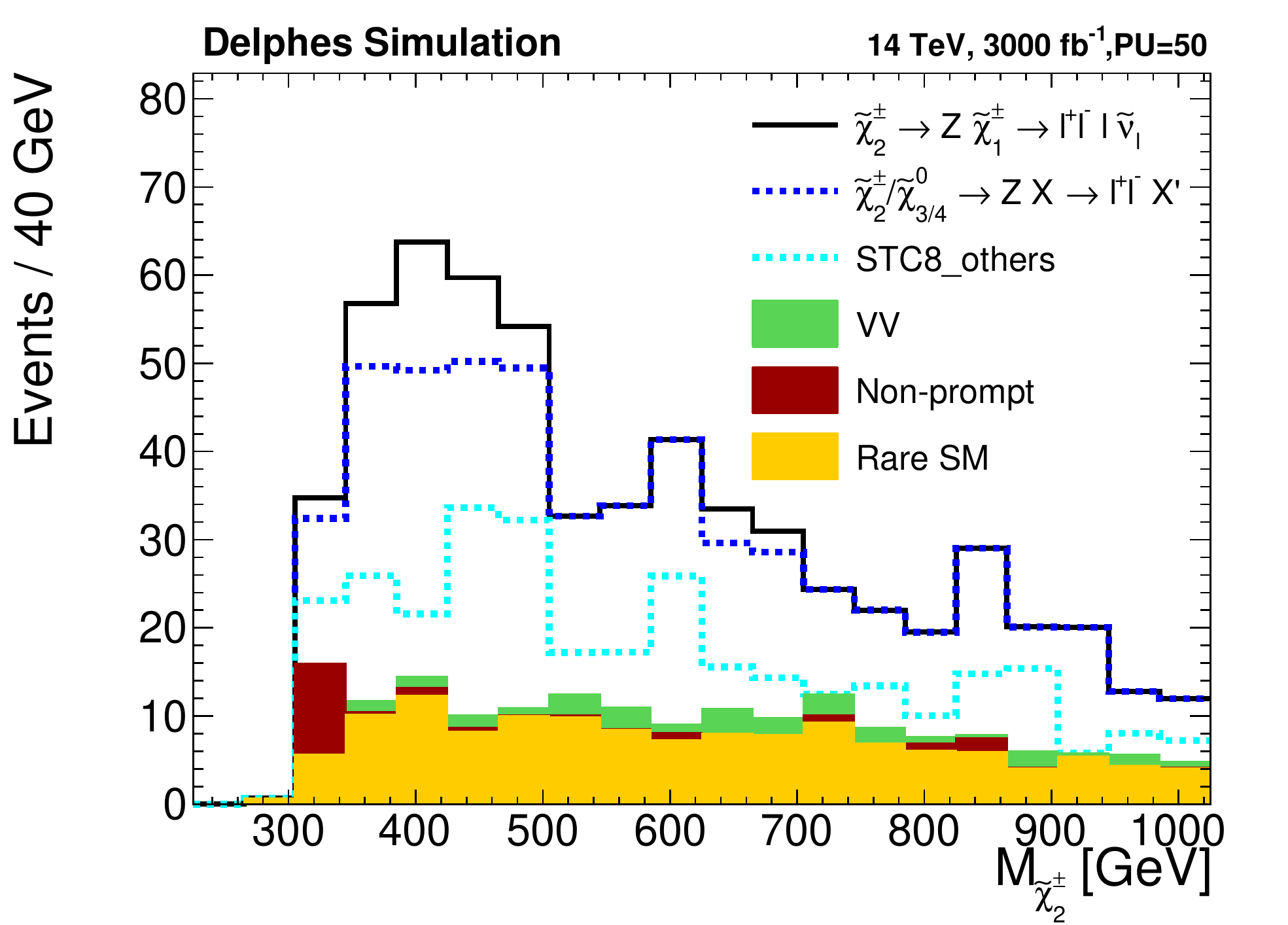} }\\
\caption{ \XPM{2} mass information at LHC when \MXC{1} and \msnu\ are known from ILC. 
(a) MC truth information of the invariant mass as well the reconstructed \MXC{2} for the same events. 
(b) Reconstructed \MXC{2} for all selected SM and SUSY events, where based on (a) the leftmost peak near
$400$\,GeV can be identified as \MXC{2}.
}
\label{fig:chargino2_mass}
\end{figure}

\subsubsection{LHC-ILC Interplay - a Natural SUSY example}
\label{subsubsec:lightNP_NatSUSY}

So far LHC has gathered only $\sim 1\%$ of its projected integrated luminosity. 
Thus, there is still an excellent chance that rather light top squarks and/or gluinos could be discovered at LHC.
Also, discovery of wino pair production~\cite{Baer:2013yha} and/or higgsino pair production\cite{Baer:2014kya} 
remains as a possibility.
Any of these discoveries could provide confirmation for a SUSY spectrum compatible with naturalness
as discussed in section~\ref{sec:BSM-SUSY-Naturalness}. 
However, similar to the case explained in Sec.~\ref{subsubsec:noNP_NatSUSY}, the direct production and
observation of all three individual higgsino states as well as their precise characterisation 
(c.f.\ section~\ref{subsec:directNP_higgsinos}) will remain an important task for the ILC.
Such measurements could help determine whether the lightest higgsino comprises all of, or just a portion of, the
dark matter in the universe.

By just assuming the higgsino property determinations of the ILC, 
plus a $10$\% measurement of the gluino mass from LHC, most weak-scale SUSY parameters can be determined.
This enterprise will  enable predictions for possibly unobserved parts of the spectrum, 
including the heavier electroweakinos, the top squarks and the heavy Higgs bosons. 
Typical precisions for mass determinations are expected at the $10$-$20$\% level\cite{Baer:2016wkz}, 
depending on the exact scenario. 
Similar to the example given in section~\ref{subsubsec:lightNP_SUSY}, this information will provide important input 
to LHC analyses, as well as for the planning of future colliders.

Furthermore, from running the weak-scale gaugino masses via renormalization group evolution to high energy 
(as shown in Fig.~\ref{fig:higgsino_unification_withLHC}), tests can be made as to the locus of the 
gaugino mass unification scale; in this case, standard GUT-scale mass unification can clearly be 
distinguished from alternatives such as e.g.\ mirage unification. 
\begin{figure}[htb]
\centering
\subfigure[]{\includegraphics[width=0.44\linewidth]{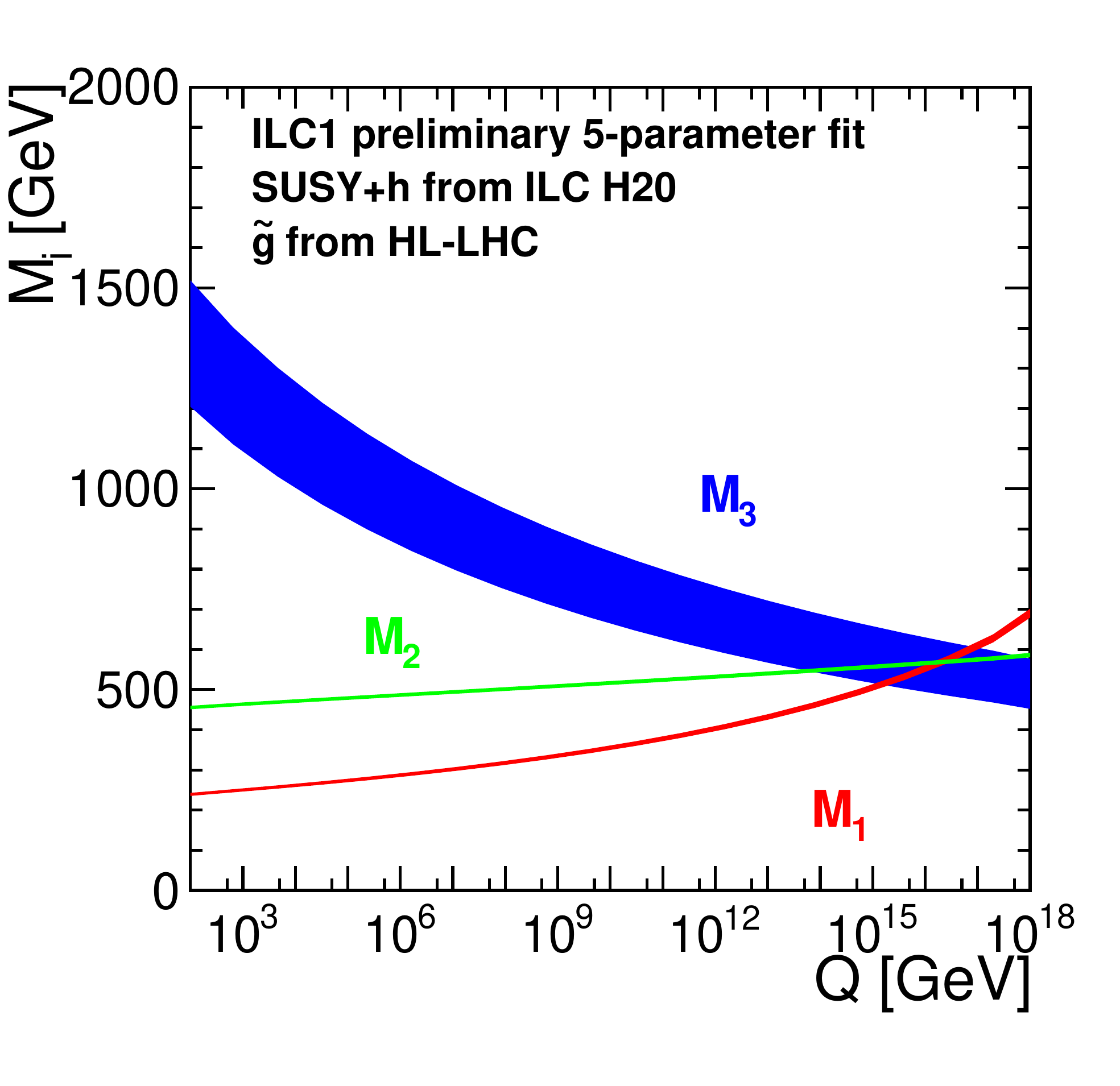}\label{fig:runnig_ILC1_withLHC}}
\hspace{0.1\linewidth}
\subfigure[]{\includegraphics[width=0.44\linewidth]{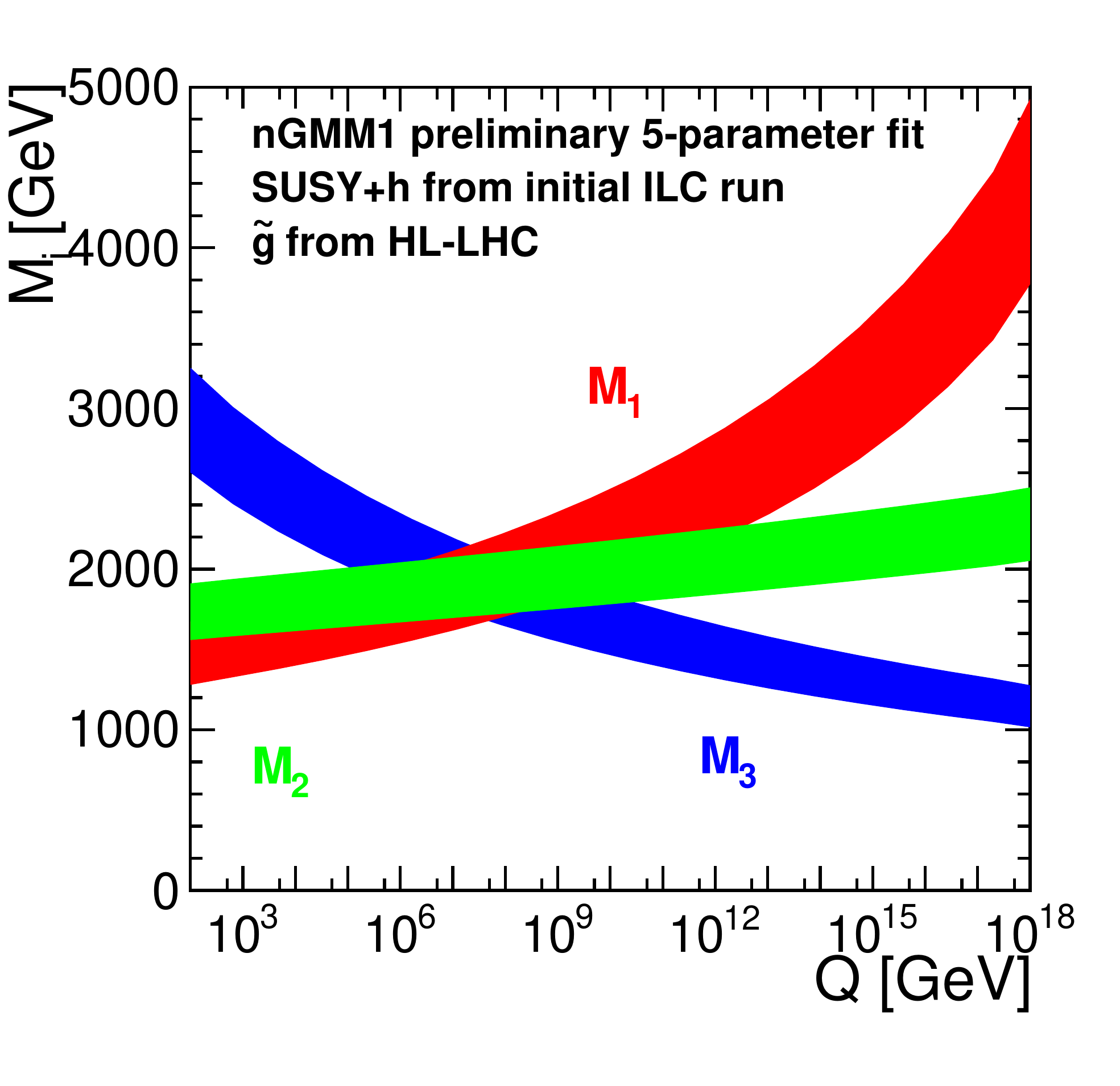}\label{fig:running_nGMM1_withLHC}}
\caption{ILC and LHC probing the GUT scale: RGE extrapolation of the gaugino mass parameters $M_1$ and $M_2$ as determined from higgsinos properties measured at the ILC (c.f.\ section~\ref{subsec:directNP_higgsinos}) and assuming a $10\%$ measurement of the gluino mass from the LHC in two different high-scale models. (a) An NUHM2-based model with mass unification at the GUT scale (ILC1, assuming the full H-20 running scenario). (b) A mirage mediation model with mass unification at an intermediate mirage scale (nGMM1, assuming only an initial ILC dataset before the luminosity upgrade).
}
\label{fig:higgsino_unification_withLHC}
\end{figure}

\subsubsection{WIMP Dark Matter}
\label{subsubsec:lightNP_WIMPs}
While the LHC already has some sensitivity to WIMP masses, it continues to extend the reach
at low masses towards smaller and smaller couplings. Thus, in principle it could still find a deviation from the standard
model in the kinematic reach of the ILC. In this case, it would be important to test whether the found WIMP also couples
to leptons, and at which strength. If it couples to leptons as well and is visible at the ILC, the fact that at a lepton
collider the initial four-momentum is known and the beam helicities can be chosen will allow a full precision characterisation
of  the WIMP and the associated mediator particle, as summarized in section~\ref{subsec:directNP_WIMP}.


%
\subsection{Scenario 3: LHC Discovers Relatively Heavy New Particles}
\label{subsec:heavyNP}
%
%
\newcommand{\lhhh}{\lambda_{hhh}}
\def\beq{\begin{equation}}
\def\eeq#1{\label{#1}\end{equation}}
\def\eeqn{\end{equation}}
\def\ee{e^+e^-}

\def\leqn#1{(\ref{#1})}

In this section, we will address the case that the LHC discovers a rather heavy new particle. This is, after all, the kinematic regime
in which the LHC is most powerful, since Standard Model backgrounds by QCD start to lower. If it had turned out to be a real signal, $X750$ would have been an excellent example in this category, for which the potential of the ILC has been discussed in detail
in Ref.~\cite{Fujii:2016raq}. Here, we will summarize and generalize the points made therein and will show that also in this
case the complementary discovery potential of the ILC will add crucial information in order to complete the picture of the new physics, which in any realistic model will consist out of more than one new particle.

\subsubsection{Extended Higgs Sectors and Higgs Coupling Measurements}
\label{subsubsec:heavyNP_SUSYHiggs}
If the discovered particle is compatible with being a spin-0 resonance, 
it could be part of an extended Higgs sector, for instance
one of the heavy $H,A$ states of a 2HDM (see section~\ref{sec:2HDM}). In this case, it would be mandatory to both search for the remaining Higgs states for which the ILC offers discovery potential complementary to the LHC. As discussed in section~\ref{subsubsec:noNP_SUSYHiggs}, the additional Higgs states could be produced directly at the ILC at some cases, while
the model-independent precision measurements of the couplings of the H(125) 
to SM particles will reliably discover further states
up to high mass scales, beyond the direct reach of the LHC.

In addition, such a new scalar could lead to very interesting enhancements in the Higgs self-coupling $\lambda$ by at least $20$\%, as they would be required for electroweak baryogenesis~\cite{Noble:2007kk}, e.g.\ as in the example of 2HDMs discussed in section~\ref{sec:2HDM}.  
Loosely speaking, the deviations on $\lambda$ scale with the mass of the new scalar. This is illustrated in Fig.~\ref{fig:delhhh} in which possible deviations of $\lambda$ (denoted as $\lhhh$) for different masses of the new scalar, denoted as $m_\Phi$, and as a function the mass parameter $M$ of the 2HDM model. As can be seen,
 the potential deviations are largely independent of the latter parameter as long as it is not too high. The contour labeled as $\phi_c/T_c$ indicates mass values above which a first order phase transition can occur. 

\begin{figure}
 \begin{center}
    \includegraphics[width=0.4\textwidth,angle=-90]{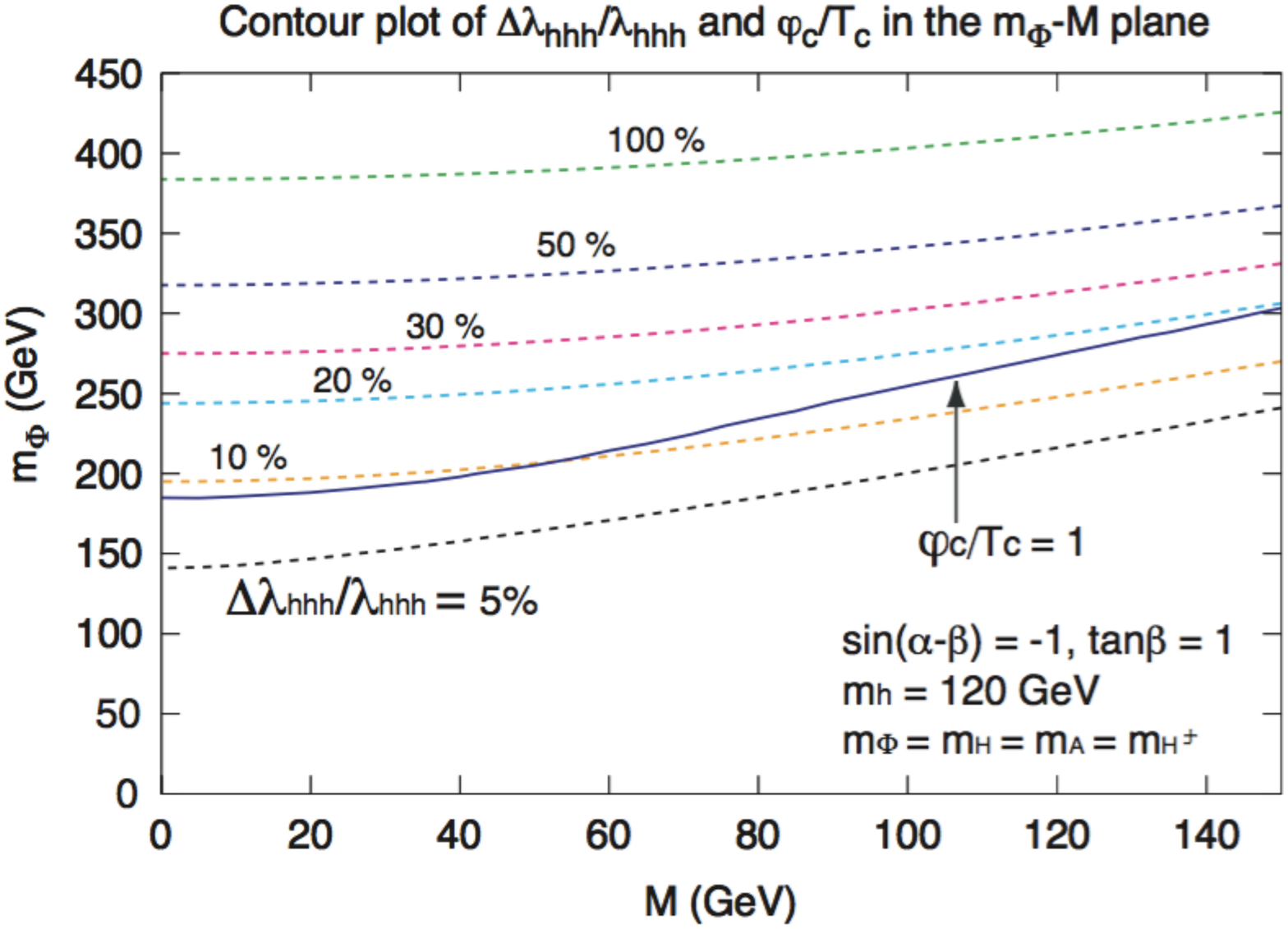}
    \caption{\label{fig:delhhh} Deviations of the Higgs self-coupling $\lhhh$ for different masses of a heavy scalar as a function of the mass parameter $M$ of the 2HDM Model~\cite{Senaha:2013fva}}.
\end{center}
\end{figure}

In case of the new particle been a radion of an extra dimension model, the mixing with the Higgs boson may vanish as e.g.\ argued in~\cite{Ahmed:2015uqt}. In this case the H(125) couplings to fermions will be SM-like, while couplings to photons and gluons  will vary by some amount. However, as discussed in section~\ref{sec:RS}, concrete realisations of such models imply the existence of additional heavy vector (Spin-1) bosons that occur e.g.\ as so-called Kaluza-Klein (KK) excitations. These vector bosons will be discussed further below in section~\ref{subsubsec:heavyNP_indirect}.
 
But also a new Spin-1 resonance, for instance as arising in Higgs compositeness models, would introduce shifts in the couplings of the Higgs boson to heavy Standard Model particles $W$, $Z$ and $t$ that are always similar to those given in Eq.~\ref{eq:kappa_SILH}. Typical shifts for new $5$\,TeV particles are $\sim8\%$  but even a $20$\,TeV particle would produce observable effects, for both $W$ and $Z$, in the ILC precision measurements~\cite{Malm:2014gha}.

In composite Higgs models as well as in RS models, effective top quark Yukawa couplings are generated. Figure~\ref{fig:ctct5RSC} shows the shifts of the Higgs boson couplings computed in \cite{Malm:2014gha} for a variety of parameters sets, as a function of the mass of the lightest KK gluon.  Note that these shifts are in general complex,
leading to $CP$ violation in the $ht\bar t$ coupling at levels observable at the ILC~\cite{Godbole:2011hw}.   
The figure shows the shifts as parameters $c_t$, $c_{t5}$ corresponding to the effective coupling
$
    \delta \L =  - {m_t\over v} \ h \  \bigl[  c_t\ \bar t t + i
    c_{t5}\  \bar t \gamma^5 t \bigr]  .
$

\begin{figure}
 \begin{center}
    \includegraphics[width=0.9\textwidth]{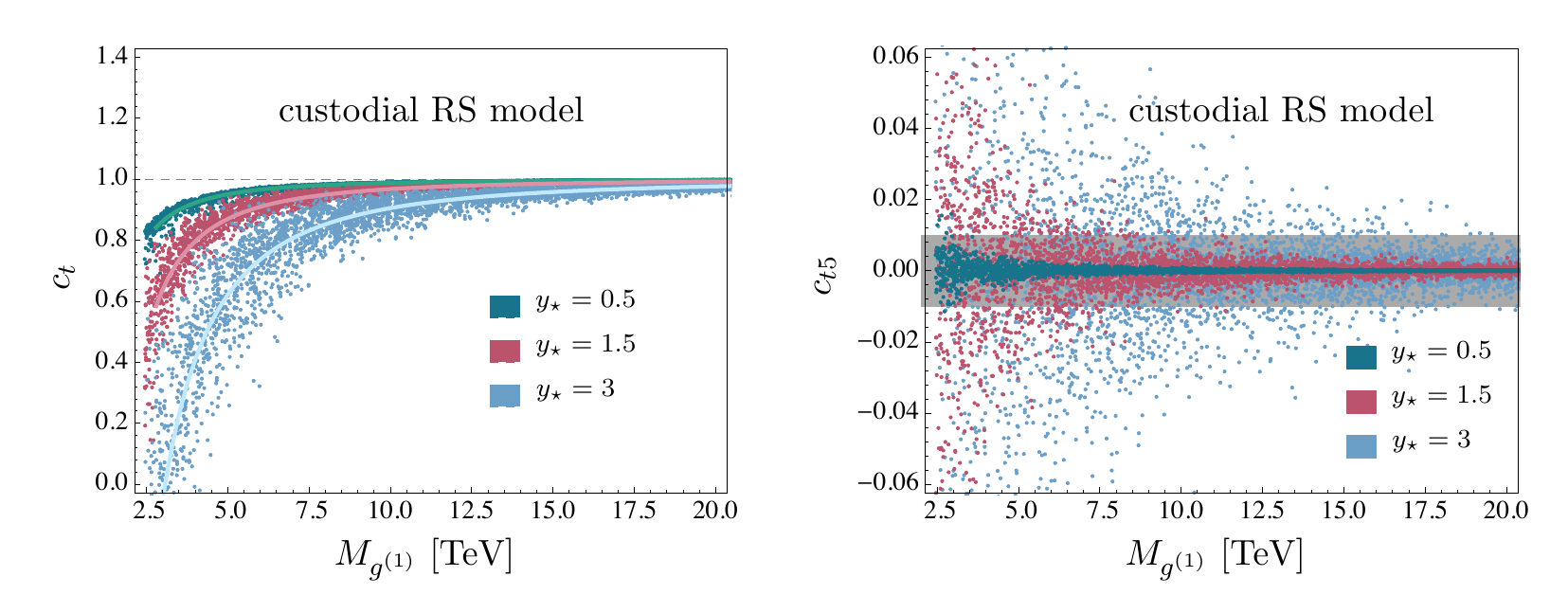}
    \caption{\label{fig:ctct5RSC}Predictions of $CP$-even (left) 
and $CP$-odd (right) Higgs couplings to the $t$ quark in a Randall-Sundrum model with a custodial
symmetry. The parameters $c_t$, $c_{t5}$ are defined in the text.  The point clouds are a scan of the space of Yukawa couplings for three values of the free parameter $y_{\ast}$. The gray band on the right hand 
side shows the experimental bound at 90\% on the $CP$-odd predictions 
derived from the electron dipole moment (EDM)~\cite{Malm:2014gha}. 
%
}
\end{center}
\end{figure}

\subsubsection{Discovery via Precise Measurements of Standard Model Parameters}
\label{subsubsec:heavyNP_indirect}
Heavy resonances will impact a large variety of SM observables. As discussed already in section~\ref{subsubsec:noNP_indirect}, the
discovery reach of the ILC precision program exceeds the discovery reach by direct production at the LHC in many cases.
If a heavy resonance will be discovered at the LHC, it gives a clear proof for the existance of BSM physics. Realistic models describing the LHC observation will, however, contain more than this one new particle, but typically partners or exitations
of many SM particles. Therefore precision measurements at the ILC will not only provide essential information to characterize
the resonance found at the LHC, but they will also offer significant discovery potential for additional states.

\paragraph{Electroweak Couplings of the Top Quark:}

As discussed in section~\ref{subsubsec:noNP_indirect}, the ILC will be able to improve the precisions on the couplings of the top quark to the $Z$ boson by about an order of magnitude, thus being sensitive to e.g.\ Kaluza-Klein resonances up to masses of several tens of TeV, as shown in Fig.~\ref{fig:topcoup-lumi}. 
In the presence of a resonance already discovered at the LHC, the capability of the ILC
to distinguish between the various BSM models as illustrated in Fig.~\ref{fig:models-rp}, is of special importance.

Note here, that the ILC will offer also a superb precision on the coupling to the $b$ quark suited to underpin observations that will be made for the $t$ quark. As the effects on the $b$ quark may be smaller than that for the $t$ quark, these measurements will benefit from the high integrated luminosity available for the H-20 scenario.

\paragraph{Two-Fermion Production and New Gauge Bosons: } 
As discussed in section~\ref{subsubsec:noNP_indirect}, the precise measurement of polarized two-fermion production cross sections
at the ILC will allow to observe new gauge bosons, for instance $Z'$ bosons, up to very high scales, which in several models
exceed significantly the reach of the LHC. In the presence of one $Z'$-like boson already discovered at the LHC, the ILC
has complementary discovery potential for additional such states, either at higher masses, or even in the case of two
degenerate new $Z'$ bosons~\cite{Barducci:2015aoa} remaining unresolved as one state at the LHC. It is shown that beam polarisation will allow for disentangling the two $Z'$ at the ILC, underlining that its versatile design is ideally suited to decipher details of new physics scenarios. 


\subsubsection{Spin-2 Resonance} 
Clearly a major breakthrough in combining relativistic quantum field theory and gravity would be the discovery of a Spin-2 particle that may henceforth be interpreted as a KK excitation of the graviton. As before it can be expected that the KK graviton will be accompanied by KK excitations of the known Standard Model vector bosons so the observations made in the previous paragraphs may hold here again. The Ref.~\cite{Fujii:2016raq} points out that it is at least not excluded that the Spin-2 particle couples to $\ee$ pairs which could turn into an avenue for the ILC if the mass is at least within the 1 TeV upgrade scenario for the ILC. In~\cite{Fujii:2016raq} it is also reminded that a Spin-2 resonance would be a strong case for considering the option to run the ILC as a $\gamma \gamma$ collider.

\subsubsection{Loophole-free Search for Light States }
\label{subsubsec:heavyNP_simplifiedSUSY}
If LHC discovers a relatively heavy particle without a clear indication for the existance of any lighter state,
this does not neccessarily mean that no lighter states exist, for the same reasons as given in section~\ref{subsubsec:noNP_loopholefree}, in particular for the 
there mentioned non-SUSY cases.
There might also be ambigous cases, where it is not clear whether the heavy new particle decays to SM particles
directly, or via some much lighter, exotic states whose decay products cannot be resolved in the detectors due to
the high boost, as it was also discussed as a possible explanation of the $750$\,GeV excess, see e.g.\ \cite{Ellwanger:2016qax}.
In any of these cases, the capability of the ILC to probe for all
existing particles in Nature with electromagnetic, hyper-charge or
electroweak quantum numbers still provides discovery potential complementary to that of the LHC.

\subsubsection{WIMP Dark Matter }
\label{subsubsec:heavyNP_WIMPs}

In the case that a heavy new particle is discovered at the LHC, it is still important to check for lighter, invisible
particles coupling to leptons, as it can be done in a generic way via the mono-photon signature (c.f.\ section~\ref{subsec:directNP_WIMP}). Discovery and characterisation, or else a rather model-independent exclusion of such a particle as enabled by ILC data would provide important and complementary information for unveiling the nature of the
heavy particle found at the LHC.


%

\section{Complementarity and Synergy of LHC and ILC}
\label{sec:ILCandLHC}

The physics program of the ILC and its resulting discovery potential show a high degree of complementarity with the LHC.
But there will be also many synergies between the two machines:

The huge production rates at LHC enable, for example, a high sensitivity to rare Higgs decays with distinctive signatures such as $h \to \gamma\gamma$, $Z\gamma$, and $\mu^+\mu^-$, though measurements of their absolute branching fractions are limited by various systematic uncertainties and the lack of a possibility for model-independent total cross-section measurement at the LHC. The systematic uncertainties can be largely cancelled by taking their ratios to $h \to ZZ^*$, and together with the ILC's model-independent precision measurement of the total Higgs strahlung cross section, we can hence achieve a percent level measurements for these rare modes, which is a notable synergy of the ILC and the LHC. Independently of the direct production of new particles at the LHC or the ILC, precision measurements of all properties of the Higgs boson (c.f.\ section~\ref{subsec:ILC_Higgs}), along with those of the top quark (c.f.\ section~\ref{subsec:ILC_top}) and the $W$ and $Z$ bosons(c.f.\ section~\ref{subsec:ILC_ew}) will tell us a lot about the -- to date unknown -- energy scale of new physics, which is essential input to the physics case of energy upgrades of the ILC itself or of higher energy hadron colliders.

However, we stress that the ILC's capability is not limited to precision measurements, but that also searches for direct production
of new particles will add important information to our picture of our universe, 
because direct searches at the LHC and the ILC (c.f.\ section~\ref{sec:ILC_NP}) cover different parts of the parameter space: the LHC has in general a higher reach for heavier, coloured states, while ILC has higher sensitivity to subtle signals from weakly coupled new particles.
%
%
A typical example of the ILC-LHC complementarity in parameter space coverage is from the generic WIMP search discussed in sections~\ref{subsec:directNP_WIMP},~\ref{subsubsec:noNP_WIMPs},~\ref{subsubsec:lightNP_WIMPs}, and~\ref{subsubsec:heavyNP_WIMPs}: The LHC is sensitive to WIMP couplings to quarks, while the ILC is sensitive to WIMP couplings to electrons.
While the LHC has a higher WIMP mass reach, the ILC has a higher sensitivity to smaller couplings, thus higher mediator mass scales.
In SUSY DM models, dark matter co-annihilation with NLSP is preferably used to reduce the DM relic density, which in turn requires a small mass difference, resulting in subtle signals. 
The ILC is capable of detecting these subtle signals as shown in sections~\ref{subsec:directNP_NoLoophole} and~\ref{subsec:directNP_SUSYDM}
Once discovered, the ILC's clean environment allows us to characterize the WIMP dark matter.

The LHC-ILC complementarity also applies to searches for other new particles including SUSY particles (c.f.\ section~\ref{subsubsec:noNP_loopholefree} and~\ref{subsubsec:noNP_NatSUSY}) or extra Higgs bosons (c.f.\ section~\ref{subsubsec:noNP_SUSYHiggs}).
While the LHC has largest sensitivity to colored SUSY particles such as gluino cascade-decaying into 
lighter uncolored SUSY particles with large energy release, the ILC is going to look directly for uncolored SUSY particles with a small mass gap and hence small energy release.
As detailed in section~\ref{subsec:SUSY}, the radiatively driven natural SUSY scenario provides a very interesting and attractive possibility, where the small mass difference is theoretically required. 
Since in this case the higgsinos reside in the blind region of the LHC in Fig.~\ref{fig:noloophole2} near the diagonal, the ILC's capability described in section~\ref{subsec:directNP_higgsinos} to cover this hole is essential to the full exploration of this scenario.
The ILC is capable not only of discovering these higgsinos but also of measuring their masses and production cross sections with high precision (c.f.\ section~\ref{subsec:directNP_higgsinos}).
These precision measurements of the higgsinos together with those of the $125$\,GeV Higgs boson will allow us to determine the model parameters such as $\mu$, $M_1$, $M_2$, etc.\ as demonstrated in section~\ref{subsubsec:noNP_NatSUSY}.

The LHC-ILC complementarity and synergy becomes most prominent in the scenario described in section~\ref{subsec:lightNP}, where the LHC experiments discover relatively light new particles. Particle masses and mixings measured at the ILC will then be important inputs to disentangle complicated cascade decays of SUSY particles at the LHC, as discussed in section~\ref{subsubsec:lightNP_SUSY}.
If a gluino is found at the LHC, we can test the gaugino unification as shown in section~\ref{subsubsec:lightNP_NatSUSY}, opening up the window to GUT scale physics, thus probing the unimaginable energy scale of $10^{16}$\,GeV.
If, however, the mass unification scale turns out to be significantly lower than the GUT scale, it would be an indication of the mirage unification scenario. If the gaugino mass unification does not happen, it would discriminate certain classes of SUSY breaking scenarios.
The predicted SUSY particle masses will again be important inputs to the physics case of energy upgrades of the ILC itself or of higher energy hadron colliders.

But also in the case that LHC discovers a relatively heavy new particle (c.f.\ section~\ref{subsec:heavyNP}), both the precision measurements and the discovery potential of the ILC are highly complemetary to the information to be expected from the LHC and remain crucial for understanding the nature of the heavy new particle -- and thus for pushing further our knowledge of the universe.



%

\section{Conclusions} 
\label{sec:conclusions}
In this paper, we have demonstrated that the ILC has significant potential to discover phenomena beyond the Standard Model, irrespective of the results that LHC will obtain in the coming years. 

Precision measurements at the ILC have in many cases sensitivity to the new physics scales that far exceeds that of the direct searches at the LHC.  But also, there are many scenarios in which the capabilities for the direct discovery of new particles exceed those of the LHC. Both capabilities rely on the well-appreciated properties of $e^+e^-$  colliders: the well-defined initial state, the clean environment without QCD backgrounds, and the democratic production of particles with electroweak charges. They also benefit from the ILC's extendability in energy and polarised beams. Together, these features lead to synergies between the physics programs of ILC and LHC. We have discussed explicitly how these synergies play out in three scenarios for discoveries at the LHC before the turn-on of the ILC. 

The current LHC Run 2 so far has not provided evidence for other new particles beyond the Higgs boson with a mass of $125$\,GeV. Though the LHC still has a significant window for the direct discovery of new particles, this window will narrow considerably if no evidence for new particles appears in this year's results  (with $30$-$40$\,fb$^{-1}$ at $13$\,TeV). On the other hand, the fundamental mysteries of particle physics will not have gone away. This puts additional emphasis on the Scenario 1 that we have discussed here and makes it especially important to prepare an alternative route for the discovery of new physics beyond the Standard Model. We hope very much that the LHC will break through to the discovery of new particles in the next few years. However, even in the most pessimistic case, the ILC offers distinct and very powerful strategies for the discovery of phenomena that lie beyond the Standard Model of particle physics, phenomena that will illuminate physics both at small distances and in the large-scale makeup of the universe.



\section*{Acknowledgements}
The authors thank the ILD and SiD detector concepts groups for providing material for this document.
We thank M.~Savoy and H.~Serce for help with some figures.
M.~Berggren, C.~Grojean, S.~Lehtinen, J.~List and J.~Reuter thankfully acknowledge the support of the SFB676 
of the Germany Science Foundation. F.~Simon acknowledges the support of the DFG cluster of excellence `Origin and Structure of the Universe'. C.~Grojean is supported by the European Commission through the Marie Curie Career Integration Grant 631962 and by the Helmholtz Association.
The work of S.~Heinemeyer is supported in part by CICYT (Grant FPA 2013-40715-P) and by the Spanish MICINN's Consolider-Ingenio 2010 Program under Grant MultiDark CSD2009-00064. The work of  T.~Barklow and M.~Peskin is supported by the U.S. Department of Energy, contract
DE-AC02-76SF00515.
Some material is based upon work supported by the U.S. Department of Energy, Office of Science, 
Office of High Energy Physics under Award Number DE-SC-0009956.
This work is also supported by the Grants-in-Aid for Science Research No. 16H02173 and 16H02176 of the Japan Society for Promotion of Science (JSPS).


\appendix
\begin{footnotesize}
\bibliographystyle{apsrev}

\begin{thebibliography}{00}

\bibitem{Barklow:2015tja}
  T.~Barklow, J.~Brau, K.~Fujii, J.~Gao, J.~List, N.~Walker and K.~Yokoya,
  ``ILC Operating Scenarios,''
  \href{http://arxiv.org/abs/1506.07830}{arXiv:1506.07830 [hep-ex]}.
  
\bibitem{LandauGinzburg}
V.L.~Ginzburg and L.D.~Landau, Zh. Eksp. Teor. Fiz.~{\bf 20}4 (1950) 1064.

\bibitem{BCS}
L.~N.~Cooper, Phys. Rev. {\bf 104} (1956) 1189;
J.~Bardeen, L.~N.~Cooper, and J.~R.~Schrieffer, Phys. Rev. {\bf 106}
(1957) 162, Phys. Rev. {\bf 108} (1957) 1175.

\bibitem{Fujii:2015jha}
  K.~Fujii {\it et al.},
  ``Physics Case for the International Linear Collider,''
  \href{http://arxiv.org/abs/1506.05992}{arXiv:1506.05992 [hep-ex]}.
  
\bibitem{Kanemura:2014bqa} 
  S.~Kanemura, K.~Tsumura, K.~Yagyu and H.~Yokoya,
  \href{http://dx.doi.org/10.1103/PhysRevD.90.075001}{Phys.\ Rev.\ D {\bf 90} (2014)  075001}
  \href{http://arxiv.org/abs/1406.3294}{arXiv:1406.3294 [hep-ph]}.
  
\bibitem{Fujii:2016raq}
{\bf LCC Physics Working Group}, K.~Fujii {\em et al.}, ``{Implications of the
  750 GeV gamma-gamma Resonance as a Case Study for the International Linear
  Collider}''
  \href{http://arxiv.org/abs/1607.03829}{arXiv:1607.03829 [hep-ph]}.

\bibitem{Giudice:2008bi}
  G.~F.~Giudice,
  ``Naturally Speaking: The Naturalness Criterion and Physics at the LHC,''
  In *Kane, Gordon (ed.), Pierce, Aaron (ed.): Perspectives on LHC physics* 155-178
  \href{http://arxiv.org/abs/0801.2562}{arXiv:0801.2562 [hep-ph]}.

\bibitem{Baer:2014eja}
  H.~Baer, K.~Y.~Choi, J.~E.~Kim and L.~Roszkowski,
  \href{http://dx.doi.org/10.1016/j.physrep.2014.10.002}{Phys.\ Rept.\  {\bf 555} (2015) 1}
  \href{http://arxiv.org/abs/1407.0017}{arXiv:1407.0017 [hep-ph]}.

\bibitem{Bae:2013bva}
K.~J.~Bae, H.~Baer and E.~J.~Chun,
  \href{http://dx.doi.org/10.1103/PhysRevD.89.031701}{Phys.\ Rev.\ D {\bf 89} (2014) no.3,  031701}
  \href{http://arxiv.org/abs/1309.0519}{arXiv:1309.0519 [hep-ph]}.

\bibitem{Dimopoulos:1981zb}
S.~Dimopoulos and H.~Georgi, ``{Softly Broken Supersymmetry and SU(5) }'' {\em
  Nucl.Phys.} {\bf B193}  150.

\bibitem{Feng:1995zd}
  J.~L.~Feng, M.~E.~Peskin, H.~Murayama and X.~R.~Tata,
  \href{http://dx.doi.org/10.1103/PhysRevD.52.1418}{Phys.\ Rev.\ D {\bf 52} (1995) 1418}
  \href{http://arxiv.org/abs/hep-ph/9502260}{arXiv:hep-ph/9502260}.

\bibitem{Baer:2012up}
  H.~Baer, V.~Barger, P.~Huang, A.~Mustafayev and X.~Tata,
  Phys.\ Rev.\ Lett.\  {\bf 109} (2012) 161802
  \href{http://dx.doi.org/10.1103/PhysRevLett.109.161802}{Phys.\ Rev.\ Lett.\  {\bf 109} (2012) 161802}
  \href{http://arxiv.org/abs/1207.3343}{arXiv:1207.3343 [hep-ph]};
H.~Baer, V.~Barger, P.~Huang, D.~Mickelson, A.~Mustafayev and X.~Tata,
  Phys.\ Rev.\ D {\bf 87} (2013) no.11,  115028
  \href{http://dx.doi.org/10.1103/PhysRevD.87.115028}{Phys.\ Rev.\ D {\bf 87} (2013) no.11,  115028}
  \href{http://arxiv.org/abs/1212.2655}{arXiv:1212.2655 [hep-ph]}.

\bibitem{Baer:2015rja}
  H.~Baer, V.~Barger and M.~Savoy,
  \href{http://dx.doi.org/10.1103/PhysRevD.93.035016}{Phys.\ Rev.\ D {\bf 93} (2016) no.3,  035016}
  \href{http://arxiv.org/abs/1212.2655}{arXiv:1509.02929 [hep-ph]}.

\bibitem{Baer:2014ica}
  H.~Baer, V.~Barger, D.~Mickelson and M.~Padeffke-Kirkland,
  Phys.\ Rev.\ D {\bf 89} (2014) no.11,  115019
  doi:10.1103/PhysRevD.89.115019
  [arXiv:1404.2277 [hep-ph]].

\bibitem{Baer:2016usl}
  H.~Baer, V.~Barger, M.~Savoy and X.~Tata,
  Phys.\ Rev.\ D {\bf 94} (2016) no.3,  035025
  doi:10.1103/PhysRevD.94.035025
  [arXiv:1604.07438 [hep-ph]].


\bibitem{Baer:2016hfa}
  H.~Baer, V.~Barger, H.~Serce and X.~Tata,
  arXiv:1610.06205 [hep-ph].

\bibitem{Baer:2000hx}
  H.~Baer, C.~Balazs, J.~K.~Mizukoshi and X.~Tata,
  Phys.\ Rev.\ D {\bf 63} (2001) 055011
  doi:10.1103/PhysRevD.63.055011
  [hep-ph/0010068].
  
\bibitem{Deppisch:2003wt}
  F.~Deppisch, H.~Pas, A.~Redelbach, R.~Ruckl and Y.~Shimizu,
  Phys.\ Rev.\ D {\bf 69} (2004) 054014
  doi:10.1103/PhysRevD.69.054014
  [hep-ph/0310053].
  
\bibitem{Murayama:1992dj}
  H.~Murayama, H.~Suzuki and T.~Yanagida,
  Phys.\ Lett.\ B {\bf 291} (1992) 418.
  doi:10.1016/0370-2693(92)91397-R

\bibitem{Bae:2014yta}
  K.~J.~Bae, H.~Baer and H.~Serce,
  Phys.\ Rev.\ D {\bf 91} (2015) no.1,  015003
  doi:10.1103/PhysRevD.91.015003
  [arXiv:1410.7500 [hep-ph]].

\bibitem{Porod:2000hv}
  W.~Porod, M.~Hirsch, J.~Romao and J.~W.~F.~Valle,
  Phys.\ Rev.\ D {\bf 63} (2001) 115004
  doi:10.1103/PhysRevD.63.115004
  [hep-ph/0011248].
  
\bibitem{Hirsch:2003fe}
  M.~Hirsch and W.~Porod,
  Phys.\ Rev.\ D {\bf 68} (2003) 115007
  doi:10.1103/PhysRevD.68.115007
  [hep-ph/0307364].

\bibitem{Kanemura:2004ch}
  S.~Kanemura, Y.~Okada and E.~Senaha,
  Phys.\ Lett.\ B {\bf 606} (2005) 361
  doi:10.1016/j.physletb.2004.12.004
  [hep-ph/0411354].
  
\bibitem{Noble:2007kk}
  A.~Noble and M.~Perelstein,
  Phys.\ Rev.\ D {\bf 78} (2008) 063518
  doi:10.1103/PhysRevD.78.063518
  [arXiv:0711.3018 [hep-ph]].

\bibitem{Randall:1999ee}
L.~Randall and R.~Sundrum, ``{A Large mass hierarchy from a small extra
  dimension}'' \href{http://dx.doi.org/10.1103/PhysRevLett.83.3370}{{\em
  Phys.Rev.Lett.} {\bf 83} (1999)  3370--3373},
\href{http://arxiv.org/abs/hep-ph/9905221}{{\tt arXiv:hep-ph/9905221
  [hep-ph]}}.

\bibitem{Goldberger:1999uk}
  W.~D.~Goldberger and M.~B.~Wise,
  Phys.\ Rev.\ Lett.\  {\bf 83} (1999) 4922
  doi:10.1103/PhysRevLett.83.4922
  [hep-ph/9907447].
  
\bibitem{Djouadi:2006rk}
A.~Djouadi, G.~Moreau, and F.~Richard, ``{Resolving the A(FB)**b puzzle in an
  extra dimensional model with an extended gauge structure}''
  \href{http://dx.doi.org/10.1016/j.nuclphysb.2007.03.019}{{\em Nucl.Phys.}
  {\bf B773} (2007)  43--64},
\href{http://arxiv.org/abs/hep-ph/0610173}{{\tt arXiv:hep-ph/0610173
  [hep-ph]}}.

\bibitem{Carena:2006bn}
M.~S.~Carena, E.~Ponton, J.~Santiago, and C.~E.~Wagner, ``{Light Kaluza Klein
  States in Randall-Sundrum Models with Custodial SU(2)}''
  \href{http://dx.doi.org/10.1016/j.nuclphysb.2006.10.012}{{\em Nucl.Phys.}
  {\bf B759} (2006)  202--227},
\href{http://arxiv.org/abs/hep-ph/0607106}{{\tt arXiv:hep-ph/0607106
  [hep-ph]}}.

\bibitem{Csaki:2015hcd}
C.~Csaki, C.~Grojean and J.~Terning,
 Rev.\ Mod.\ Phys.\  {\bf 88} (2016) no.4,  045001
 doi:10.1103/RevModPhys.88.045001
 [arXiv:1512.00468 [hep-ph]].

\bibitem{Panico:2015jxa}
  G.~Panico and A.~Wulzer,
  Lect.\ Notes Phys.\  {\bf 913} (2016) pp.1
  doi:10.1007/978-3-319-22617-0
  [arXiv:1506.01961 [hep-ph]].
  
\bibitem{Csaki:2008zd}
  C.~Csaki, A.~Falkowski and A.~Weiler,
  JHEP {\bf 0809} (2008) 008
  doi:10.1088/1126-6708/2008/09/008
  [arXiv:0804.1954 [hep-ph]].

\bibitem{Matsedonskyi:2012ym}
  O.~Matsedonskyi, G.~Panico and A.~Wulzer,
  JHEP {\bf 1301} (2013) 164
  doi:10.1007/JHEP01(2013)164
  [arXiv:1204.6333 [hep-ph]].
  
\bibitem{LHCHiggsCrossSectionWorkingGroup:2012nn}
  A.~David {\it et al.} [LHC Higgs Cross Section Working Group Collaboration],
  arXiv:1209.0040 [hep-ph].
  
\bibitem{Agashe:2006at}
  K.~Agashe, R.~Contino, L.~Da Rold and A.~Pomarol,
  Phys.\ Lett.\ B {\bf 641} (2006) 62
  doi:10.1016/j.physletb.2006.08.005
  [hep-ph/0605341].

\bibitem{Grojean:2013qca}
  C.~Grojean, O.~Matsedonskyi and G.~Panico,
  JHEP {\bf 1310} (2013) 160
  doi:10.1007/JHEP10(2013)160
  [arXiv:1306.4655 [hep-ph]].

\bibitem{ArkaniHamed:2001nc} 
  N.~Arkani-Hamed, A.~G.~Cohen and H.~Georgi,
  Phys.\ Lett.\ B {\bf 513}, 232 (2001)
  doi:10.1016/S0370-2693(01)00741-9
  [hep-ph/0105239].

\bibitem{ArkaniHamed:2002qx} 
  N.~Arkani-Hamed, A.~G.~Cohen, E.~Katz, A.~E.~Nelson, T.~Gregoire and J.~G.~Wacker,
  JHEP {\bf 0208}, 021 (2002)
  doi:10.1088/1126-6708/2002/08/021
  [hep-ph/0206020].

\bibitem{ArkaniHamed:2002qy} 
  N.~Arkani-Hamed, A.~G.~Cohen, E.~Katz and A.~E.~Nelson,
  JHEP {\bf 0207}, 034 (2002)
  doi:10.1088/1126-6708/2002/07/034
  [hep-ph/0206021].

\bibitem{Schmaltz:2004de} 
  M.~Schmaltz,
  JHEP {\bf 0408}, 056 (2004)
  doi:10.1088/1126-6708/2004/08/056
  [hep-ph/0407143].
  
\bibitem{Kilian:2004pp} 
  W.~Kilian, D.~Rainwater and J.~Reuter,
  Phys.\ Rev.\ D {\bf 71}, 015008 (2005)
  doi:10.1103/PhysRevD.71.015008
  [hep-ph/0411213].

\bibitem{Chacko:2005pe}
  Z.~Chacko, H.~S.~Goh and R.~Harnik,
  Phys.\ Rev.\ Lett.\  {\bf 96} (2006) 231802
  doi:10.1103/PhysRevLett.96.231802
  [hep-ph/0506256].

\bibitem{Chacko:2005un}
  Z.~Chacko, H.~S.~Goh and R.~Harnik,
  JHEP {\bf 0601} (2006) 108
  doi:10.1088/1126-6708/2006/01/108
  [hep-ph/0512088].

\bibitem{Craig:2015pha}
  N.~Craig, A.~Katz, M.~Strassler and R.~Sundrum,
  JHEP {\bf 1507} (2015) 105
  doi:10.1007/JHEP07(2015)105
  [arXiv:1501.05310 [hep-ph]].

\bibitem{Baer:2013cma}
  H.~Baer {\it et al.},
  ``The International Linear Collider Technical Design Report - Volume 2: Physics,''
  arXiv:1306.6352 [hep-ph].
  
\bibitem{Moortgat-Picka:2015yla}
  G.~Moortgat-Pick {\it et al.},
  Eur.\ Phys.\ J.\ C {\bf 75} (2015) no.8,  371
  doi:10.1140/epjc/s10052-015-3511-9
  [arXiv:1504.01726 [hep-ph]].
  
  
\bibitem{Djouadi:2007ik}
  G.~Aarons {\it et al.} [ILC Collaboration],
  ``International Linear Collider Reference Design Report Volume 2: Physics at the ILC,''
  arXiv:0709.1893 [hep-ph].
  
\bibitem{Asner:2013psa}
  D.~M.~Asner {\it et al.},
  ``ILC Higgs White Paper,''
  arXiv:1310.0763 [hep-ph].
  
\bibitem{Baer:2013vqa}
  H.~Baer, M.~Berggren, J.~List, M.~M.~Nojiri, M.~Perelstein, A.~Pierce, W.~Porod and T.~Tanabe,
  ``Physics Case for the ILC Project: Perspective from Beyond the Standard Model,''
  arXiv:1307.5248 [hep-ph].

\bibitem{Baak:2013fwa}
  M.~Baak {\it et al.},
  arXiv:1310.6708 [hep-ph].

\bibitem{Agashe:2013hma}
  K.~Agashe {\it et al.} [Top Quark Working Group Collaboration],
  arXiv:1311.2028 [hep-ph].
  
  
\bibitem{Vos:2016til}
  M.~Vos {\it et al.},
  arXiv:1604.08122 [hep-ex].

\bibitem{CMS:2013xfa}
  [CMS Collaboration],
  ``Projected Performance of an Upgraded CMS Detector at the LHC and HL-LHC: Contribution to the Snowmass Process,''
  arXiv:1307.7135 [hep-ex].
  
\bibitem{bib:ATLAS_HLLHC_Higgs}
  ATLAS Collaboration,
  LHCC-I-023, CERN-LHCC-2012-022.

\bibitem{Jeans:2016}
 D.~Jeans, "Measuring the $CP$ state of tau lepton pairs
2 from Higgs decay at the ILC", paper in preparation.
  
\bibitem{Harnik:2013aja}
  R.~Harnik, A.~Martin, T.~Okui, R.~Primulando and F.~Yu,
  Phys.\ Rev.\ D {\bf 88} (2013) no.7,  076009
  doi:10.1103/PhysRevD.88.076009
  [arXiv:1308.1094 [hep-ph]].

\bibitem{BhupalDev:2007ftb}
  P.~S.~Bhupal Dev, A.~Djouadi, R.~M.~Godbole, M.~M.~Muhlleitner and S.~D.~Rindani,
  Phys.\ Rev.\ Lett.\  {\bf 100} (2008) 051801
  doi:10.1103/PhysRevLett.100.051801
  [arXiv:0707.2878 [hep-ph]].

\bibitem{Godbole:2011hw}
  R.~M.~Godbole, C.~Hangst, M.~Muhlleitner, S.~D.~Rindani and P.~Sharma,
  Eur.\ Phys.\ J.\ C {\bf 71} (2011) 1681
  doi:10.1140/epjc/s10052-011-1681-7
  [arXiv:1103.5404 [hep-ph]].
  
\bibitem{Hagiwara:2016rdv}
  K.~Hagiwara, K.~Ma and H.~Yokoya,
  JHEP {\bf 1606} (2016) 048
  doi:10.1007/JHEP06(2016)048
  [arXiv:1602.00684 [hep-ph]].
  
\bibitem{Duerig:2016dvi}
  C.~F.~D\"urig,
  ``Measuring the Higgs Self-coupling at the International Linear Collider,''
  DESY-THESIS-2016-027, \url{http://inspirehep.net/record/1493742/files/phd_thesis_duerig.pdf}.
  
\bibitem{Tian:2013}
J.~Tian and K.~Fujii, "Study of Higgs self-coupling at the ILC 
based on the full detector simulation 
at $\sqrt{s}$ = 500 GeV and 1 TeV", 
LC-REP-2013-003.

\bibitem{Kurata:2013}
M.~Kurata {\it et al.}, "The Higgs self-coupling analysis 
using the events containing $H\to WW^*$ decay"
LC-REP-2013-025.

\bibitem{Tian:2015}
J.~Tian and K.~Fujii, "ZHH cross section analysis", 
talk at the LCWS15, paper in preparation.

\bibitem{Seidel:2013sqa} 
  K.~Seidel, F.~Simon, M.~Tesar and S.~Poss,
  Eur.\ Phys.\ J.\ C {\bf 73}, 2530 (2013)
  [arXiv:1303.3758 [hep-ex]].
  
  \bibitem{Beneke:2015kwa} 
  M.~Beneke {\it et al.}, 
  Phys.\ Rev.\ Lett.\  {\bf 115}, 192001 (2015)
  [arXiv:1506.06864 [hep-ph]].

\bibitem{Simon:2016htt} 
  F.~Simon,
  arXiv:1603.04764 [hep-ex].
  
  \bibitem{Simon:2016pwp} 
  F.~Simon,
  arXiv:1611.03399 [hep-ex].

\bibitem{Fuster:2015jva} 
  J.~Fuster, I.~Garc\'ia, P.~Gomis, M.~Perell\'o, E.~Ros and M.~Vos,
  Eur.\ Phys.\ J.\ C {\bf 75}, 223 (2015)
  [arXiv:1411.2355 [hep-ex]].
  
  \bibitem{Simon:2014hna} 
  F.~Simon,
  arXiv:1411.7517 [hep-ex].

\bibitem{Marquard:2015qpa} 
  P.~Marquard, A.~V.~Smirnov, V.~A.~Smirnov and M.~Steinhauser,
  Phys.\ Rev.\ Lett.\  {\bf 114}, no. 14, 142002 (2015) 
  [arXiv:1502.01030 [hep-ph]].
  
\bibitem{Amjad:2013tlv}
M.~Amjad, M.~Boronat, T.~Frisson, I.~Garcia, R.~Poschl, {\em et al.}, ``{A
  precise determination of top quark electro-weak couplings at the ILC
  operating at $\sqrt{s}=500$ GeV}''
\href{http://arxiv.org/abs/1307.8102}{{\tt arXiv:1307.8102 [hep-ex]}}.

\bibitem{Amjad:2015mma}
M.~S.~Amjad {\em et al.}, ``{A precise characterisation of the top quark
  electro-weak vertices at the ILC}''
  \href{http://dx.doi.org/10.1140/epjc/s10052-015-3746-5}{{\em accepted by
  European Journal of Physics C} (2015)  },
\href{http://arxiv.org/abs/1505.06020}{{\tt arXiv:1505.06020 [hep-ex]}}.

\bibitem{Baur:2004uw}
U.~Baur, A.~Juste, L.~Orr, and D.~Rainwater, ``{Probing electroweak top quark
  couplings at hadron colliders}''
  \href{http://dx.doi.org/10.1103/PhysRevD.71.054013}{{\em Phys.Rev.} {\bf D71}
  (2005)  054013},
\href{http://arxiv.org/abs/hep-ph/0412021}{{\tt arXiv:hep-ph/0412021
  [hep-ph]}}.

\bibitem{Baur:2005wi}
U.~Baur, A.~Juste, D.~Rainwater, and L.~Orr, ``{Improved measurement of $ttZ$
  couplings at the CERN LHC}''
  \href{http://dx.doi.org/10.1103/PhysRevD.73.034016}{{\em Phys.Rev.} {\bf D73}
  (2006)  034016},
\href{http://arxiv.org/abs/hep-ph/0512262}{{\tt arXiv:hep-ph/0512262
  [hep-ph]}}.

\bibitem{Khachatryan:2015sha}
{\bf CMS}, V.~Khachatryan {\em et al.}, ``{Observation of top quark pairs
  produced in association with a vector boson in pp collisions at sqrt(s) = 8
  TeV}''
\href{http://arxiv.org/abs/1510.01131}{{\tt arXiv:1510.01131 [hep-ex]}}.

 
  
\bibitem{bib:panico-priv}
{G.~Panico, A.~Wulzer, private communication, Possible deviations of couplings
  in framework described in~\cite{Grojean:2013qca}}.

\bibitem{Berger:2005ht}
C.~Berger, M.~Perelstein, and F.~Petriello, ``{Top quark properties in little
  Higgs models}''
\href{http://arxiv.org/abs/hep-ph/0512053}{{\tt arXiv:hep-ph/0512053
  [hep-ph]}}.

\bibitem{Pomarol:2008bh}
A.~Pomarol and J.~Serra, ``{Top Quark Compositeness: Feasibility and
  Implications}'' \href{http://dx.doi.org/10.1103/PhysRevD.78.074026}{{\em
  Phys.Rev.} {\bf D78} (2008)  074026},
\href{http://arxiv.org/abs/0806.3247}{{\tt arXiv:0806.3247 [hep-ph]}}.

\bibitem{Cui:2010ds}
Y.~Cui, T.~Gherghetta, and J.~Stokes, ``{Fermion Masses in Emergent Electroweak
  Symmetry Breaking}'' \href{http://dx.doi.org/10.1007/JHEP12(2010)075}{{\em
  JHEP} {\bf 1012} (2010)  075},
\href{http://arxiv.org/abs/1006.3322}{{\tt arXiv:1006.3322 [hep-ph]}}.

\bibitem{Barducci:2015aoa}
D.~Barducci, S.~De~Curtis, S.~Moretti, and G.~M.~Pruna, ``{Top pair production
  at a future $e^+e^-$ machine in a composite Higgs scenario}''
\href{http://arxiv.org/abs/1504.05407}{{\tt arXiv:1504.05407 [hep-ph]}}.




\bibitem{Richard:2014upa} 
  F.~Richard,
  ``Present and future constraints on top EW couplings,''
  arXiv:1403.2893 [hep-ph].

\bibitem{Khiem:2015ofa}
P.~Khiem, E.~Kou, Y.~Kurihara, and F.~L.~Diberder, ``{Probing New Physics using
  top quark polarization in the $e^+ e^- \rightarrow t \bar{t}$ process at
  future Linear Colliders}''
\href{http://arxiv.org/abs/1503.04247}{{\tt arXiv:1503.04247 [hep-ph]}}.

\bibitem{Freitas:2013xga} 
  A.~Freitas, K.~Hagiwara, S.~Heinemeyer, P.~Langacker, K.~Moenig, M.~Tanabashi and G.~W.~Wilson,
  arXiv:1307.3962.

\bibitem{Osland:2009dp}
  P.~Osland, A.~A.~Pankov and A.~V.~Tsytrinov,
  Eur.\ Phys.\ J.\ C {\bf 67} (2010) 191
  [arXiv:0912.2806 [hep-ph]].

\bibitem{Godfrey:2005pm}
  S.~Godfrey, P.~Kalyniak and A.~Tomkins,
  hep-ph/0511335.

\bibitem{Godfrey:2000hc}
  S.~Godfrey, P.~Kalyniak, B.~Kamal and A.~Leike,
  Phys.\ Rev.\ D {\bf 61} (2000) 113009
  [hep-ph/0001074].

\bibitem{Baer:2001ia}
  H.~Baer and A.~Belyaev,
  eConf C {\bf 010630} (2001) P336
  [hep-ph/0111017].

\bibitem{Birkedal:2004xn}
  A.~Birkedal, K.~Matchev and M.~Perelstein,
  Phys.\ Rev.\ D {\bf 70} (2004) 077701
  [hep-ph/0403004].
  
\bibitem{Dreiner:2012xm}
  H.~Dreiner, M.~Huck, M.~Kr\"amer, D.~Schmeier and J.~Tattersall,
  Phys.\ Rev.\ D {\bf 87} (2013) 075015
  [arXiv:1211.2254 [hep-ph]].
  
\bibitem{Chae:2012bq}
  Y.~J.~Chae and M.~Perelstein,
  ``Dark Matter Search at a Linear Collider: Effective Operator Approach,''
  arXiv:1211.4008 [hep-ph].

\bibitem{Bartels:2012ex}
  C.~Bartels, M.~Berggren and J.~List,
  Eur.\ Phys.\ J.\ C {\bf 72} (2012) 2213
  doi:10.1140/epjc/s10052-012-2213-9
  [arXiv:1206.6639 [hep-ex]].

\bibitem{bib:chaus}
  A.~Chaus, ``Searches for Dark Matter particules and development of a pixellized readout of the Time Projection Chamber for
the International Linear Collider (ILC),'' PhD thesis, 2014, Universit\'e Paris 11
\url{http://www.theses.fr/2014PA112300}.

\bibitem{bib:habermehl}
 M.~Habermehl {\it et al.}, ``WIMP Searches at the International Linear Collider,'' PoS(ICHEP2016)155, \url{https://pos.sissa.it/archive/conferences/282/155/ICHEP2016_155.pdf}.

\bibitem{Baltz:2006fm}
  E.~A.~Baltz, M.~Battaglia, M.~E.~Peskin and T.~Wizansky,
  Phys.\ Rev.\ D {\bf 74} (2006) 103521
  doi:10.1103/PhysRevD.74.103521
  [hep-ph/0602187].

\bibitem{Berggren:2013vna}
M.~Berggren, ``{Simplified SUSY at the ILC }''
\href{http://arxiv.org/abs/1308.1461}{{\tt arXiv:1308.1461 [hep-ph]}}.

\bibitem{deVries:2015hva}
  K.~J.~de Vries {\it et al.},
  Eur.\ Phys.\ J.\ C {\bf 75} (2015) no.9,  422
  doi:10.1140/epjc/s10052-015-3599-y
  [arXiv:1504.03260 [hep-ph]].
  
  
\bibitem{Lehtinen:2016qis}
  S.~L.~Lehtinen, M.~Berggren and J.~List,
  ``Dark matter relic density from observations of supersymmetry at the ILC,''
  arXiv:1602.08439 [hep-ph].
  

                                                                           \bibitem{Bechtle:2009em}
 P.~Bechtle  {\it et al.},
   \href{http://dx.doi.org/10.1103/PhysRevD.82.055016}{ Phys. Rev.
   {\bfseries D82} (2010) 055016},
 [\href{http://arxiv.org/abs/0908.0876}{{\ttfamily arXiv:0908.0876 [hep-ex]}}].                           

\bibitem{Berggren:2015qua}                                                 M.~Berggren  {\it et al.},                                                 
  Eur.\ Phys.\ J.\ C {\bf 76} (2016) no.4,  183
[\href{http://arxiv.org/abs/1508.04383}{{\ttfamily arXiv:1508.04383 [hep-ph]}}].                          
\bibitem{Berggren:2013vfa}
  M.~Berggren, F.~Br\"ummer, J.~List, G.~Moortgat-Pick, T.~Robens, K.~Rolbiecki and H.~Sert,
  Eur.\ Phys.\ J.\ C {\bf 73} (2013) no.12,  2660
  doi:10.1140/epjc/s10052-013-2660-y
  [arXiv:1307.3566 [hep-ph]].

\bibitem{Baer:2016new}
  H.~Baer, M.~Berggren, K.~Fujii, S.~L.~Lehtinen, J.~List, T.~Tanabe and J.~Yan,
  ``Naturalness and light higgsinos: A powerful reason to build the ILC,''
  arXiv:1611.02846 [hep-ph], extended version in preparation.

\bibitem{Yan:2016LCWS}
  H.~Baer, M.~Berggren, K.~Fujii, S.~L.~Lehtinen, J.~List, T.~Tanabe and J.~Yan,
  ``Search for Light Higgsinos with compressed mass spectrum at ILC center-of-mass energy 500 GeV based on full detector simulation,'' presentation at LCWS 2016, 
 \url{http://agenda.linearcollider.org/event/7371/contributions/37853/}.

\bibitem{Potter:2015wsa} 
  C.~T.~Potter,
  Eur.\ Phys.\ J.\ C {\bf 76}, no. 1, 44 (2016)
  doi:10.1140/epjc/s10052-015-3867-x
  [arXiv:1505.05554 [hep-ph]].

\bibitem{Potter:ECFALC2016} 
  C.~T.~Potter,
  ``The NMSSM Singlet at the ILC,''
  presented at ECFA-LC 2016.

\bibitem{List:2013dga}
  B.~Vormwald and J.~List,
  Eur.\ Phys.\ J.\ C {\bf 74} (2014) 2720
  doi:10.1140/epjc/s10052-014-2720-y
  [arXiv:1307.4074 [hep-ex]].
  
  
\bibitem{ICFAtoMEXT}  
  J.~Mnich {\it et al.}, ``The International Linear Collider: Comments on Scientific Significance
  and Technical Issues'' - in response to ``Summary of the ILC Advisory Panel's discussion to date (June 25 2015)'', December 15, 2015.

\bibitem{Heinemeyer:2007bw}
  S.~Heinemeyer, W.~Hollik, A.~M.~Weber and G.~Weiglein,
  JHEP {\bf 0804} (2008) 039
  [arXiv:0710.2972 [hep-ph]].
 
\bibitem{Heinemeyer:2013dia}
  S.~Heinemeyer, W.~Hollik, G.~Weiglein and L.~Zeune,
  JHEP {\bf 1312} (2013) 084
  [arXiv:1311.1663 [hep-ph]].
  
  
\bibitem{Degrassi:2012ry} 
  G.~Degrassi, S.~Di Vita, J.~Elias-Miro, J.~R.~Espinosa, G.~F.~Giudice, G.~Isidori and A.~Strumia,
  JHEP {\bf 1208}, 098 (2012)
  doi:10.1007/JHEP08(2012)098
  [arXiv:1205.6497 [hep-ph]].

\bibitem{Carena:2013ytb}
  M.~Carena, S.~Heinemeyer, O.~St{\aa}l, C.~E.~M.~Wagner and G.~Weiglein,
  Eur.\ Phys.\ J.\ C {\bf 73} (2013) no.9,  2552
  doi:10.1140/epjc/s10052-013-2552-1
  [arXiv:1302.7033 [hep-ph]].
    
\bibitem{CMS:pas-hig-16-007}
CMS Collaboration, ``Summary results of high mass BSM Higgs searches using CMS run-I data ,''  CMS-PAS-HIG-16-007, \url{https://cds.cern.ch/record/2142432}. 

\bibitem{bib:atlas_higgsself_bbgg}
ATLAS Collaboration, ``Prospects for measuring Higgs pair production in the channel $H(\to \gamma\gamma)H(\to b\bar{b})$ using the ATLAS detector at the HL-LHC,'' ATLAS-PHYS-PUB-2014-019 (2014), \url{http://cds.cern.ch/record/1956733}.

\bibitem{bib:atlas_higgsself_bbtt}
ATLAS Collaboration, ``Higgs pair production in the $H(\to \tau\tau)H(\to b\bar{b})$ channel at the High-Luminosity LHC,'' ATLAS-PHYS-PUB-2015-046 (2015), \url{http://cds.cern.ch/record/2065974}.

\bibitem{bib:cms_higgsself}
CMS Collaboration, ``Technical Proposal for the Phase-II Upgrade of the Compact Muon Solenoid,'' CERN-LHCC-2015-010 (2015), \url{http://cds.cern.ch/record/2020886}.

\bibitem{Aad:2014vma}
{\bf ATLAS Collaboration}, G.~Aad {\em et al.}, ``{Search for direct production
  of charginos, neutralinos and sleptons in final states with two leptons and
  missing transverse momentum in $pp$ collisions at $\sqrt{s} =$ 8 TeV with the
  ATLAS detector }'' {\em JHEP} {\bf 1405}  071.

\bibitem{ATLAS:2013hta}
{\bf ATLAS Collaboration}, ``{Physics at a High-Luminosity LHC with ATLAS }''
\href{http://arxiv.org/abs/1307.7292}{{\tt arXiv:1307.7292 [hep-ex]}}.

\bibitem{Heister:2002mn}
{\bf ALEPH Collaboration}, A.~Heister {\em et al.}, ``{Search for charginos
  nearly mass degenerate with the lightest neutralino in e+ e- collisions at
  center-of-mass energies up to 209-GeV }'' {\em Phys.Lett.} {\bf B533}
  223--236.

\bibitem{Abdallah:2003xe}
{\bf DELPHI Collaboration}, J.~Abdallah {\em et al.}, ``{Searches for
  supersymmetric particles in e+ e- collisions up to 208-GeV and interpretation
  of the results within the MSSM }'' {\em Eur.Phys.J.} {\bf C31}  421--479.

\bibitem{Abbiendi:2002vz}
{\bf OPAL Collaboration}, G.~Abbiendi {\em et al.}, ``{Search for nearly mass
  degenerate charginos and neutralinos at LEP }'' {\em Eur.Phys.J.} {\bf C29}
  479--489.

\bibitem{bib:ATLAS_bRPV}
ATLAS Collaboration, ``Search for supersymmetry at sqrts=7~TeV~in final states with large jet multiplicity, missing transverse momentum and one isolated lepton with the ATLAS detector,'' ATLAS-CONF-2012-140, \url{http://cds.cern.ch/record/1483511}.

\bibitem{Baer:2013ula}                                                     
H.~Baer and J.~List, 
\href{http://dx.doi.org/10.1103/PhysRevD.88.055004}{ Phys. Rev. {\bfseries D88} (2013) 055004}, 
[\href{http://arxiv.org/abs/1307.0782}{{\ttfamily arXiv:1307.0782 [hep-ph]}}].                            

\bibitem{Baer:2013yha}
  H.~Baer, V.~Barger, P.~Huang, D.~Mickelson, A.~Mustafayev, W.~Sreethawong and X.~Tata,
  Phys.\ Rev.\ Lett.\  {\bf 110} (2013) no.15,  151801
  doi:10.1103/PhysRevLett.110.151801
  [arXiv:1302.5816 [hep-ph]].

\bibitem{Baer:2014kya}
  H.~Baer, A.~Mustafayev and X.~Tata,
  Phys.\ Rev.\ D {\bf 90} (2014) no.11,  115007
  doi:10.1103/PhysRevD.90.115007
  [arXiv:1409.7058 [hep-ph]].

\bibitem{Baer:2016wkz}
  H.~Baer, V.~Barger, J.~S.~Gainer, P.~Huang, M.~Savoy, D.~Sengupta and X.~Tata,
  arXiv:1612.00795 [hep-ph].
 
 \bibitem{Ahmed:2015uqt} 
  A.~Ahmed, B.~M.~Dillon, B.~Grzadkowski, J.~F.~Gunion and Y.~Jiang,
  arXiv:1512.05771 [hep-ph].

\bibitem{Senaha:2013fva}
  E.~Senaha,
  arXiv:1305.1563 [hep-ph].

 

\bibitem{Malm:2014gha}
R.~Malm, M.~Neubert, and C.~Schmell, ``{Higgs Couplings and Phenomenology in a
  Warped Extra Dimension}''
  \href{http://dx.doi.org/10.1007/JHEP02(2015)008}{{\em JHEP} {\bf 02} (2015)
  008},
\href{http://arxiv.org/abs/1408.4456}{{\tt arXiv:1408.4456 [hep-ph]}}.


\bibitem{Ellwanger:2016qax}
  U.~Ellwanger and C.~Hugonie,
  JHEP {\bf 1605} (2016) 114
  doi:10.1007/JHEP05(2016)114
  [arXiv:1602.03344 [hep-ph]].
  

\end{thebibliography}

\end{footnotesize}

\end{document}